\documentclass[12pt,epsf]{article}
\usepackage[english]{babel}
\usepackage{graphics}
\usepackage{amssymb}
\usepackage{epsf}
\usepackage{epsfig}
\usepackage{rotating}
\usepackage{axodraw}
\usepackage{pstricks}

\def\d{\delta}
\def\w{\omega}

\def\a{\alpha}
\def\b{\beta}

\def\g{\gamma}

\def\m{\mu}

\def\ds#1{#1\kern-1ex\hbox{/}}
\def\dsh{h\kern-1.2ex /}

\newcommand{\bea}{\begin{eqnarray}}
\newcommand{\eea}{\end{eqnarray}}

\def\nn{\nonumber}
\def\beq{\begin{equation}}
\def\eeq{\end{equation}}

\def\beqn{\begin{eqnarray}}
\def\eeqn{\end{eqnarray}}
\def\ba{\begin{eqnarray}}
\def\ea{\end{eqnarray}}

\def\m{{\tt -}}

\def\p{{\tt +}}

\def\slash#1{#1\hskip-6pt/\hskip6pt}

\newcommand{\beqa}{\begin{eqnarray}}
\newcommand{\eeqa}{\end{eqnarray}}

\newcommand{\la}{\lambda}

\begin{document}

\begin{center}
\vspace{1.cm}
{\bf\large
 Trilinear  Anomalous Gauge Interactions from Intersecting Branes \\
 and the Neutral Currents Sector\\}
\vspace{1.5cm}
{\bf\large $^a$Roberta Armillis, $^{a, b}$Claudio Corian\`{o} and $^{a}$Marco Guzzi}

\vspace{1cm}

{\it $^a$Dipartimento di Fisica, Universit\`{a} del Salento \\
and  INFN Sezione di Lecce,  Via Arnesano 73100 Lecce, Italy}\\
\vspace{.5cm}
{\it $^b$ Department of Physics and Institute of Plasma Physics \\
University of Crete, 71003 Heraklion, Greece}\\

\begin{abstract}
We present a study of the trilinear gauge
interactions in extensions of the Standard Model (SM) with several anomalous
extra $U(1)$'s, identified in various constructions,
from special vacua of string theory to large extra dimensions.
In these models  an axion and generalized Chern-Simons interactions for anomalies cancellation are present.
We derive generalized Ward identities for these vertices and discuss their structure in the St\"uckelberg and Higgs-St\"uckelberg phases. We give their explicit expressions in all the relevant cases,
which can be used for phenomenological studies of these models at the LHC.
\end{abstract}
\end{center}
\newpage
\section{Introduction}

Models of intersecting branes (see \cite{Kiritsis} for an overview) have been under an intense theoretical scrutiny in the last several years. The motivations for studying this class of theories are manifolds, being them obtained
from special vacua of string theory, for instance from the orientifold construction \cite{AKRT,BL,Ibanez}. Their generic gauge structure is of the form $SU(3)\times SU(2)\times U(1)_Y\times U(1)^p$, where the symmetry of the Standard Model (SM) is enlarged with a certain number of extra abelian factors $(p)$. Several phenomenological studies \cite{CIK,CIM1,CIM2,CI,GIIQ,Leontaris} have allowed to characterize their general structure, whose string origin has been analyzed at an increasing level of detail \cite{ABDK,Pascal2} down to more direct issues, connected with their realization as viable theories beyond the SM.
Related studies of the St\"uckelberg field \cite{KN1} in a non-anomalous context have clarified this mechanism of mass generation and analyzed some of its implications at colliders both in the SM and in its supersymmetric extensions.

In scenarios with extra dimensions where the interplay between anomaly cancellations in the bulk and on the boundary branes is critical for their consistency, very similar models could be obtained following the construction of
\cite{Hill}, with a suitable generalization in order to generate at low energy a non abelian gauge structure.

Specifically, the role played by the extra $U(1)$'s at low energy in theories of this type after electroweak symmetry breaking has been addressed in \cite{CIK,CIM1,CIM2}, where some of the quantum features of their effective action have been clarified. These, for instance, concern the phases
of these models, from their defining phase, the St\"uckelberg phase, being the anomalous
$U(1)$ broken at low energy but with a gauge symmetry  restored by shifting (St\"uckelberg)
axions,  down to the electroweak phase - or Higgs-St\"uckelberg phase, (HS) - where the vev's of the Higgs
of the SM combine with the St\"uckelberg axions to produce a physical axion \cite{CIK} and a certain
number of goldstone modes. The axion in the low energy effective action is interesting both for collider
physics and for cosmology \cite{CI}, working as a modified Peccei-Quinn (PQ) axion. In this respect some interesting proposals to explain an anomaly in gamma ray propagation as seen by MAGIC \cite{Roncadelli} using a pseudoscalar (axion-like) has been presented recently, while more experimental searches of effects of this type are planned for the future by several collaborations using Cerenkov telescopes (see \cite{Roncadelli} for more details and references). Other
interesting revisitations of the traditional Weinberg-Wilczek axion \cite{Zurab1} to evade the astrophysical constraints  and
in the context of Grand Unification/mirror worlds  \cite{Zurab} may well deserve attention in the future and be analyzed within the framework that we outline below. At the same time, comparisons between anomalous and non anomalous string
constructions of models with extra $Z^{\prime}$s should also be part of this analysis \cite{Alon}.

The presence of axion-like particles in effective theories is, in general, connected to an anomalous gauge structure, but for reasons which may be  rather different and completely unrelated, as discussed in \cite{CI}. For the rest, though, the study of the perturbative expansion in theories of this type is rather general and shows some interesting features that deserve a careful analysis. In \cite{CIM1,CIM2} several steps in the analysis of the perturbative expansion have been performed. In particular
it has been shown how to organize the loop expansion in a gauge-invariant way in $1/M_1$, where $M_1$ is the
St\"uckelberg mass. A way to address this point is to use a typical $R_\xi$ gauge and follow the pattern of cancellation of the gauge parameter in order to characterize it. This has been done up to 3-loop level in a simple $U(1)\times U(1)$ model where one of the two $U(1)$'s is anomalous.

The St\"uckelberg symmetry is responsible for rendering
the anomalous gauge bosons massive (with a mass $M_1$) before electroweak symmetry breaking. A second scale $M$ controls the interaction of the axions with the gauge fields but is related to the first by a condition of gauge invariance in the effective action \cite{CI}. In general, for a theory with several $U(1)$'s, there is an independent mass scale for each St\"uckelberg field.

In the case of a complete extension of the SM incorporating anomalous
$U(1)$'s, all the neutral current sectors, except for the photon current, acquire an anomalous contribution that modifies the trilinear (chiral) gauge
interactions. For the $Z$ gauge boson this anomalous component
decouples as $M_1$ gets large, though it remains unspecified. For instance, in theories containing extra dimensions it could even be of the order of 10 TeV's or so, in general being of the order of $1/R$, where $R$ is the radius of compactification. In other constructions \cite{Ibanez} based on toroidal compactifications with branes wrapping around the extra dimensions, their masses and couplings are expressed in terms of a string scale $M_s$ and of the integers characterizing the wrappings \cite{GIIQ}.
Beside the presence of the extra neutral currents, which are common to all the models with extra abelian gauge structures, here, in addition,
the presence of chiral anomalies leaves some of the trilinear interactions to contribute even in the massless fermion (chiral) limit, a feature which is completely absent  in the SM, since in
the chiral limit these vertices vanish.

 As we are going to see, the analysis of these vertices
 is quite delicate, since their behaviour is essentially controlled by the mass differences within a given fermion generation \cite{CIM2}, and for this reason they are sensitive
 both to spontaneous and to chiral symmetry breaking. The combined role played by these sources of breaking is not unexpected, since any
 pseudoscalar induced in an anomalous theory feels both the structure of the QCD
 vacuum and of the electroweak sector, as in the case of the Peccei-Quinn (PQ) axion.
In this work we are going to proceed with a general analysis of these vertices, extending the discussion in \cite{CIM2}. Our analysis here is performed at a field theory level, leaving the phenomenological discussion to a companion work.
Our analysis is organized as follows.

After a brief summary on the structure
of the effective action, which has been included to make our treatment self-contained,
we analyze the Slavnov-Taylor identities of the theory,
focusing our attention on the trilinear gauge boson vertices. Then we characterize the structure
of the $Z\gamma\gamma$ and $Z Z \gamma$ vertices away from the chiral limit, extending
the discussion presented in \cite{CIM2}. In particular we clarify when the CS terms can be absorbed by a re-distribution of the anomaly before moving
away from the chiral limit. In models containing several anomalous $U(1)$'s different theories are identified by the different partial anomalies associated to the trilinear gauge interactions involving at least three extra $Z^{\prime}$s.
In this case the CS terms are genuine components which are specific for a given model and are
accompanied by  a specific set of axion counterterms. Symmetric distributions of the partial anomalies are sufficient to exclude all the CS terms, but these particular assignments may not be general enough.

Away from the chiral limit, we show how the mass dependence
of the vertices is affected by the external Ward identity, which is a generic
feature of anomalous interactions for nonzero fermion masses. This point is worked
out using chiral projectors and counting the mass insertions into each vertex. On the basis of
this study we are able to formulate general and simple rules which allow to handle quite
straightforwardly all the vertices of the theory. We conclude with some phenomenological comments concerning the possibility of future studies of these theories at the LHC. In an appendix we present the Faddeev-Popov lagrangean of the model, which has not been given before, and that can be useful for further studies of these theories.

\subsection{Construction of the effective action}
The construction of the effective action, from the field theory point of
view, proceeds as follows \cite{CIK,CIM2}.

\begin{figure}[h]
{\centering \resizebox*{12cm}{!}{\rotatebox{0}
{\includegraphics{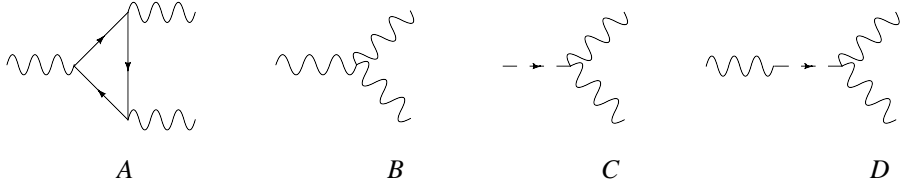}}}\par}
\caption{Counterterms allowed in the low energy effective action in the chiral limit: anomalous contributions (A), CS interaction (B), WZ term (C) and $B-b$ mixing contribution (D). In particular the bilinear mixing of the axions with the gauge fields is vanishing only for on-shell vertices and is removed in the $R_\xi$ gauge in the WZ case. A discussion of this term and its role in the GS mechanism can be found in \cite{CGM}.}
\label{counter}
\end{figure}

One introduces a set of counterterms in the form of CS and WZ operators
and requires that the effective action is gauge invariant at 1-loop. Each anomalous $U(1)$ is accompanied by an axion, and every gauge variation of the anomalous gauge field can be cancelled by the corresponding WZ term. The remaining  anomalous gauge variations are cancelled by CS counterterms. A list of typical vertices and counterterms is shown in
Fig. \ref{counter}.

We consider the simplest anomalous extension of the SM with a gauge structure of the form $SU(3)\times SU(2) \times U(1)_Y \times U(1)_B$ model with a single anomalous $U(1)_B$. The anomalous contributions are those involving the $B$ gauge boson and involve the trilinear (triangle) vertices $BBB$, $BYY,$ $BBY,$ $BWW$ and $BGG$, where $W$'s and the $G$'s are the $SU(2)$ and $SU(3)$ gauge bosons respectively. All the remaining trilinear interactions mediated by fermions are anomaly-free and therefore vanish in the massless limit.
Therefore the axion ($b$) associated to $B$ appears in abelian counterterms
of the form $b F_B\wedge F_B, b F_B \wedge F_Y,b F_Y \wedge F_Y $ and in the analogous non-abelian ones $b Tr W\wedge W$ and $b Tr G\wedge G$. In the absence of a kinetic
term for the axion $b$, its role is unclear: it allows to ``cancel''
the anomaly but can be gauged away. As emphasized by Preskill \cite{Preskill}, the role of
the WZ term is, at this stage, just to allow a consistent
power counting in the perturbative expansion, hinting that an anomalous
theory is non-renormalizable, but, for the rest, unitary below a certain scale.
Theories of this type are in fact characterized by a unitarity
bound since local a counterterm is not sufficient to erase the
bad high energy behaviour of the anomaly \cite{CGM}.
Although the structure
of the vertices constructed in this work is identified using the
WZ effective action at the lowest order (using only the axion counterterm),
their extension to the Green-Schwarz case is straightforward.
In this second case the vertices here defined need to be modified with
the addition of extra massless poles on the external gauge lines.

The b field remains unphysical even in the presence of a St\"uckelberg mass
term for the B field, $\sim(\partial b - M B)^2$
since the gauge freedom remains and it is then natural to interpret $b$ as a Nambu-Goldstone mode. In a physical gauge it can be set to vanish.

Things change drastically when the B field mixes with the other scalars
of the Higgs sector of the theory. In this case a linear combination of
$b$ and the remaining CP-odd phases (goldstones) of the Higgs
doublets becomes physical and is called the axi-Higgs. This happens only in specific potentials
characterized  also by a global $U(1)_{PQ}$ symmetry ($V_{PQ}$) \cite{CIK} which are, however,
sufficiently general. In the absence of Higgs-axion mixing the CP odd goldstone modes of the broken theory, after electroweak symmetry breaking, are just linear combinations of the St\"uckelberg and of the goldstone mode of the Higgs potential and no physical axion appears in the spectrum. For potentials that allow a physical axion, even in the massless case, the axion mass can be lifted by the QCD vacuum due to instanton effects exactly as for the Peccei-Quinn axion, but now the spectrum allows an axion-like particle.

\subsection{\bf Anomaly cancellation in the interaction eigenstate basis, CS terms and regularizations}

The anomalies of the model are cancelled in the interaction eigenstate basis of
$(b,A_Y,B, W)$ and the CS and WZ terms are fixed at this stage. The B field
is massive and mixes with the axion, but the gauge symmetry is still
intact. The Ward identities of the theory for the triangle diagrams assume a nontrivial form due to the $B\partial b$ mixing. In the case
of on-shell trilinear vertices one can show that these mixing terms vanish.

The CS counterterms are necessary in order to cancel the gauge variations
of the $Y, W$ and $G$ gauge bosons in anomalous diagrams involving the interaction with $B$. These are the diagrams mentioned before. The role of these
terms is to render vector-like at 1-loop all the currents which become
anomalous in the interaction with the $B$ gauge boson. For instance,
in a triangle such as $YBB$, the $A_Y B\wedge F_B$ CS term effectively
``moves'' the chiral projector from the Y vertex to the B vertex
symmetrically on the two B's, assigning the anomalies to the B vertices.
These will then be cancelled by the axion $b$ via a suitable WZ term ($b F_B\wedge F_Y$).

The effective action has the structure given by
\beqn
{\mathcal S} &=&   {\mathcal S}_0 +{\mathcal S}_{an} + {\mathcal S}_{WZ} + {\mathcal S}_{CS}
\label{defining}
\eeqn

where ${\mathcal S}_0$ is the classical action. It is a canonical gauge theory with dimension-4 operators whose explicit structure can be found in \cite{CIM2}. In Eq. (\ref{defining}) the anomalous contributions coming from the 1-loop triangle diagrams involving abelian and non-abelian gauge interactions are summarized by the expression
\beqn
{\mathcal S}_{an}&=& \frac{1}{2!} \langle T_{BWW} BWW \rangle +  \frac{1}{2!} \langle T_{BGG} BGG \rangle
 + \frac{1}{3!} \langle T_{BBB} BBB \rangle     \nonumber\\
&&+ \frac{1}{2!} \langle T_{BYY} BYY \rangle + \frac{1}{2!} \langle T_{YBB} YBB \rangle,
\eeqn
where the symbols $\langle \rangle$ denote integration \cite{CIM1}. In the same notations
the Wess Zumino (WZ) counterterms are given by
\beqn
{\mathcal S}_{WZ}&=& \frac{C_{BB}}{M} \langle b  F_{B} \wedge F_{B}  \rangle
+ \frac{C_{YY}}{M} \langle b F_{Y} \wedge F_{Y}  \rangle + \frac{C_{YB}}{M} \langle b F_{Y} \wedge F_{B}  \rangle \nonumber\\
&&+ \frac{F}{M} \langle b Tr[F^W \wedge F^W]  \rangle  +  \frac{D}{M} \langle b Tr[F^G \wedge F^G] \rangle,
\eeqn
and the gauge dependent CS abelian and non abelian counterterms \cite{Pascal2} needed to cancel
the mixed anomalies involving a B line with any other gauge interaction of the
SM take the form
\beqn
{\mathcal S}_{CS}&=&+ d_{1} \langle BY \wedge F_{Y} \rangle + d_{2} \langle YB \wedge F_{B} \rangle  \nonumber\\
&&+ c_{1} \langle \epsilon^{\mu\nu\rho\sigma} B_{\mu} C^{SU(2)}_{\nu\rho\sigma} \rangle
+ c_{2} \langle \epsilon^{\mu\nu\rho\sigma} B_{\mu} C^{SU(3)}_{\nu\rho\sigma} \rangle.
\eeqn

Explicitly
\ba
\langle T_{B W W} B W W \rangle&\equiv& \int dx\, dy \, dz
  T^{\lambda \mu \nu, ij}_{BWW}(z,x,y) B^{\lambda}(z) W^{\mu}_{i}(x)
W^{\nu}_{j}(y)
\ea
and so on.

The non-abelian CS forms are given by
\beqn
C^{SU(2)}_{\mu \nu \rho} &=&  \frac{1}{6} \left[ W^{i}_{\mu} \left( F^W_{i,\,\nu \rho} + \frac{1  }{3} \, g^{}_{2}
\, \varepsilon^{ijk} W^{j}_{\nu} W^{k}_{\rho}  \right) + cyclic   \right]              ,    \\
C^{SU(3)}_{\mu \nu \rho} &=&  \frac{1}{6} \left[ G^{a}_{\mu} \left( F^G_{a,\,\nu \rho} + \frac{1 }{3} \, g^{}_{3}
\, f^{abc} G^{b}_{\nu} G^{c}_{\rho}  \right) + cyclic  \right].
\eeqn

In our conventions, the field strengths are defined as
\beqn
F^W_{i, \,\mu \nu} &=& \partial_{\mu} W^{i}_{\nu} - \partial_{\nu} W^{i}_{\mu}
-  g^{}_{2} \varepsilon_{ijk} W^{j}_{\mu} W^{k}_{\nu}
=  \hat{F}^{W}_{i,\, \mu \nu}-  g^{}_{2} \varepsilon_{ijk} W^{j}_{\mu} W^{k}_{\nu} \\
F^G_{a,\,\mu \nu} &=& \partial_{\mu} G^{a}_{\nu} - \partial_{\nu} G^{a}_{\mu}
 -  g^{}_{3} f_{abc} G^{b}_{\mu} G^{c}_{\nu} = \hat{F}^{G}_{a,\, \mu \nu} -  g^{}_{3} f_{abc} G^{b}_{\mu} G^{c}_{\nu},
\eeqn
whose variations under non-abelian gauge transformations are
\beqn
\delta_{SU(2)} C^{SU(2)}_{\mu \nu \rho} &=& \frac{1}{6}  \left[ \partial^{}_{\mu} \theta^{i} \,( \hat{F}^{W}_{i, \,\nu \rho})
+ cyclic \right],      \\
\delta_{SU(3)} C^{SU(3)}_{\mu \nu \rho} &=& \frac{1}{6} \left[ \partial^{}_{\mu} \vartheta^{a} \,( \hat{F}^{G}_{a,\,\nu \rho})
+ cyclic \right],
\eeqn

where $\hat{F}$ denotes the ``abelian'' part of the non-abelian field strength.

Coming to the formal definition of the effective action,  interpreted as the generator of the 1-particle irreducible diagrams with external classical fields, this is defined, as usual,  as a linear combination of correlation functions with an arbitrary number of external lines of the form $A_Y, B, W, G$,
that we will denote conventionally as $\mathcal{W}(Y,B,W)$. It is given by
\beqn
W[Y,B,W,G] &=&
 \sum^{\infty}_{n_{1} = 1}  \sum^{\infty}_{n_{2} = 1} \frac{i^{\,n_1 + n_2}}{n_{1}! n_{2}!} \int dx_{1}...dx_{n_{1}}
dy_{1}...dy_{n_{2}} T^{\lambda_{1}...\lambda_{n_{1}} \mu_{1}...\mu_{n_{2}}}(x_{1}...x_{n_1}, y_1...y_{n_2})   \nonumber\\
&& \hspace{3.5cm}  B^{\lambda_1}(x_1)...B^{\lambda_{n_1}}(x_{n_1})
A_{Y\mu_1}(y_1)...A_{Y\mu_{n_2}}(y_{n_2}) + ... \nonumber
\eeqn
where we have explicitly written only its abelian part and the ellipsis refer to the additional non abelian
or mixed (abelian/non-abelian) contributions. We will be using the invariance of the effective action
under re-parameterizations of the external fields to obtain information on the trilinear vertices of the
theory away from the chiral limit. Before coming to that point, however, we show how to fix the structure of the counterterms exploiting its BRST symmetry. This will allow to derive simple STI's for the action involving the anomalous vertices.

\section{BRST conditions in the St\"uckelberg and HS phases}
We show in this section how to fix the counterterms of the effective action by imposing directly the STI's on its anomalous vertices
in the two broken phases of the theory, thereby removing the Higgs-axion mixing of the low energy effective theory. As we have already mentioned, the lagrangean of the St\"uckelberg phase contains a coupling of the St\"uckelberg field to the gauge field which is typical of a goldstone mode. In \cite{CIM1,CIM2} this mixing has been removed and the WZ counterterms have been computed in a particular gauge, which is a typical $R_\xi$ gauge with $\xi=1$. Here we start by showing that this way of fixing the counterterms is equivalent to require that the
trilinear interactions of the theory in the St\"uckelberg phase satisfy a generalized Ward identity (STI).

After electroweak symmetry breaking, in general one would be needing a second gauge choice, since the new breaking would again re-introduce bilinear derivative couplings of the new goldstones to the gauge fields. So the question to ask is if the STI's of the first phase, which fix completely the counterterms of the theory and remove the b-B mixing, are compatible with the STI's of the second phase, when we remove the coupling of the gauge bosons to their goldstones. The reason for asking these questions is obvious: it is convenient to fix the counterterms once and for all in the effective lagrangeans and this can be more easily done in the
 St\"uckelberg phase or in the HS phase depending on whether we need the effective action either expressed in terms of interactions or of  mass eigenstates respectively. In both cases we need generalized Ward identities which are {\em local}.
 The presence of bilinear mixings on the external lines of the 3-point functions would render the analysis of these interactions more complex and essentially non-local.

 This point is also essential in our identification of the effective vertices of the physical gauge bosons since, as we will discuss below, the definition of these vertices is entirely based on the possibility of parameterizing the anomalous effective action, at the same time, in the interaction base and in the mass eigenstate basis. We need these mixing terms to disappear in both cases. This happens, as we are going to show, if both in the St\"uckeberg phase and in the HS phase we perform a gauge choice of $R_\xi$ type (we will choose $\xi=1$).
 These technical points are easier to analyze in a simple abelian model, following the lines of \cite{CIM1}.
In this model the  $B$ is a vector-axial vector (${\bf V-A}$) anomalous gauge boson and $A $ is vector-like and anomaly-free.

We will show that in this model we can fix the counterterms in the first phase, having removed the b-B mixing and then proceed to
determine the effective action in the HS phase, with its STI's which continue to be valid also in this phase.

\begin{figure}[t]
  \begin{center}
\includegraphics[width=13cm]{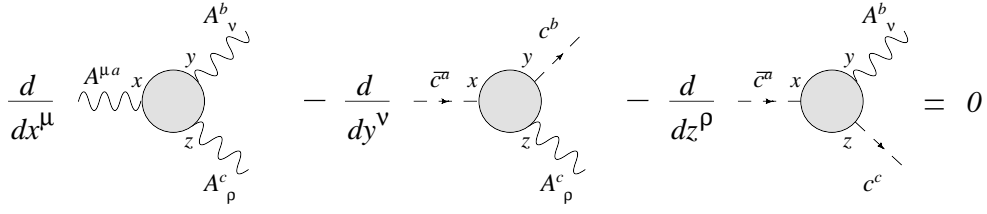}\\
\caption{ Graphical representation of Eq. (\ref{STI}) at any perturbative order.}
  \label{STIdentity}
  \end{center}
\end{figure}

Let's illustrate this point in some detail.
We recall that for an ordinary (non abelian) gauge theory in the exact (non-broken) phase the derivation of the conditions of BRST invariance follow from the well known BRST variations in the $R_\xi$ gauge

\bea
\delta_{BRST} \, A_{\mu}^a &\equiv& s A_{\mu}^a = \omega \mathcal{D}_{\mu}^{ab} c_b
\label{YMBRSTA}\\
\delta_{BRST} \, c^a &\equiv& s c^a = - \frac{1}{2} \omega g f^{abc} c_b c_c
\label{YMBRSTghost}\\
\delta_{BRST} \, \bar{c}\,^a &\equiv& s \bar{c}\,^a = \frac{\omega}{\xi } \partial_{\mu} A^{\mu\, a }.
\label{YMBRSTantighost}
\eea
These involve the nonabelian gauge field $A_\mu^a$, the ghost ($c^a$) and antighost
($\bar{c}^a$) fields, with $\omega$ being a Grassmann parameter.
We will be interested in trilinear correlators whose STI's are arrested at 1-loop level and which involve anomalous diagrams.
For instance we could use the invariance of a specific correlator ($\bar{c} AA$ ) under a BRST transformation in order to obtain the generalized WI's for trilinear gauge interactions
\bea
s \, \langle 0 | T\, \bar{c}^a(x) A^b_{\nu}(y) A^c_{\rho}(z) | 0 \rangle =0.
\eea
These are obtained from the relations (\ref{YMBRSTantighost}) rather straightforwardly

\bea
s \, \langle 0 | T\, \bar{c}^a(x) A^b_{\nu}(y) A^c_{\rho}(z) | 0 \rangle &=&
\langle 0 | T\, (s \bar{c}^a(x)) A^b_{\nu}(y) A^c_{\rho}(z) | 0 \rangle +\nonumber \\
+ \langle 0 | T\, \bar{c}^a(x) (s A^b_{\nu}(y)) A^c_{\rho}(z) | 0 \rangle
& + & \langle 0 | T\, \bar{c}^a(x) A^b_{\nu}(y) (s A^c_{\rho}(z)) | 0 \rangle =0.\nonumber \\
\eea
In fact, by using Eqs. (\ref{YMBRSTA}) and (\ref{YMBRSTantighost}) we obtain
\bea
s \, \langle 0 | T\, \bar{c}^a(x) A^b_{\nu}(y) A^c_{\rho}(z) | 0 \rangle &=&
\frac{1}{\xi} \langle 0 | T\, \omega \partial_{\mu} A^{\mu\, a } A^b_{\nu}(y) A^c_{\rho}(z) | 0 \rangle +\nonumber \\
+ \langle 0 | T\, \bar{c}^a(x) \omega \mathcal{D}_{\nu}^{bl}c_l(y) A^c_{\rho}(z) | 0 \rangle
&+& \langle 0 | T\, \bar{c}^a(x) A^b_{\nu}(y) \omega\mathcal{D}_{\rho}^{cm}c_m(z) | 0 \rangle =0.\nonumber \\
\eea
Choosing $\xi=1$ we get
\bea
&& \frac{\partial}{\partial x^{\mu}} \langle 0 | T\, A^{\mu\, a }(x) A^b_{\nu}(y) A^c_{\rho}(z) | 0 \rangle \nonumber \\
&+& \langle 0 | T\, \bar{c}^a(x)  [\delta^{bl} \partial_{\nu} - g f^{bld} A_{\nu\,d}(y) ]c_l(y) A^c_{\rho}(z) | 0 \rangle
\nonumber \\
&+& \langle 0 | T\, \bar{c}^a(x) A^b_{\nu}(y) [\delta^{cm} \partial_{\rho} - g f^{cmr} A_{\rho\,r}(z) ]c_m(z) | 0 \rangle
 =0. \nonumber \\
\eea
The two fields
 $A_{\nu\,d}(y) c_l(y)$ e $A_{\rho\,r}(z) c_m(z)$ on the same spacetime point do not contribute on-shell and integrating by parts on the second and third term we obtain
\bea
\frac{\partial}{\partial x^{\mu}} \langle 0 | T\, A^{\mu\, a } A^b_{\nu}(y) A^c_{\rho}(z) | 0 \rangle -
\frac{\partial}{\partial y^{\nu}} \langle 0 | T\, \bar{c}^a(x)  c^b(y) A^c_{\rho}(z) | 0 \rangle
- \frac{\partial}{\partial z^{\rho}} \langle 0 | T\, \bar{c}^a(x) A^b_{\nu}(y) c^c(z) | 0 \rangle
 =0, \nonumber \\
\label{STI}
\eea
which is described diagrammatically in Fig. \ref{STIdentity}.
%
Let's now focus our attention on the A-B model of
\cite{CIM1} where we have an anomalous generator $Y_B$. This model describes quite well many of the properties of the abelian sector of the general model discussed in \cite{CIM2} with a single anomalous $U(1)$.
It is an ordinary gauge theory of the form $U(1)_A\times U(1)_B$ with $B$ made massive at tree level by the St\"uckelberg term
\bea
\mathcal{L}_{St}=\frac{1}{2}(\partial_{\mu} b + M_1 B_{\mu})^2.
\eea
This term introduces a mixing $M_1 B_{\mu}\partial^{\mu} b$ which signals the presence of a broken phase in the theory. Introducing the gauge fixing lagrangean
\bea
\mathcal{L}_{gf} = - \frac{1}{2\xi_B}(\mathcal{F}^S_B[B_{\mu}])^2,
\eea

\bea
\mathcal{F}^S_B[B_{\mu}] \equiv\partial_{\mu} B^{\mu} - \xi_B M_1 b,
\eea
we obtain the partial contributions (mass term plus gauge fixing term)
to the total action
\bea
\mathcal{L}_{St} + \mathcal{L}_{gf} = \frac{1}{2} \biggl [(\partial_{\mu} b)^2 + M_1^2 B_{\mu} B^{\mu}
- (\partial_{\mu} B^{\mu})^2 - \xi_B M_1^2 b^2  \biggr]
\label{LStueckelberg}
\eea
and the corresponding Faddeev-Popov lagrangean
\bea
\mathcal{L}_{FP} = \bar{c}_B \, \frac{ \delta \mathcal{F}_B }{ \delta \theta_B } \,c_B
 = \bar{c}_B \, \biggl[ \partial_{\mu} \frac{ \delta B^{\mu} }{ \delta \theta_B }
 - \xi_B M_1 \, \frac{ \delta b }{ \delta \theta_B } \, \biggr ] c_B,
 \label{Faddeev}
\eea
with $c_B$ and $\bar{c}_B$ are the anticommuting ghost/antighosts fields.  It can be written as
\bea
\mathcal{L}_{FP} = \bar{c}_B \, (\square + \xi_B M_1^2) \, c_B,
\eea
having used the shift of the axion under a gauge transformation
\bea
\delta b = - M_1 \theta.
\label{deltab}
\eea
In the following we will choose $\xi_B=1$. The anomalous sector is described by
\beqa
\mathcal{S}_{an} &=& \mathcal{S}_1 + \mathcal{S}_3 \nonumber \\
\mathcal{S}_{1} &=&  \int d x \, d y \, d z \, \left( \frac{g^{}_{B} \, g^{2}_{A}}{2!} \,
T_{\bf AVV}^{\la\mu\nu}(x,y,z) B_{\la}(z) A_{\mu}(x)A_{\nu}(y)
 \right)\nonumber \\
\mathcal{S}_3 &=&  \int d x \, d y \, d z \, \left( \frac{ g^{3}_{B}}{3!} \,
 T_{\bf AAA}^{\la\mu\nu}(x,y,z) B_\la(z) B_\mu(x) B_\nu(y) \right),
\nonumber \\
\eeqa
where we have collected all the anomalous diagrams of the form ({\bf AVV} and {\bf AAA}) and whose gauge variations are
\beqa
 \frac{1}{2!} \delta_B  \left[ T_{\bf AVV} BAA \right] &=&  \frac{i}{2!} a_3(\beta) \frac{1}{4} \left[ F_A \wedge F_A
\theta_B \right] \nonumber\\
 \frac{1}{3!} \delta_B\left[ T_{\bf AAA} BBB \right] &=&  \frac{i}{3!} \frac{a_n}{3} \frac{3}{4} \langle  F_B \wedge F_B
\theta_B \rangle,
\eeqa
having left open the choice over the parameterization of the loop momentum,
denoted by the presence of the arbitrary parameter $\beta$ with
\beq
a_3(\beta)=- \frac{i}{4 \pi^2} + \frac{i}{2 \pi^2} \beta\qquad a_3\equiv \frac{a_n}{3}=-\frac{i}{6 \pi^2},
\eeq
while
\beqa
\frac{1}{2!}\delta_A\left[ T_{\bf AVV} BAA\right] =  \frac{i}{2!}  a_1(\beta) \frac{2}{4} \left[ F_B \wedge F_A
\theta_A \right].
\eeqa
We have the following equations for the anomalous variations
\beqa
\delta_B \mathcal{L}_{an} &=&  \frac{i g^{}_{B} g^{\,2}_{A}}{2!} \, a_3(\beta) \frac{1}{4}  F_A \wedge F_A  \theta_B
+  \frac{ i g^{\,3}_{B} }{3!}\, \frac{a_n}{3}  \frac{3}{4} F_B \wedge F_B \theta_B
\nonumber \\
\delta_A \mathcal{L}_{an} &=&  \frac{i g^{}_{B} g^{\,2}_{A}}{2! } \, a_1(\beta) \frac{2}{4} F_B \wedge F_A \theta_A,
\eeqa

while $\mathcal{L}_{b,c}$, the axionic contributions (Wess-Zumino terms) needed to
restore the gauge symmetry violated at 1-loop level, are given by
\beq
\mathcal{L}_{b} =  \frac{C^{}_{AA}}{M} b \, F_A \wedge F_A  + \frac{C^{}_{BB}}{M} b \, F_B \wedge F_B.
\eeq
The gauge invariance on $A$ requires that $\beta= -1/2\equiv \beta_0$ and is equivalent to a vector current conservation (CVC) condition.
By imposing gauge invariance under B gauge transformations, on the other hand, we obtain
\beqa
\delta_B \left(  \mathcal{L}_b + \mathcal{L}_{an}  \right) = 0
\eeqa
which implies that
\beq
C^{}_{AA} =  \frac{i \,g^{}_{B}  g^{\,2}_{A} }{2!} \frac{1}{4} \, a_3(\beta_0) \, \frac{M}{M_1},
\qquad C^{}_{BB} =  \frac{i g^{\,3}_{B}}{3!} \frac{1}{4} \, a_n \, \frac{M}{M_1}.
\label{CAA}
\eeq
This procedure, as we are going to show, is equivalent to the imposition of the STI on the corresponding anomalous vertices of
the effective action. In fact the counterterms $C_{AA}$ and $C_{BB}$ can be determined formally from a BRST analysis.

\begin{figure}[t]
\begin{center}
\includegraphics[width=9cm]{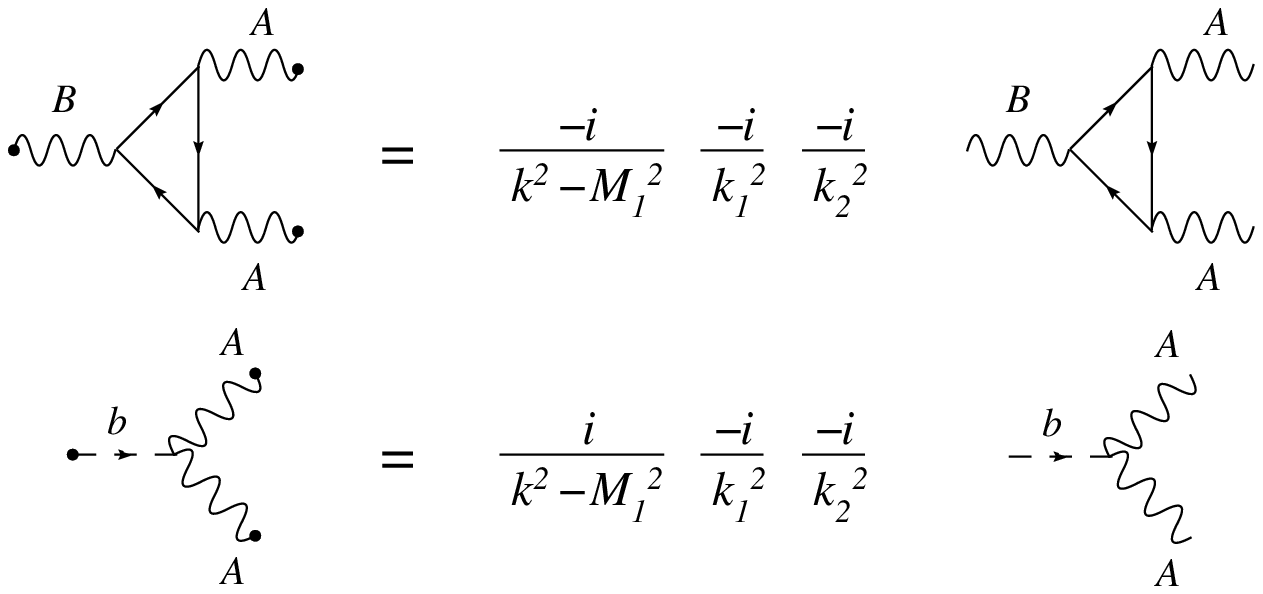}
\caption{Relation between a correlator with non amputated external lines (left) used in a STI and an amputated one (right) used in the effective action  for a triangle vertex and for a CS term.}
\label{amputated}
\end{center}
\end{figure}

\begin{figure}[t]
\begin{center}
\includegraphics[width=13cm]{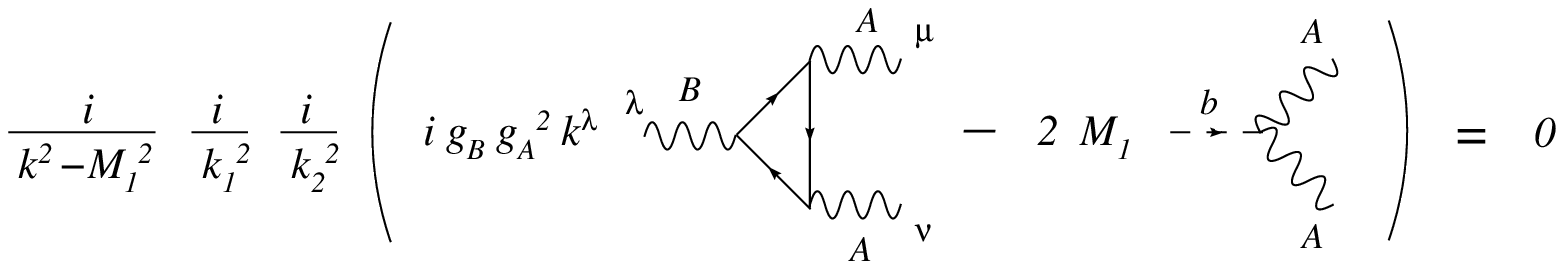}
\caption{Representation in terms of Feynman diagrams in momentum space of the Slavnov-Taylor identity obtained in the St\"uckelberg phase for the anomalous triangle $BAA$. Here we deal with correlators with non-amputated external lines. A CS term has been absorbed to ensure the conserved vector current (CVC) conditions on the A lines. }
\label{STI_St_BAAa}
\end{center}
\end{figure}

In fact, the BRST variations of the model are defined as
\bea
\delta_{BRST} \, B_{\mu} &=& \omega \partial_{\mu} c_B \nonumber \\
\delta_{BRST} \, b &=& - \omega M_1 c_B \nonumber \\
\delta_{BRST} \, A_{\mu} &=& \omega \, \partial_{\mu} c_A \nonumber \\
\delta_{BRST} \, c_B &=&0 \nonumber \\
\delta_{BRST} \, \bar{c}_B &=& \frac{\omega}{\xi_B} \mathcal{F}^S_B
= \frac{\omega}{\xi_B} (\partial_{\mu} B^{\mu} - \xi_B M_1 b). \nonumber \\
\eea
To derive constraints on the 3-linear interactions  involving 2 abelian (vector-like) and one
vector-axial vector gauge field,  that we will encounter in our analysis below, we require the BRST invariance of a specific correlator such as
\bea
\delta_{BRST} \, \langle 0 | T\, \bar{c}_B(z) A_{\mu}(x) A_{\nu}(y) | 0 \rangle =0,
\label{CAA}
\eea
(Fig.~\ref{amputated} shows the difference between the non-amputated and the amputated correlators)
and applying the BRST operator we obtain
\bea
\frac{\omega}{\xi_B} \, \langle 0 | T\, [\partial_{\la} B^{\la}(z) - \xi_B M_1 b(z)] A_{\mu}(x) A_{\nu}(y) | 0 \rangle
&+&  \langle 0 | T\, \bar{c}_B(z) \omega \partial_{\mu} c_A(x) A_{\nu}(y) | 0 \rangle
\nonumber \\
&+& \langle 0 | T\, \bar{c}_B(z) A_{\mu}(x) \omega \partial_{\nu} c_A(y) | 0 \rangle  =0,
\eea
with the last two terms being trivially zero. Choosing $\xi_B=1$ we
obtain the STI (see Fig.~\ref{STI_St_BAAa}) involving only the WZ term and the
anomalous triangle diagram $BAA$. This reads
\bea
\frac{\partial}{\partial z ^{\la}} \langle 0 | T\,  B^{\la}(z) A_{\mu}(x) A_{\nu}(y) | 0 \rangle
- M_1 \langle 0 | T \,  b(z) A_{\mu}(x) A_{\nu}(y) | 0 \rangle = 0.
\label{BAA_ST}
\eea

A similar STI holds for the BBB vertex and its counterterm
\bea
\frac{\partial}{\partial z ^{\la}} \langle 0 | T\,  B^{\la}(z) B_{\mu}(x) B_{\nu}(y) | 0 \rangle
- M_1 \langle 0 | T \,  b(z) B_{\mu}(x) B_{\nu}(y) | 0 \rangle =0.
\label{BBB_ST}
\eea
These two equations can be rendered explicit. For instance, to extract from (\ref{BAA_ST}) the corresponding expression in momentum space and the constraint on
$C_{AA}$, we work at the lowest order in the perturbative expansion obtaining
\bea
\frac{1}{2!}\frac{\partial}{\partial z ^{\la}} \langle 0 | T\,  B^{\la}(z)
A_{\mu}(x) A_{\nu}(y) \left[J_5 B\right] \left[J A\right]^2 | 0 \rangle
- M_1 \langle 0 | T \,  b(z) A_{\mu}(x) A_{\nu}(y)
\left[ b F_A\wedge F_A\right]| 0 \rangle =0, \nn \\
\eea
where we have introduced the notation $\left[\,\, \right] $ to denote the
spacetime integration of the vector ($J$) and axial current ($J_5$) to their
corresponding gauge fields
\bea
J \, A &=& - g_A \bar{\psi} \gamma^{\mu} \psi A_{\mu}, \\
J_5 \, B &=& - g_B \bar{\psi} \gamma^{\mu} \gamma^5 \psi B_{\mu}\\
\tilde{J}_5 \, G_B &=& 2i g_B \frac{m_f}{M_B}\bar{\psi}\gamma^5 \psi G_B\,,
\eea
where $M_B$ is the mass of the $B$ gauge boson in the Higgs-St\"uckelberg phase
that we will analyze in the next sections.

In momentum space the STI represented in Fig.~\ref{STI_St_BAAa} becomes $(\xi_B=1)$

\bea
&& \frac{1}{2!}\, 2 \, \left[i k^{\la'} \right] \,
\left[ - \frac{i g_{\la\la'}}{k^2 - M_1^2}\right] \,
\left[ - \frac{i g_{\mu\mu'}}{k_1^2} \right] \,
\left[ - \frac{i g_{\nu\nu'}}{k_2^2} \right] \,
\left[ -g_B g_A^2 \right] \, \Delta^{\la\mu\nu}(k_1,k_2) \nn \\
&-& 2 \, M_1 \left[ \frac{i }{k^2 - M_1^2} \right] \,
\left[ - \frac{i g_{\mu\mu'}}{k_1^2} \right] \,
\left[ - \frac{i g_{\nu\nu'}}{k_2^2} \right] \,
 V^{\mu\nu}_A(k_1,k_2) =0,
\label{BAA_STp}
\eea
%
%
where the factor  $\frac{1}{2!}$ comes from the presence in the effective action of a diagram with 2 identical external lines, in this case two $A$ gauge bosons, and the factor $2$, present in both  terms, comes from the possible contractions with the external fields.
Using in (\ref{BAA_STp}) the corresponding anomaly equation
\bea
k_{\la} \Delta^{\la\mu\nu}(k_1,k_2) = a_3(\b_0) \epsilon^{\mu\nu\alpha\beta} k_{1\alpha}k_{2\beta}
\eea
and the expression of the vertex  $V^{\mu\nu}_A(k_1,k_2)$
\bea
V^{\mu\nu}_A (k_1,k_2)=\frac{4 C_{AA}}M \epsilon^{\mu\nu\alpha \beta}k_{1\alpha}k_{2\beta}
\label{GSvertex}
\eea
we obtain
\bea
\left[ \frac{i }{k^2 - M_1^2} \right]  \,
\left[ - \frac{i g_{\mu\mu'}}{k_1^2} \right] \,
\left[ - \frac{i g_{\nu\nu'}}{k_2^2} \right] \,
 \biggl[ i \, g_B g_A^2 a_3(\b_0) \epsilon^{\mu\nu\alpha \beta}k_{1\alpha}k_{2\beta}
- 2 \, M_1 \frac{4 C_{AA}}M \epsilon^{\mu\nu\alpha \beta}k_{1\alpha}k_{2\beta} \biggr] = 0, \nn \\
\label{BAA_ST2}
\eea
from which we get
\bea
i \,  g_B g_A^2 a_3(\b_0) =  2 \, M_1 \frac{4 C_{AA}}M \qquad \Rightarrow \qquad
C_{AA}= \frac{i \,g^{}_{B}  g^{\,2}_{A} }{2} \frac{1}{4} \, a_3(\beta_0) \, \frac{M}{M_1}.
\eea
This condition determines $C_{AA}$ at the same value as before in (\ref{CAA}), using the constraints
of gauge invariance, having brought the anomaly on the B vertex $(\beta_0=-1/2)$.

\begin{figure}[t]
\begin{center}
\includegraphics[width=13cm]{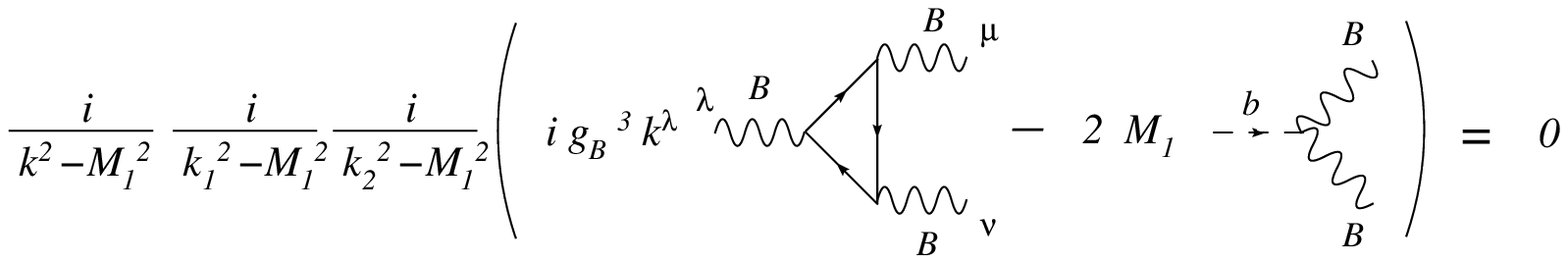}
\caption{ Diagrammatic representation of (\ref{ST_BBB_3})
in the St\"uckelberg  phase, determining the counterterm  $C_{BB}$.}
\label{BBB_S}
\end{center}
\end{figure}

In the case of the second STI given in (\ref{BBB_ST}), expanding this equation at the lowest relevant order we get
\bea
 \frac{1}{3!}\frac{\partial}{\partial z ^{\la}} \langle 0 | T\,  B^{\la}(z)
B_{\mu}(x) B_{\nu}(y) \left[J_5 B\right]^3 | 0 \rangle
 - M_1 \langle 0 | T \,  b(z) B_{\mu}(x) B_{\nu}(y)
\left[ b F_B\wedge F_B\right]| 0 \rangle =0.
\label{ST_BBBx_4}
\eea
%
Also in this case, setting $\xi_B=1$,  we re-express (\ref{ST_BBBx_4}) as
\bea
&& \frac{1}{3!}\, 3! \,  \left[ i k^{\la'} \right] \,
\left[ - \frac{i g_{\la\la'}}{k^2 - M_1^2} \right] \,
\left[ - \frac{i g_{\mu\mu'}}{k_1^2 - M_1^2} \right] \,
\left[ - \frac{i g_{\nu\nu'}}{k_2^2 - M_1^2} \right] \,
\left[-g_B^3 \right] \, \Delta^{\la\mu\nu}(k_1,k_2) \nn \\
&-& 2 \, M_1 \left[ \frac{i }{k^2 - M_1^2} \right] \,
\left[ - \frac{i g_{\mu\mu'}}{k_1^2 - M_1^2} \right] \,
\left[ - \frac{i g_{\nu\nu'}}{k_2^2 - M_1^2} \right] \,
 V^{\mu\nu}_B(k_1,k_2) =0,
\label{BBB_3}
\eea
where, similarly to $BAA$, the factor $\frac{1}{3!}$ comes from the 3 identical gauge B bosons on the external lines, the coefficient $3!$ in the first term counts all the contractions between the vertex $\Delta^{\la\mu\nu}$
and the propagators of the $B$ gauge bosons, while the coefficient  $2$  comes from the contractions of  $V^{\mu\nu}_B$ with the external lines.  From Eq. (\ref{BBB_3}) we get
\bea
\left[  \frac{i}{k^2 - M_1^2} \right] \,
\left[ - \frac{i g_{\mu\mu'}}{k_1^2 - M_1^2} \right] \,
\left[ - \frac{i g_{\nu\nu'}}{k_2^2 - M_1^2} \right] \,
\biggl[ i g_B^3 \,  k_\la  \Delta^{\la\mu\nu}(k_1,k_2)
- 2 \, M_1 V^{\mu\nu}_B(k_1,k_2) \biggr]=0\,,
\label{ST_BBB_3}
\eea
as depicted in Fig.~\ref{BBB_S}.

The anomaly equation for $BBB$ distributes the total anomaly  $a_n$ equally among the three $B$ vertices, therefore
\bea
k_\lambda \Delta^{\lambda\mu\nu}(k_1,k_2)= \frac{a_n}{3}\epsilon^{\mu\nu\alpha\beta}k_{1\alpha} k_{2\beta},
\label{BBB_anom}
\eea
and for the $V^{\mu\nu}_B (k_1,k_2)$ vertex we have
\bea
V^{\mu\nu}_B (k_1,k_2)=\frac{4 C_{BB}}M \epsilon^{\mu\nu\alpha \beta}k_{1\alpha}k_{2\beta}.
\label{V_BB}
\eea
Inserting (\ref{BBB_anom}), (\ref{V_BB}) into (\ref{ST_BBB_3}) we obtain
\bea
i\, g_B^3 \, \frac{a_n}{3} =  2 \, M_1 \, \frac{4 C_{BB}} {M} \qquad \Rightarrow \qquad
C_{BB}= \frac{i \,g^{3}_{B} }{2} \frac{1}{4} \, \frac{a_n}{3} \, \frac{M}{M_1},
\eea
in agreement with (\ref{CAA}).
Therefore we have shown that if we gauge-fix the effective lagrangean in the S\"tuckelberg phase to remove the b-B mixing and fix the CS counterterms so that  the anomalous variations of the trilinear vertices  are absent, we are actually imposing generalized Ward identities or STI's on the effective action.  On this gauge-fixed axion the b-B mixing is
completely absent also off-shell and the structure  of the trilinear vertices is rather simple. We need to check
that these STI's are compatible with those obtained after electroweak symmetry breaking, so that the mixing is
absent off-shell also in the physical basis.

\subsection{The Higgs-St\"uckelberg phase (HS)}
Now consider the same effective action of the previous model after electroweak symmetry breaking. If we interpret the gauge-fixed action derived above as a completely determined theory where the counterterms have been found by the procedure that we have just illustrated, once we expand the fields around the Higgs vacuum we encounter a new mixing of the goldstones with
the gauge fields. Due to Higgs-axion mixing \cite{CIM1}  the goldstones of this theory are extracted by a suitable rotation that allows to separate physical from unphysical degrees of freedom.  In fact the St\"uckelberg is decomposed into a physical axi-Higgs and a genuine goldstone. It is then natural to
ask whether we could have just worked out  the lagrangean {\em directly} in this phase by keeping the coefficients in front of the counterterms of the theory free, and had them fixed by imposing directly generalized WI's in this phase, bypassing completely the first construction. As we are now going to show in this model the counterterms are determined consistently also in this case  at the same values
given before.

Let's see how this happens. In this phase the mixing that needs to be eliminated
is of the form $B^{\mu}\partial_{\mu}G_B$, where $G_B$ is the goldstone of the HS phase. In this case we use the gauge-fixing lagrangean

\bea
\mathcal{L}_{gf}= -\frac{1}{2\xi_B}(\mathcal{F}^H_B)^2=
-\frac{1}{2\xi_B}\left( \partial_{\mu} B^{\mu} - \xi_B M_B G^{}_{B}\right),
\eea
and the BRST transformation of the antighost field $\bar{c}_B$ is given by
\bea
\delta_{BRST} \, \bar{c}_B = \frac{\omega}{\xi_B} \mathcal{F}^H_B
= \frac{\omega}{\xi_B} \left( \partial_{\mu} B^{\mu} - \xi_B M_B G^{}_{B}\right).\eea
Also in this case we use the 3-point function in Eq.~(\ref{CAA}) and
$\xi_B=1$ to obtain the STI
\bea
\frac{\partial}{\partial z ^{\la}} \langle 0 | T\,  B^{\la}(z) A_{\mu}(x) A_{\nu}(y) | 0 \rangle
- M_B \langle 0 | T \,  G_B(z) A_{\mu}(x) A_{\nu}(y) | 0 \rangle=0.
\label{BAA_STHS1}
\eea
\begin{figure}[t]
\begin{center}
\includegraphics[width=10cm]{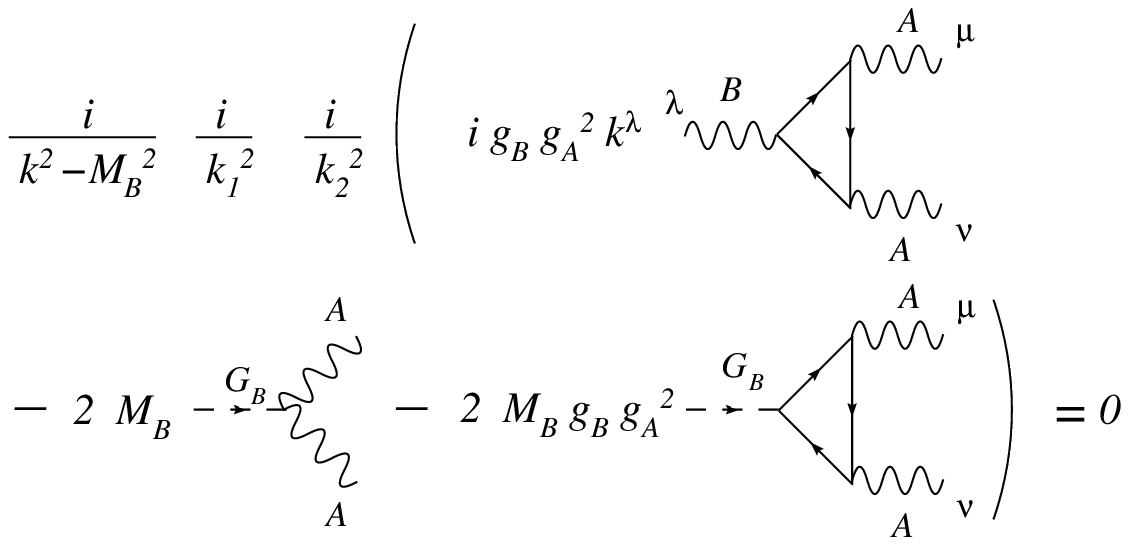}
\caption{Diagrammatic representation of Eq. (\ref{XXXX})
 in the HS phase, determining the counterterm  $C_{AA}$. A CS term has been absorbed by the CVC conditions on the $A$ gauge bosons.}
\label{BAA_HS}
\end{center}
\end{figure}

To get insight into this equation we expand perturbatively (\ref{BAA_STHS1}) and obtain
\bea
&&\frac{1}{2!}\, \frac{\partial}{\partial z ^{\la}} \, \langle 0 | T\,  B^{\la}(z)
A_{\mu}(x) A_{\nu}(y) \left[J_5 B\right] \left[J A\right]^2 | 0 \rangle \nn \\
&& - \, M_B \,\langle 0 | T \,  G_B(z) A_{\mu}(x) A_{\nu}(y)
\left[ G_B F_A\wedge F_A\right]| 0 \rangle  \nn \\
&& - \, M_B \,\langle 0 | T \,  G_B(z) A_{\mu}(x) A_{\nu}(y) \left[\tilde{J}_5 G_B\right]
\left[J A\right]^2 | 0 \rangle =0,
\label{XXXX}
\eea
where the first term is the usual triangle diagram with the $BAA$ gauge bosons
on the external lines, the second is a WZ vertex with $G_B$ on the exernal line and the third term,
which is absent in the St\"uckelberg phase, is a triangle diagram involving the $G_B$ gauge boson that couples to the fermions by a Yukawa coupling (see Fig.~\ref{BAA_HS}).
In the St\"uckelberg phase there is no analogue of this third contribution in the
cancellation of the anomalies for this vertex, since $b$ does not couple to the fermions.

Notice that the STI now contains a vertex derived from the $b F_A\wedge F_A$ counterterm, but projected
on the interaction $G_B F_A\wedge F_A$ via the factor $M_1/M_B$.
This factor is generated by the rotation matrix that allows
the change of variables $(\phi^{}_{2},b) \to (\chi^{}_{B},G^{}_{B})$ and is given by
%
\beq
U=\left(
\begin{array}{ll}
 -\cos \theta^{}_{B} & \sin \theta^{}_{B} \\
 \sin  \theta^{}_{B} & \cos \theta^{}_{B}
\end{array}
\right)
\eeq
%
with $\theta^{}_{B}={\arccos} ({M_1/M_B})={\arcsin}(q^{}_{B} g^{}_{B} v/ M_B)$.
We recall \cite{CIM1} that the axion $b$ can be expressed as a linear combination of the rotated fields $\chi^{}$ and $G^{}_{B}$ of the form
\beqa
b = \alpha_1 \chi^{}_{B} + \alpha_2 G^{}_{B} = \frac{q^{}_{B} g^{}_{B} v}{M_B} \chi^{}_{B} + \frac{M_1}{M_B} G^{}_{B},
\label{projection}
\eeqa
where $\chi$ is the physical axion and $G_B$ the Goldstone boson;
we also recall that the gauge field $B^{}_\mu$ gets its mass $M^{}_B$ through the combined Higgs-St\"{u}ckelberg
mechanism
\beq
M_B=\sqrt{M_1^2 + (q^{}_{B} g^{}_{B} v)^2}.
\eeq

Now we express the STI given in (\ref{XXXX}) choosing $\xi_B=1$
\bea
&& \frac{1}{2!}\, 2 \, \left[i k^{\la'} \right] \,
\left[ - \frac{i g_{\la\la'}}{k^2 - M_B^2}\right] \,
\left[ - \frac{i g_{\mu\mu'}}{k_1^2} \right] \,
\left[ - \frac{i g_{\nu\nu'}}{k_2^2} \right] \,
\left[ -g_B g_A^2 \right] \, \Delta^{\la\mu\nu}(m_f, k_1,k_2) \nn \\
&-&  \, M_B \left[ \frac{i }{k^2 - M_B^2} \right] \,
\left[ - \frac{i g_{\mu\mu'}}{k_1^2} \right] \,
\left[ - \frac{i g_{\nu\nu'}}{k_2^2} \right] \,
  \biggl\{ 2 \, \frac{M_1}{M_B} \, V^{\mu\nu}_A(k_1,k_2)
  \nn \\ && \hspace{3.5cm}
  + \frac{1}{2!} \,2\,i\, g_B g_A^2 \, \left(2i \, \frac{m_f}{M_B} \right)\,
 \Delta^{\mu\nu}_{G_B AA}(m_f, k_1,k_2) \biggr\} =0, \nn \\
\label{BAA_STprime}
\eea
where the $\left[ G_B F_A\wedge F_A\right]$ interaction has been
obtained from the $\left[ b F_A\wedge F_A\right]$ vertex
by projecting the $b$ field on the field $G_B$, and the coefficient
$2im_f/M_B$ comes from the coupling of $G_B$ with the massive fermions \cite{CIM1}. The remaining coefficient $M_1/M_B$ rotates the $V^{\mu\nu}_A(k_1,k_2)$ vertex
as in Eq. (\ref{BAA_STprime}).


Replacing in (\ref{BAA_STprime}) the WI obtained for a massive AVV vertex
\ba
k_\lambda\Delta^{\lambda\mu\nu}(\beta, m_f, k_1,k_2)&=& a_3(\beta) \varepsilon^{\mu\nu\alpha\beta}
k_1^\alpha k_2^\beta +  2 m_{f} \Delta^{\mu \nu}(m_f, k_1,k_2),
\ea
where
\ba
&&\Delta^{\mu \nu}(m_f, k_1,k_2)=m_f \varepsilon^{\alpha\beta\mu\nu}k_{1,\alpha}k_{2,\beta}
\left(\frac{1}{2\pi^2}\right)I(m_f)
\nonumber\\\nonumber\\
&&I(m_f)\equiv -\int_0^1\int_0^{1-x}dx dy \frac{1}{m_f^2+(x-1)x k_1^2+(y-1)y k_2^2-2 xyk_1\cdot k_2},
\ea
and the expression for  the $V^{\mu\nu}_A(k_1,k_2)$ vertex
\bea
V^{\mu\nu}_A (k_1,k_2)=\frac{4 C_{AA}}M \epsilon^{\mu\nu\alpha \beta}k_{1\alpha}k_{2\beta},
\eea
we get
\bea
&& \left[  \frac{i g_{\la\la'}}{k^2 - M_B^2}\right] \,
\left[  \frac{i g_{\mu\mu'}}{k_1^2} \right] \,
\left[  \frac{i g_{\nu\nu'}}{k_2^2} \right] \,
\biggl\{ i \, g_B g_A^2 \, a_3(\b_0) \, \epsilon^{\mu\nu\alpha \beta}k_{1\alpha}k_{2\beta} \nn \\
&&   + 2\, i \, g_B g_A^2 \, m_f \, \Delta^{\mu\nu}(m_f, k_1,k_2)
- 2 \, M_B \, \frac{4 C_{AA}}{M} \epsilon^{\mu\nu\alpha \beta}k_{1\alpha}k_{2\beta}
\nn \\ &&
- 2 \,i  g_B g_A^2 \, M_B \, \frac{m_f}{M_B} \, \Delta^{\mu\nu}_{G_BAA}(m_f, k_1,k_2) \biggr\} = 0.
\label{BAA_STHS3}
\eea
Since $\Delta^{\mu\nu}_{G_BAA}=\Delta^{\mu\nu}$, Eq.(\ref{BAA_STHS3})  yields the same condition obtained by fixing $C_{AA}$ in the St\"uckelberg phase, that is
\bea
i \,  g_B g_A^2 a_3(\b_0) =  2 \, M_1 \frac{4 C_{AA}}M \qquad \Rightarrow \qquad
C_{AA}= \frac{i \,g^{}_{B}  g^{\,2}_{A} }{2} \frac{1}{4} \, a_3(\beta_0) \, \frac{M}{M_1}.
\eea
%
%
%
%

A similar STI can be derived for the $BBB$ vertex in this phase, obtaining
\bea
\frac{\partial}{\partial z ^{\la}} \langle 0 | T\,  B^{\la}(z) B_{\mu}(x) B_{\nu}(y) | 0 \rangle
- M_B \langle 0 | T \,  G_B(z) B_{\mu}(x) B_{\nu}(y) | 0 \rangle =0 .
\label{BBB_STHS1}
\eea

Expanding perturbatively (\ref{BBB_STHS1}) we obtain

\bea
&&\frac{1}{3!}\, \frac{\partial}{\partial z ^{\la}} \, \langle 0 | T\,  B^{\la}(z)
B_{\mu}(x) B_{\nu}(y) \left[J_5 B \right]^3 | 0 \rangle \nn \\
&& - \, M_B \,\langle 0 | T \,  G_B(z) B_{\mu}(x) B_{\nu}(y)
\left[ G_B F_B \wedge F_B \right]| 0 \rangle  \nn \\
&& - \, M_B \,\langle 0 | T \,  G_B(z) B_{\mu}(x) B_{\nu}(y) \left[\tilde{J}_5 G_B \right]
\left[J_5 B \right]^2 | 0 \rangle =0,
\label{BBB_STHS2}
\eea
that gives
\bea
&& \frac{1}{3!}\, 3! \, \left[i k^{\la'} \right]
\left[ - \frac{i g_{\la\la'}}{k^2 - M_B^2}\right]
\left[ - \frac{i g_{\mu\mu'}}{k_1^2 - M_B^2} \right]
\left[ - \frac{i g_{\nu\nu'}}{k_2^2 - M_B^2} \right]
\left[ -g_B^3 \right]  \Delta^{\la\mu\nu}(m_f, k_1,k_2)  \nn \\
&& -   \, M_B \left[ \frac{i }{k^2 - M_B^2} \right] \,
\left[ - \frac{i g_{\mu\mu'}}{k_1^2 - M_B^2} \right] \,
\left[ - \frac{i g_{\nu\nu'}}{k_2^2 - M_B^2} \right] \,
  \biggl\{ 2 \, \frac{M_1}{M_B} \, V^{\mu\nu}_B(k_1,k_2) \nn \\
 && + \frac{1}{2!} \, 2 \, i \, g_B^3 \, \left(2i \, \frac{m_f}{M_B} \right)\,
 \Delta^{\mu\nu}_{G_B BB}(m_f, k_1,k_2) \biggr\} =0, \nn \\
\label{BBB_STp}
\eea
where we have defined
\ba
&&\Delta^{\mu\nu}_{G_B BB}=\int \frac{ d^4 q }{ ( 2 \pi )^4 }
\frac{ Tr \left[ \gamma^{5} ( \slash{q} - \slash{k} + m_f ) \gamma^{\nu}\g^5 ( \slash{q} -
\slash{k_1} + m_f)  \gamma^{\mu}\g^5 ( \slash{q} + m_f ) \right] }
{ \left[ q^2 - m^2_f \right] \left[ ( q - k )^2 - m^{2}_{f} \right]
\left[ (q - k_1)^2 - m^{2}_{f} \right] }
\nonumber\\
&&\hspace{2cm}+ \left\{\mu \leftrightarrow \nu,k_1\leftrightarrow k_2\right\}\,.
\label{diagr5V5V5}
\ea
Since this contribution is finite, it gives
\ba
\Delta^{\mu\nu}_{G_B BB}=2\int\frac{d^4q}{(2\pi)^4}\int_{0}^1 \int_0^{1-x} d x d y
\frac{2 m 4 i\varepsilon^{\mu\nu\alpha\beta}k_{1,\alpha}k_{2,\beta}}{\left[
q^2 -k_2^2(y-1)y -k_1^2(x-1)x + 2 x y - m_f^2\right]^3}
\ea
and we obtain again
\ba
\Delta^{\mu\nu}_{G_B BB}=\Delta^{\mu\nu}=\varepsilon^{\alpha\beta\mu\nu}k_{1,\alpha}k_{2,\beta}
m_f\left(\frac{1}{2\pi^2}\right)I(m_f)\,,
\ea

Using the anomaly equations in the chirally broken phase
\bea
k_\lambda \Delta_{3}^{\lambda\mu\nu}(k_1,k_2)&=& \frac{a_n}{3} \varepsilon^{\mu\nu\alpha\beta}
k_1^\alpha k_2^\beta +   2 m_{f} \Delta^{\mu \nu}
\label{bbshift}
\eea
and the expression of the vertex
\bea
V^{\mu\nu}_B (k_1,k_2)=\frac{4 C_{BB}}M \epsilon^{\mu\nu\alpha \beta}k_{1\alpha}k_{2\beta},
\eea
we obtain
\bea
C_{BB}= \frac{i \,g^{3}_{B} }{2} \frac{1}{4} \, \frac{a_n}{3} \, \frac{M}{M_1}.
\eea

Expanding to the lowest nontrivial order this identity we obtain
\beq
i\left( \frac{a_n}{3} \epsilon^{\mu\nu\alpha \beta}k_{1\alpha}k_{2\beta} + 2 m_f \Delta^{\mu\nu}\right)
- 2 M_B \left( \frac{4}{M}C_{BB} \frac{M_1}{M_B}\right) \epsilon^{\mu\nu\alpha \beta}k_{1\alpha}k_{2\beta}
- M_B \left(2 i \frac{m_f}{M_B}\right) \Delta^{\mu\nu}_{G_B BB}=0,
\eeq
which can be easily solved for $C_{BB}$, thereby determining $C_{BB}$ exactly at the same value
inferred from the St\"uckelberg phase, as discussed above.


\subsection{Slavnov-Taylor Identities and  BRST symmetry in the complete model}

\begin{figure}[t]
{\centering \resizebox*{13cm}{!}{\rotatebox{0}
{\includegraphics{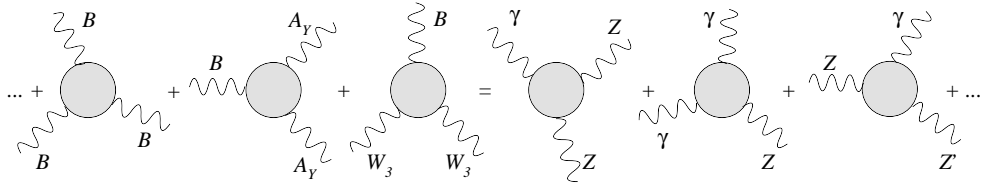}}}\par}
\caption{The anomalous effective action in the
two basis in the $R_{\xi}$ gauge where we have eliminated the mixings on the external lines in both basis.}
\label{EA1}
\end{figure}

It is obvious, from the analysis presented above, that a similar treatment is possible also in the non-abelian case,
though the explicit analysis is more complex. The objective of this investigation, however, is by now clear:
we need to connect the anomalous effective action of the general model in the interaction basis and in the mass eigenstate basis keeping into account that {\em both} phases are broken phases. In Fig. \ref{EA1} this point is shown
pictorially.  In both cases the bilinear mixings of the goldstones with the corresponding gauge fields, $Z\partial G_Z, Z^{\prime}\partial G_{Z^{\prime}}$ have been
removed and the counterterms in the eigenstate basis have been fixed as in \cite{CIM2},
where we have just shown it for the A-B model. Equivalently, we can fix the counterterms in the HS phase by imposing the STI's directly at this stage, thereby defining the anomalous effective action plus WZ terms completely.  For this we need the BRST transformation of the fundamental fields. As usual, in the gauge sector these can be obtained by replacing the gauge parameter in their
gauge variations with the corresponding ghost fields times a Grassmann parameter $\omega$.
Denoting by  $s$ the BRST operator, these are given by

\bea
s A^{\g}_{\mu}  =  \w \, \partial_{\mu} c_{\g}+ i\, O^{A}_{11}  \, g_2  \, \w \left( c^{-} W^{+}_{\mu}
 - c^{+} W^{-}_{\mu}  \right),   \\
s Z_{\mu} =  \w \, \partial_{\mu} c_{Z} + i \, O^{A}_{21} \,  g_2 \, \w \, \left(c^{-} W^{+}_{\mu}
 - c^{+} W^{-}_{\mu} \right), \\
s  Z^{\prime}_{\mu}  =  \w \, \partial_{\mu} c_{Z^\prime} + i  \,  O^{A}_{31}  \,  g_2 \, \w \left( c^{-} W^{+}_{\mu}
- c^{+} W^{-}_{\mu}  \right)
\eea
\bea
s \, W^{\p}_{\mu} &=& \w \,\partial_{\mu} c^{\p} - i g_{2} W^{\p}_{\mu}\, \w \left( O^{A}_{11}
c_{\g} + O^{A}_{21} c_{Z} + O^{A}_{31} c_{Z^{\prime}} \right) \nn \\
&+& i g_2 \, \left ( O^{A}_{11}
A_{\g \mu} + O^{A}_{21} Z_{\mu} + O^{A}_{31} Z^{\prime}_{\mu} \right)  \w c^{\p},  \\
s W^{\m}_{\mu} &=& \w \partial_{\mu} c^{\m} + i g_{2} W^{\m}_{\mu} \w \, \left( O^{A}_{11}
c_{\g} + O^{A}_{21} c_{Z} + O^{A}_{31} c_{Z^{\prime}} \right) \nn \\
&-&  i g_2\left( O^{A}_{11}
A_{\g \mu} + O^{A}_{21} Z_{\mu} + O^{A}_{31} Z^{\prime}_{\mu} \right) \w c^{\m}, 
\eea
where the $O^{A}_{ij}$ are matrix elements defined  exactly as in Eq. (\ref{matmat}) below.
To determine the transformations rules for the ghost/antighost fields we recall that the  gauge-fixing lagrangeans in the $R_\xi$ gauge are given by

\bea
\mathcal{L}_{gf}^{Z}&=& - \frac{1}{2\xi_{Z}} \mathcal{F}[Z,G^Z]^2=
- \frac{1}{2\xi_{Z}}(\partial_{\mu}Z^{\mu} - \xi_{Z}M_{Z}G^Z)^2, \\
\mathcal{L}_{gf}^{Z'}& =& - \frac{1}{2\xi_{Z'}} \mathcal{F}[Z',G^{Z'}]^2=
- \frac{1}{2\xi_{Z'}}(\partial_{\mu}Z'^{\mu} - \xi_{Z'}M_{Z'}G^{Z'})^2, \\
\mathcal{L}_{gf}^{A_\gamma}&=&- \frac{1}{2\xi_A} \mathcal{F}[A_\gamma]^2=
- \frac{1}{2\xi_{A}}(\partial_{\mu}A_{\gamma}^{\mu })^2, \\
\mathcal{L}_{gf}^{W}&=& - \frac{1}{\xi_W} \mathcal{F}[W^+,G^+]\mathcal{F}[W^-,G^-]=  \nonumber \\
& =& - \frac{1}{\xi_{W}}(\partial_{\mu}W^{+ \mu } + i \xi_{W}M_{W}G^+)
(\partial_{\mu}W^{- \mu } - i \xi_{W}M_{W}G^-),
\eea

where $G^Z$, $G^{Z'}$, $G^+$ and $G^-$ are the goldstones of  $Z$, $Z'$,
$W^+$ and $W^-$ respectively. \\

In particular, the FP (ghost) part of the lagrangean is canonically given by
\beq
\mathcal{L}_{FP}=- \bar{c}^{a} \frac{\d \mathcal{F}^a[Z,z]}{\d \theta^b} c^b,
\eeq
where the sum over $a$ and $b$  runs over the fields $Z$, $Z'$, $A_{\g}$,
$W^+$ e $W^-$ and is explicitly given in the appendix.  For the BRST variations of the antighosts we obtain
\bea
s \, \bar{c}_a= - \frac{i}{\xi_a} \, \w \, \mathcal{F}^a \qquad \qquad a=Z,Z',\g,+,-
\eea
and in particular
\bea
s \, \bar{c}_Z&=& - \frac{i}{\xi_Z} \, \w \, \left( \partial_{\mu} Z^{\mu} - \xi_Z M_Z G^Z \right) \\
s \, \bar{c}_{Z'}&=& - \frac{i}{\xi_{Z'}} \, \w \, \left( \partial_{\mu} Z'^{\mu} - \xi_{Z'} M_{Z'} G^{Z'} \right) \\
s \, \bar{c}_{\g}&=& - \frac{i}{\xi_{\g}} \, \w \, \left( \partial_{\mu} A_{\g}^{\mu} \right)\\
s \, \bar{c}_+ &=& - \frac{i}{\xi_W} \, \w \, \left( \partial_{\mu} W^{+ \mu}+ i \xi_W M_W G^+ \right) \\
s \, \bar{c}_- &=& - \frac{i}{\xi_W} \, \w \, \left( \partial_{\mu} W^{- \mu} -i  \xi_W M_W G^- \right),
\eea

giving typically the STI
\bea
\frac{\partial}{\partial z ^{\la}} \langle 0 | T\,  Z^{\la}(z) A_{\mu}(x) A_{\nu}(y) | 0 \rangle
- M_Z \langle 0 | T \,  G_Z(z) A_{\mu}(x) A_{\nu}(y) | 0 \rangle = 0,
\label{modif}
\eea
and a similar one for the $Z^\prime$ gauge boson.

\begin{figure}[t]
{\centering \resizebox*{12cm}{!}{\rotatebox{0}
{\includegraphics{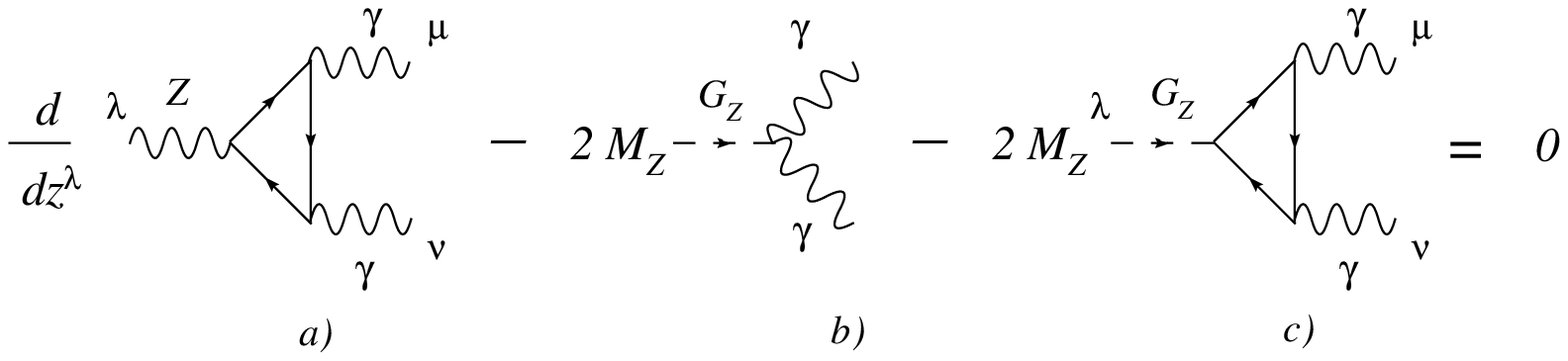}}}\par}
\caption{The general STI for the $Z\gamma\gamma$ vertex in our anomalous model away from the chiral limit. The analogous STI for the SM case consists of only diagrams a) and c).}
\label{anomdiag4}
\vspace{0.5cm}
{\centering \resizebox*{10cm}{!}{\rotatebox{0}
{\includegraphics{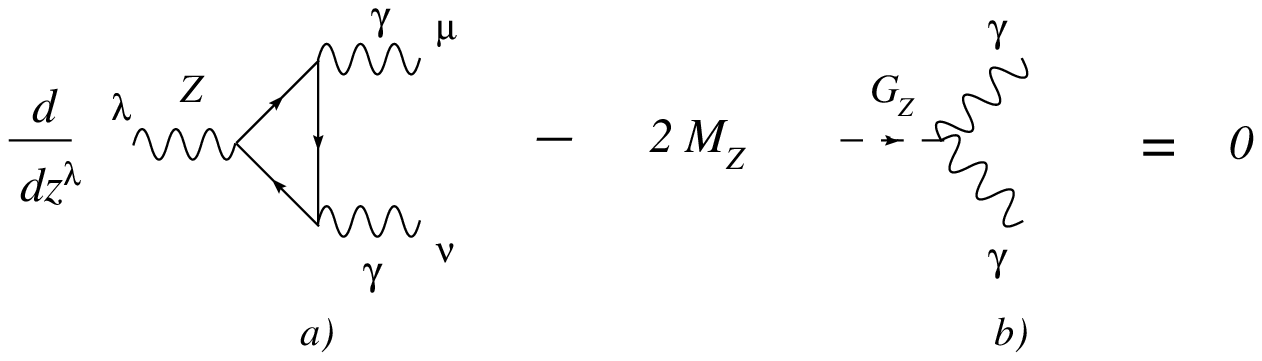}}}\par}
\caption{The STI for the $Z\gamma\gamma$ vertex for our anomalous model
and in the chiral phase. The analogous STI in the SM consists of only diagram a).}
\label{anomdiag3}
\end{figure}

We pause for a moment to emphasize the difference between this
STI and the corresponding one in the SM.  In this latter case the structure of the STI is

\beqn
k_{\rho} \, G^{\rho \nu \mu}  &=&  (k_1 + k_2 )_{\rho} \,  G^{\,  \rho \nu \mu }     \nonumber\\
&=& \frac{ e^2 g }{ \pi^{2} \cos \theta_{W}}  \sum_{f}  g^{f}_{A}  Q^{2}_{f} \,  \epsilon^{\, \nu \mu \alpha \beta }
k_{1 \alpha} k_{2 \beta } \, \left[  - m^{2}_{f} \int^{1}_{0} d x_1 \int^{1 - x_1 }_{0}
d x_2  \,  \frac{1}{ \Delta }    \right],
\label{ABJ_anomaly}
\eeqn
where $G^{\rho \nu \mu}$ is the gauge boson vertex,
which is shown pictorially in Fig. \ref{anomdiag4} (diagrams a and c). Notice that the goldstone contribution is the factor in square brackets in the expression above, being the coupling of the Goldstone proportional to $m_f^2/M_Z$. In the chiral limit the STI of the $Z \gamma \gamma$ vertex of the Standard Model becomes an ordinary Ward identity, as in the photon case. In
Fig. \ref{anomdiag4} the modification
due to the presence of the WZ term is evident. In fact, expanding (\ref{modif}) in the anomalous case we have
\beqn
k_{\rho} \, G^{\rho \nu \mu}  &=&  (k_1 + k_2 )_{\rho} \,  G^{\,  \rho \nu \mu }     \nonumber\\
&=& \frac{ e^2 g }{ \pi^{2} \cos \theta_{W}}  \sum_{f}  g^{f}_{A}  Q^{2}_{f} \,  \epsilon^{\, \nu \mu \alpha \beta }
k_{1 \alpha} k_{2 \beta } \, \left[  \frac{1}{2} - m^{2}_{f} \int^{1}_{0} d x_1 \int^{1 - x_1 }_{0}
d x_2  \,  \frac{1}{ \Delta }    \right],
\label{ABJ_anomaly}
\eeqn
where the first term in the square brackets is now the WZ contribution and the second the usual goldstone contribution, as in the SM case. Notice that the factor  $\sum_{f}  g^{f}_{A}  Q^{2}_{f}$ is in fact proportional to the total chiral asymmetry of the $Z$ vertex, which is mass independent and appears as a factor in front of the WZ counterterm. In the chiral limit the anomalous STI is represented in Fig. \ref{anomdiag3}.

At this point we are ready to proceed with a more general analysis of the trilinear gauge interactions to derive the
expressions of all the anomalous vertices of a given theory in the mass eigenstate basis and {\em away from the chiral limit}. The reason for stressing this aspect has to do with the
way the chiral symmetry breaking effects appear in the SM and in the anomalous models. In particular, we will start by extending the analysis presented in \cite{CIM2} for the derivation of the $Z\gamma\gamma$ vertex, which is here presented in far more detail. Compared
to \cite{CIM2} we show some unobvious features of the derivation which are essential in order to formulate general rules for the computation of these vertices. We rotate the fields from the interaction eigenstate basis to the physical basis and
the CS counterterms are partly absorbed and the anomaly is moved
from the anomaly-free gauge boson vertices to the anomalous ones. This analysis is then extended to other trilinear vertices and we finally provide general rules to handle these types of interactions for a generic number of $U(1)$'s.

Before we come to the analysis of this vertex, we recall that the neutral current sector of the model is defined as \cite{CIM2}
\beqn
- {\mathcal L}_{NC} = \overline{\psi}^{}_{f} \gamma^\mu \mathcal{F} \psi^{}_{f},
\eeqn
with
\beq
\mathcal{F}=  g^{}_{2} W^{3}_{\mu} T^{\,3}
+  g^{}_{Y} Y A^{Y}_{\mu}
+ g^{}_{B} Y^{}_{B} B^{}_{\mu}
\eeq
expressed in the interaction eigenstate basis. Equivalently it can be re-expressed as
\beqn
\mathcal{F} &=& g^{}_{Z} Q^{}_{Z} Z^{}_{\mu}
+ g^{}_{Z^\prime} Q^{}_{Z^\prime} Z^{\prime}_{\mu} + e\, Q A^{\gamma}_{\mu},
 \eeqn
where $Q = T^{3} + Y$.
The physical fields $A^\g, Z, Z^\prime$ and $W_3, A^Y, B$ are related by the rotation matrix $O^A$ to the
interaction eigenstates
\ba
\pmatrix{A^\g \cr Z \cr {{Z^\prime}}} =
O^A\, \pmatrix{W_3 \cr A^Y \cr B}  \label{OA}
\ea
or equivalently
\beqn
W^{3}_{\mu} &=& O^{A}_{W_{3} \gamma} A^{\gamma}_{\mu} + O^{A}_{W_{3} Z} Z_{\mu} + O^{A}_{W_{3} Z^\prime} Z^{\prime}_{\mu}  \\
A^{Y}_{\mu} &=& O^{A}_{Y \gamma} A^{\gamma}_{\mu} + O^{A}_{Y Z} Z_{\mu} + O^{A}_{Y Z^\prime} Z^{\prime}_{\mu}  \\
B_{\mu} &=& O^{A}_{B Z} Z_{\mu} + O^{A}_{B Z^\prime} Z^{\prime}_{\mu}.
\label{matmat}
\eeqn
Substituting these transformations in the expression of the bosonic operator  $\mathcal{F}$ and reading the
coefficients of the fields  $Z_\mu, Z^\prime_\mu$ and $ A^\g_\mu$ we obtain this
set of relations for the coupling constants and the generators in the two basis, given here in a chiral form
\beqn
g_Z Q_Z^L &=& g_2 T^{3L} O^A_{W_3 Z} + g_Y Y^L O^A_{Y Z} + g_B Y_B^L O^A_{B Z} \label{gZQZL}\\
g_Z Q_Z^R &=& g_Y Y^R O^A_{Y Z} + g_B Y_B^R O^A_{B Z} \\
g_{Z^\prime} Q_{Z^\prime}^L &=& g_2 T^{3L} O^A_{W_3 Z^\prime} + g_Y Y^L O^A_{Y Z^\prime} + g_B Y_B^L O^A_{B Z^\prime}\\
g_{Z^\prime} Q_{Z^\prime}^R &=& g_Y Y^R O^A_{Y Z^\prime} + g_B Y_B^R O^A_{B Z^\prime}\\
e Q^L &=& g_2 T^{3L} O^A_{W_3 A} + g_Y Y^L O^A_{Y A} = g_Y Y^R O^A_{Y A} = e Q^R \label{eQLeQR}.
\eeqn
\section{General analysis of the $Z\gamma\gamma$ vertex}
Let's now come to a brief analysis of this vertex, stressing on the general features of its derivation,
which has not been detailed in \cite{CIM2}. In particular we highlight the general approach to follow in order to derive these vertices and apply it to the case when several anomalous $U(1)$'s are present.
 We will exploit the invariance of the
anomalous part of the effective action under transformations of the external classical fields. This is illustrated in \mbox{Fig. \ref{EA1}}. More formally we can set
\beq
W_{anom}(B,W, A_Y)= W_{anom}(Z,Z^{\prime},A_{\gamma}),
\eeq
where we limit our analysis to the anomalous contributions.

\begin{figure}[t]
{\centering \resizebox*{10cm}{!}{\rotatebox{0}
{\includegraphics{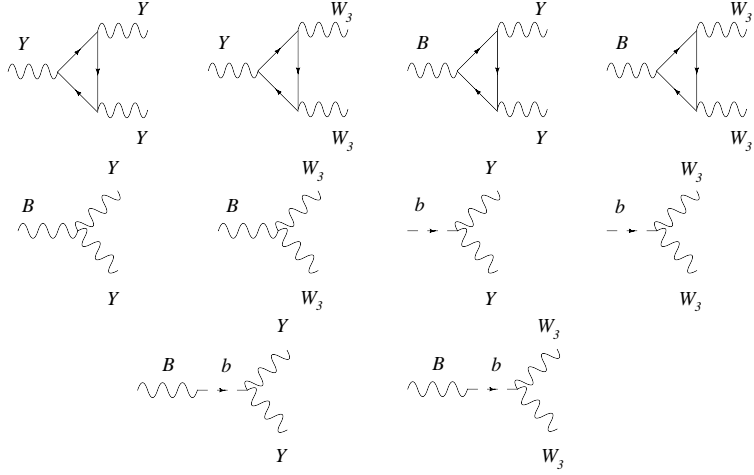}}}\par}
\caption{All the triangle diagrams and the possible CS and WZ counterterms present in the model (chiral phase).
Not all these diagrams project on $Z \rightarrow \gamma\gamma$ in the mass eigenstate basis.}
\label{anomdiag}
\end{figure}

In the chiral phase, the triangle diagrams projecting on this vertex are the following:
$YYY$, $YW_3W_3$, $BYY$ and $BW_3W_3$. They are
represented in Fig. \ref{anomdiag}, where we have added the corresponding counterterms.

The first two are SM-like and hence anomaly-free by charge assignment.
The diagrams involving the $B$ gauge boson are typical of these models,
are anomalous, and require suitable counterterms in order to cancel their anomalies.
All the possible counterterms are shown in Fig. \ref{anomdiag}. The WZ terms of the form $bYY$ or
$bW_3W_3$ will project both on a $G_Z\gamma\gamma$  and a  $\chi\gamma\gamma$ interactions, the first one being relevant for the STI of the vertex.
The main issue to be addressed is that of the distribution of the anomaly among
the triangular vertices. These points have been discussed in
\cite{CIM1} and \cite{CIM2} working in the chiral limit, when the fermion masses are removed from the diagrams.
\begin{figure}[t]
{\centering \resizebox*{10cm}{!}{\rotatebox{0}
{\includegraphics{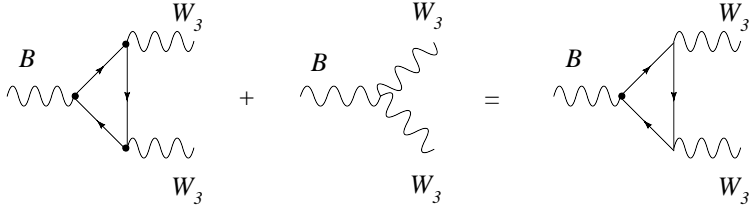}}}\par}
\caption{The routing of the anomaly and the absorption of the CS 
term into the anomalous $B$ gauge boson. The anomaly is distributed among the vertices with the black dot.}
\label{anomdiag1}
\end{figure}

The procedure can follow, equivalently, two directions: we can start from the $B Y W_3 $ basis and project onto the vertices $Z \gamma \gamma , ZZ\gamma $...,
rotating the fields (not the charges) or, equivalently, start from the $Z, Z' \gamma$ basis and rotate the charges (but not the fields) and the generators onto the interaction eigenstate basis $B Y W_3$. We obtain two equivalent descriptions of the various vertices. In the interaction basis the CS terms are absorbed and the anomaly is moved from the $Y$ or $W$ vertices into the B vertex, where it is cancelled by the axion (see Fig. \ref{anomdiag1}). This is the meaning of the STI's shown above.
Therefore it is clear that most of the CS terms do not appear explicitly if we use this approach. On the other hand, if we work in the mass eigenstate
basis  they can be kept explicit, but one has to be careful because in this case also the remaining vertices containing the generator of the electric charge  $Q \sim Y + T_3$ have partial anomalies. The two approaches, as we are going to see, can be combined in a very economical way in some special cases, for instance for the
$Z\gamma \gamma$ vertex, where one can attach all the anomaly to the
Z gauge boson and add only the $G_Z \gamma \gamma$ counterterm.
Similarly, for other interactions such as the $ZZ \gamma $ vertex, the total anomaly has to be equally distributed between the two $Z^\prime$s, since only the
B generator carries an anomaly in the chiral limit, if we choose to absorb the CS terms. For other vertices such as $Z Z Z^{\prime}$ etc, all the vertices contribute to the total anomaly and their partial contributions can be identified by decomposing the corresponding triangle in the $Y B W_3 $
basis with some CS terms left over.

\section{The $\langle Z_l \g\g  \rangle$ vertex}

In this section we begin our technical discussion of the method.
Since the most general case is encountered when at least 3 anomalous $U(1)$'s
are present in the theory, we will consider for definiteness a model with three of them,
say $B_j=\{B_1,B_2,B_3\}$. We can write the field transformation from interaction
eigenstates basis to the mass eigenstates basis as
\ba
&&W_3=O^{A}_{W_3\g}A_{\g}+\sum_{l=0}^{3}O^{A}_{W_3 Z_l}Z_{l}
\nonumber\\
&&Y=O^{A}_{Y\g}A_{\g}+\sum_{l=0}^{3}O^{A}_{Y Z_l}Z_{l}
\nonumber\\
&&B_j=O^{A}_{B_j\g}A_{\g}+\sum_{l=0}^{3}O^{A}_{B_j Z_l}Z_{l},
\ea
with $j=1,2,3$, where for $l=0$ we have the $Z_0$ belonging
to the SM and $Z_1,Z_2,Z_3$ are the anomalous ones. As in \cite{CIM2} we rotate
the external field of the anomalous interactions from one base to the other,
selecting the projections over the $Z_l\g\g$ vertex (the ellipsis indicate additional
contributions that have no projection on the vertex that we consider)

\ba
&&\frac{1}{3!}Tr\left[Q_{Y}^3\right]\langle YYY \rangle=\frac{1}{3!} Tr\left[Q_{Y}^3\right]
R^{YYY}_{Z_l\g\g } \langle Z_l\g\g \rangle+\dots
\nonumber\\
&&\frac{1}{2!}Tr\left[Q_{Y} T_{3}^2\right]\langle YWW \rangle=\frac{1}{2!}Tr\left[Q_{Y} T_{3}^2\right]
R^{YWW}_{Z_l\g\g } \langle Z_l\g\g \rangle+\dots
\nonumber\\
&&\frac{1}{2!}Tr\left[Q_{B_j} Q_{Y}^2\right]\langle B_jYY\rangle=\frac{1}{2!}Tr\left[Q_{B_j} Q_{Y}^2\right]
R^{B_{j}YY}_{Z_l\g\g } \langle Z_l\g\g \rangle+\dots
\nonumber\\
&&\frac{1}{2!}Tr\left[Q_{B_j} T_{3}^2\right]\langle B_jWW\rangle=\frac{1}{2!}Tr\left[Q_{B_j} T_{3}^2\right]
R^{WWB_{j}}_{Z_l\g\g } \langle Z_l\g\g \rangle+\dots
\nonumber\\
\ea
where the rotation coefficients
$R^{YYY}_{Z_l\g\g},R^{YWW}_{Z_l\g\g},R^{B_i YY}_{Z_l\g\g},R^{B_i WW}_{Z_l\g\g}$
containing several products of the elements of the rotation matrix $O^{A}$ are given by
\ba
&&R^{YYY}_{Z_l\g\g}=3\left[(O^{A})_{Y Z_l}(O^{A})_{Y\g}^{2}\right]\nonumber\\
&&R^{YWW}_{Z_l\g\g}=\left[2(O^{A})_{W_3\g}(O^{A})_{YZ_l}(O^{A})_{Y\g}
+(O^{A})_{W_3\g}^{2}(O^{A})_{Y Z_l}\right]\nonumber\\
&&R^{WWW}_{Z_l\g\g}=\left[3(O^{A})_{B_i Z_l}(O^{A})^{2}_{W_3\g}\right]\nonumber\\
&&R^{YYW}_{Z_l\g\g}=\left[2(O^{A})_{Y Z_l}(O^{A})_{Y\g}(O^{A})_{W_3\g}
+(O^{A})_{W_3 Z_l}(O^{A})_{Y\g}^{2}\right]\nonumber\\
&&R^{B_i YY}_{Z_l\g\g}=(O^{A})_{Y\g}^{2}(O^{A})_{B_i Z_l}\nonumber\\
&&R^{B_i WW}_{Z_l\g\g}=\left[(O^{A})_{W_3\g}^{2}(O^{A})_{B_i Z_l}\right]\nonumber\\
&&R^{B_i YW}_{Z_l\g\g}=\left[2(O^{A})_{B_i Z_l}(O^{A})_{W_3\g}(O^{A})_{Y\g}\right]\,.\nonumber\\
\ea
It is important to note that in the chiral phase the $YYY$ and $YWW$ contributions vanish because of the
SM charge assignment.
As we move to the $m_f\neq 0$ phase we must include (together with $YYY$ and $YWW$) 
the other contributions listed below
\ba
&&\frac{1}{3!}Tr\left[Q_{W}^3\right]\langle WWW \rangle=\frac{1}{3!} Tr\left[T_{3}^3\right]
R^{WWW}_{Z_l\g\g } \langle Z_l\g\g \rangle+\dots
\nonumber\\
&&Tr\left[Q_{B_j} Q_{Y} T_{3}\right]\langle B_j Y W\rangle=
Tr\left[Q_{B_j} Q_{Y} T_{3}\right] R^{B_{j}YW}_{Z_l\g\g } \langle Z_l\g\g \rangle+\dots
\nonumber\\
&&\frac{1}{2!}Tr\left[Q_{Y}^2 T_{3}\right]\langle Y Y W\rangle=
\frac{1}{2!}Tr\left[Q_{Y}^2 T_{3}\right] R^{YYW}_{Z_l\g\g } \langle Z_l\g\g \rangle+\dots
\ea

More details on the approach will be given below. For the moment we just mention that
the structure of the CS term can be computed by rotating the WZ counterterms into the 
physical basis, having started with a symmetric distribution of the anomaly in all the triangle diagrams.
The CS terms in this case take the form
\ba
V_{CS}=\frac{a_n}{3}\varepsilon^{\lambda\mu\nu\alpha}(k_{1,\alpha}-k_{2,\alpha})
\frac{1}{8}\sum_j\sum_f\left[ g_{B_j}g_{Y}^2\theta^{B_{j}YY}_{f}R^{B_jYY}_{Z_l\g\g}
+g_{B_j}g_{2}^2\theta^{B_{j}WW}_{f}R^{B_jWW}_{Z_l\g\g}\right]Z_l^{\lambda}A_{\g}^{\mu}A_{\g}^{\nu}\,,
\nonumber\\
\ea
and they are rotated into the physical basis together with the anomalous interactions \cite{CIM2}. We have defined the following chiral asymmetries
\ba
&&\theta^{B_{j}YY}_{f}=Q_{B_j,f}^{L}(Q_{Y,f}^{L})^2-Q_{B_j,f}^{R}(Q_{Y,f}^{R})^2
\nonumber\\
&&\theta^{B_{j}WW}_{f}=Q_{B_j,f}^{L}(T^{3}_{L,f})^2\,.
\ea
We can show that the equations of the vertices in the momentum space can be obtained following a procedure
similar to the case of a single $U(1)$ \cite{CIM2}, that we are now going to generalize.
In particular we will try to absorb all the CS
terms that we can, getting as close as possible to the SM result. This is in general possible for diagrams that have specific Bose symmetries or conserved electromagnetic currents, but some of the details of this construction are quite subtle especially as we move away from the chiral limit.

\subsection{Decomposition in the interaction basis
and in the mass eigenstates basis of the $Z_l \gamma \gamma$ vertex}

As we have mentioned, the anomalous effective action, composed of the triangle diagrams
plus their CS counterterms can be expressed either in the base of the mass eigenstates or
in that of the interaction eigenstates.

We start by keeping all the pieces of the 1-loop effective action in the interaction basis
in the $m_f\neq 0$ phase and rotate the external (classical) fields on the physical basis
taking all the contribution to the $\langle Z_l \g\g\rangle$ vertex.

\begin{figure}[h]
{\centering \resizebox*{10cm}{!}{\rotatebox{0}
{\includegraphics{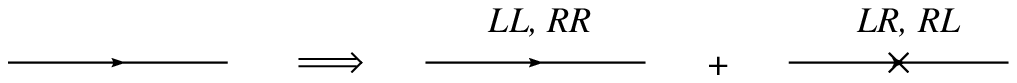}}}\par}
\caption{Chiral decomposition of the fermionic propagator after a mass insertion.}
\label{DEC}
\end{figure}
\begin{figure}[h]
{\centering \resizebox*{12cm}{!}{\rotatebox{0}
{\includegraphics{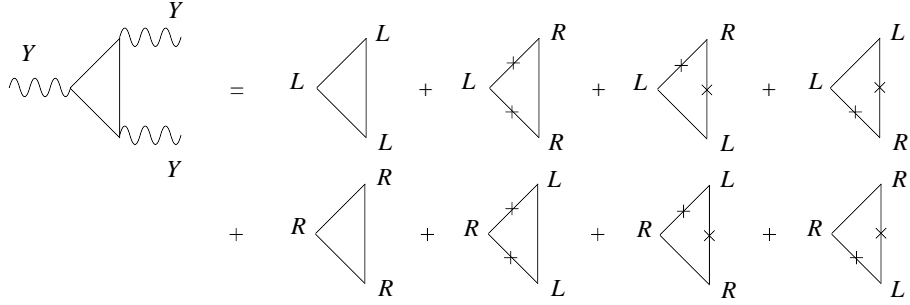}}}\par}
\caption{Chiral triangle contributions to the $YYY$ vertex.
The same decomposition holds for the $B_iYY$ case.}
\label{YYY}
\end{figure}
A given vertex is first decomposed into
its chiral contributions and then rotated into the physical gauge boson eigenstates. For instance, let's start with the non anomalous $YYY$ vertex see Figs.~(\ref{DEC},\ref{YYY}). Actually, in this specific case the sums over each fermion generation are actually zero in the chiral limit, but we will impose this condition at the end and prefer
to follow the general treatment as for other (anomalous) vertices.  We write this vertex in terms
of chiral projectors (L/R), where $L/R\equiv 1 \mp \gamma_5 $, and the diagrams contain a massive fermion of mass $m_f$. The structure of the vertex is

\ba
\langle LLL\rangle|_{m_f\neq0}=
\int\frac{d^4q}{(2\pi)^4}\frac{Tr\left[(\slash{q}+m_f)\g^{\lambda}P_{L}(\slash{q}+\slash{k}+m_f)
\g^{\nu}P_{L}(\slash{q}+\slash{k_1}+m_f)\g^{\mu}P_{L}\right]}{(q^2-m_f^2)\left[(q+k)^2-m_f^2\right]
\left[(q+k_1)^2-m_f^2\right]}+ \, exch.
\nonumber\\
\ea
The vertices of the form $LLR$, $RRL$, and so on, are obtained from the expression above just by
substituting the corresponding chiral projectors. Notice that for loops of fixed chirality we have no mass contributions from the trace
in the numerator and we easily derive the identity
\ba
\langle LLL\rangle|_{m_f\neq0}=-\langle RRR\rangle|_{m_f\neq0}.
\ea
At this point we start decomposing each diagram in the interaction basis
\ba
&&\, \langle Y Y Y \rangle \,g_Y^3 \, Tr [Q_Y^3]=\nonumber\\
&&\hspace{2.5cm}\sum_f\left[g_Y^3(Q_{Y,f}^{L})^3\langle LLL \rangle^{\lambda\mu\nu}
+g_Y^3 (Q_{Y,f}^{R})^3\langle RRR \rangle^{\lambda\mu\nu}
\right.\nonumber\\
&&\hspace{2.5cm}\left.+g_Y^3Q_{Y,f}^{L}(Q_{Y,f}^{R})^2 \langle LRR \rangle^{\lambda\mu\nu}
+g_Y^3 Q_{Y,f}^{L}Q_{Y,f}^{R}Q_{Y,f}^{L}\langle LRL \rangle^{\lambda\mu\nu}
\right.\nonumber\\
&&\hspace{2.5cm}\left.+g_Y^3(Q_{Y,f}^{L})^2 Q_{Y,f}^{R}\langle LLR \rangle^{\lambda\mu\nu}
+g_Y^3 Q_{Y,f}^{R}(Q_{Y,f}^{L})^2 \langle RLL \rangle^{\lambda\mu\nu}
\right.\nonumber\\
&&\hspace{2.5cm}\left.+g_Y^3 Q_{Y,f}^{R}Q_{Y,f}^{L}Q_{Y,f}^{R}\langle RLR \rangle^{\lambda\mu\nu}
+g_Y^3 (Q_{Y,f}^{R})^2 Q_{Y,f}^{L}\langle RRL \rangle^{\lambda\mu\nu}
\right]
\frac{1}{8}Z^{\lambda}_{l} A_{\g}^{\mu}A_{\g}^{\nu}R^{YYY}_{Z_l\g\g}
+\dots \nonumber\\
\ea
where the factor of $1/8$ comes from the chiral projectors and the dots indicate
all the other contributions of the type $Z_l Z_m\g,Z_l Z_m Z_r$ and so on, which do not contribute to the $Z_l\gamma\gamma$ vertex.
This {\em projection} contains chirality conserving and chirality flipping terms. The two combinations which are chirally conserving are $LLL$ and
$RRR$ while the remaining ones need to have 2 chirality flips to be nonzero (ex. $LLR$ or $RRL$) and are therefore proportional to $m_f^2$.

\begin{figure}[t]
{\centering \resizebox*{7.5cm}{!}{\rotatebox{0}
{\includegraphics{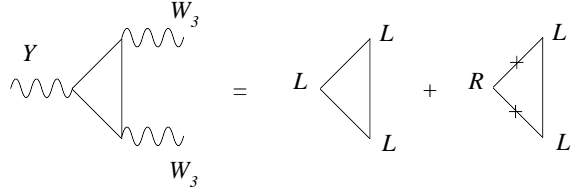}}}\par}
\caption{Chiral triangle contributions to the $YWW$ vertex. The same decomposition holds for the $B_iWW$ case.}
\label{YWW}
\end{figure}

We repeat this procedure for all the other vertices in the interaction eigenstate basis that project on the vertex we are interested in. For instance, in the case of the $\langle YWW\rangle$ vertex the structure is simpler because the generator associated to $W_3$ is left-chiral (Fig.~\ref{YWW})
\ba
&&\langle Y W W \rangle \,g_Y g_2^2 \, Tr [Q_Y (T^{3})^2]=
\sum_f\left[g_Y g_2^2Q_{Y,f}^{L}(T^{3}_{L,f})^2\langle LLL \rangle^{\lambda\mu\nu}
\right.\nonumber\\
&&\left.\hspace{2.5cm}
+g_Y g_2^2Q_{Y,f}^{R}(T^{3}_{L,f})^2\langle RLL \rangle^{\lambda\mu\nu}
\right]\frac{1}{8}Z^{\lambda}_{l} A_{\g}^{\mu}A_{\g}^{\nu}R^{YWW}_{Z_l\g\g}
+\dots \nonumber\\
\ea
Similarly, all the pieces $B_iYY$ and $B_iWW$ for $i=1,2,3$, give the projections
\ba
&& \langle B_i Y Y \rangle \,g_B g_Y^2 \, Tr [Q_{B_i} Q_Y^2]=
\sum_f\left[g_{B_i} g_Y^2 Q_{B_i,f}^{L}(Q_{Y,f}^{L})^2\langle LLL \rangle^{\lambda\mu\nu}
+g_{B_i} g_Y^2 Q_{B_i,f}^{R}(Q_{Y,f}^{R})^2\langle RRR \rangle^{\lambda\mu\nu}
\right.\nonumber\\
&&\hspace{1.5cm}\left.+g_{B_i} g_Y^2 Q_{B_i,f}^{L}(Q_{Y,f}^{R})^2 \langle LRR \rangle^{\lambda\mu\nu}
+g_{B_i} g_Y^2 Q_{B_i,f}^{L}Q_{Y,f}^{R}Q_{Y,f}^{L}\langle LRL \rangle^{\lambda\mu\nu}
\right.\nonumber\\
&&\hspace{1.5cm}\left.+g_{B_i} g_Y^2 Q_{B_i,f}^{L}Q_{Y,f}^{L}Q_{Y,f}^{R}\langle LLR \rangle^{\lambda\mu\nu}
+g_{B_i} g_Y^2 Q_{Y,f}^{R}(Q_{Y,f}^{L})^2 \langle RLL \rangle^{\lambda\mu\nu}
\right.\nonumber\\
&&\hspace{1.5cm}\left.+g_{B_i} g_Y^2 Q_{B_i,f}^{R}Q_{Y,f}^{L}Q_{Y,f}^{R}\langle RLR \rangle^{\lambda\mu\nu}
+g_{B_i} g_Y^2 Q_{B_i,f}^{R} Q_{Y,f}^{R}Q_{Y,f}^{L}\langle RRL \rangle^{\lambda\mu\nu}
\right]\frac{1}{8}Z^{\lambda}_{l} A_{\g}^{\mu}A_{\g}^{\nu}R^{B_iYY}
+\dots \nonumber\\
\ea
and
\ba
&&\langle B_i W W \rangle \,g_Y g_2^2 \, Tr [Q_{B_i} (T^{3})^2]=
\sum_f\left[g_{B_i} g_2^2 Q_{B_i,f}^{L}(T^{3}_{L,f})^2\langle LLL \rangle^{\lambda\mu\nu}
\right.\nonumber\\
&&\left.\hspace{5cm}
+g_{B_i} g_2^2 Q_{B_i,f}^{R}(T^{3}_{L,f})^2\langle RLL \rangle^{\lambda\mu\nu}
\right]\frac{1}{8} Z^{\lambda}_{l} A_{\g}^{\mu}A_{\g}^{\nu} R^{B_iWW}_{Z_l\g\g}
+\dots \nonumber\\
\ea
We obtain similar expressions for the terms $WWW$, $YYW$, $B_iYW$, etc. 
which appear in the $m_f\neq 0$ phase.

\subsubsection{The $m_f = 0$ phase}

To proceed with the analysis of the amplitude we start from the chirally symmetric phase ($m_f=0$).
The terms of mixed chirality (such as $\langle LRR\rangle$ and so on) vanish in this limit, leaving only the chiral preserving interactions LLL and RRR. In this limit we can formally impose the relation
\ba
\langle LLL\rangle^{\lambda\mu\nu}(m_f= 0)=-4\Delta_{AAA}(0)
\label{smargia}
\ea
that will be used extensively in all the work. This relation or other similar relations are just the starting point of the entire construction. The final expressions of the anomalous vertices are obtained using the generalized Ward identities of the
theory. What really defines the theories are the distributions of the partial anomalies. We will attach an equal anomaly on each axial-vector vertex in diagrams of the form
$AAA$ and we will compensate this equal distribution with additional CS interactions - so to bring these diagrams to the desired form $AVV$ or $VAV$ or $VVA$ - whenever a non anomalous $U(1)$ appears at a given vertex. For models
where a single anomalous $U(1)$ is present this does not bring-in any ambiguity. For instance, conservation of the
$Y$ current in $B_iYY$ will allow us to move the anomaly from the $Y$'s to the $B_i$ vertices and this is implicitly done using a CS term. We say that this procedure is allowing us to {\em absorb} a CS interaction.
Moving to the $YYY$ vertex, this vanishes identically in the chiral limit since we factorize left- and
right-handed modes for each generation by an anomaly-free  charge assignment

\ba
 (YYY): \qquad g_{Y}^3 Tr[Q_{Y}^{3}]=0,
\ea
\ba
 (YWW):\qquad g_{Y} g_{2}^2 Tr[Q_{Y} (T_{3}^{L})^2]=0.
\ea
At this point we pause to show how the re-distribution of the anomaly goes in the case at hand.
We have the contribution
\beq
V_{CS}^{B_iYY}= d_i \langle B_i Y\wedge F_Y\rangle
\eeq
and the BRST conditions in the St\"uckelberg phase give
\beq
{d}_i=-ig_{B_i}g_Y^2\frac{2}{3}a_n D_{B_iYY};\hspace{2cm} D_{B_iYY}=\frac{1}{8}Tr[Q_{B_i} Q_Y^2].
\eeq
Also these terms are projected on the vertex to give
\ba
&&V^{B_iYY}_{CS}=d_i \langle B_i Y\wedge F_Y\rangle=(-i)d_i\varepsilon^{\lambda\mu\nu\alpha}(k_{1\alpha}-k_{2\alpha})
\left[(O^{A})_{Y\g}^{2}(O^{A})_{B_iZ_l}\right]Z^{\lambda}_{l} A_{\g}^{\mu}A_{\g}^{\nu}+\dots
\nonumber\\
&&V^{B_iWW}_{CS}=c_i \langle \varepsilon^{\mu\nu\rho\sigma} B_{\mu,i}C_{\nu\rho\sigma}^{Abelian}\rangle=
(-i)c_i\varepsilon^{\lambda\mu\nu\alpha}(k_{1\alpha}-k_{2\alpha})
\left[(O^{A})_{W_3\g}^{2}(O^{A})_{B_iZ_l}\right]Z^{\lambda}_l A_{\g}^{\mu}A_{\g}^{\nu}+\dots
\nonumber\\\nonumber\\
\ea
In general, a vertex such as $B_iYY$ is changed into an ${\bf AVV}$,
while vertices of the form $YB B$ and $YB_i B_j$ which appear in the computation of the $\gamma ZZ$
$\gamma Z_lZ_m$ interactions are changed into {\bf VAV} + {\bf VVA}.
This procedure is summarized by the equations
\ba
&&\Delta_{AAA}^{\lambda\mu\nu}(m_f=0,k_1,k_2)-\frac{a_n}{3}\varepsilon^{\lambda\mu\nu\alpha}(k_{1,\alpha}-k_{2,\alpha})
=\Delta_{AVV}^{\lambda\mu\nu}(m_f=0,k_1,k_2)\nonumber\\\nonumber\\
&&\Delta_{AAA}^{\mu\nu\lambda}(m_f=0,k_2,-k)
-\frac{a_n}{3}\varepsilon^{\mu\nu\lambda\alpha}(k_{1,\alpha}+2 k_{2,\alpha})=
\nonumber\\
&&\Delta_{AVV}^{\mu\nu\lambda}(m_f=0,k_2,-k)=\Delta_{VAV}^{\lambda\mu\nu}(m_f=0,k_1,k_2)
\nonumber\\\nonumber\\
&&\Delta_{AAA}^{\nu\lambda\mu}(m_f=0,-k,k_1)
-\frac{a_n}{3}\varepsilon^{\nu\lambda\mu\alpha}(-2 k_{1,\alpha}-k_{2,\alpha})=
\nonumber\\
&&\Delta_{AVV}^{\nu\lambda\mu}(m_f=0,-k,k_1)=\Delta_{VVA}^{\lambda\mu\nu}(m_f=0,k_1,k_2)
\nonumber\\\nonumber\\
&&\Delta_{AAA}^{\lambda\mu\nu}(m_f=0,k_1,k_2)+\frac{a_n}{6}
\varepsilon^{\lambda\mu\nu\alpha}(k_{1,\alpha}-k_{2,\alpha})=
\nonumber\\
&&\frac{1}{2}\left[(\Delta_{VAV}^{\lambda\mu\nu}(m_f=0,k_1,k_2)+\Delta_{VVA}^{\lambda\mu\nu}(m_f=0,k_1,k_2)\right]\,,
\nonumber\\
\ea
where the last relation can be proved in a simple way by summing the second and the third contributions.
Defining $k_3^{\lambda}=-k^{\lambda}$, one can combine together the {\bf AAA}
plus the counterterms into a unique expression for each case
\ba
{\bf V}^{\lambda \mu \nu}_{B_iYY} &=&4  D^{}_{B_iYY} \, g^{}_{B_i} g^{\,2}_{Y}
\, \Delta^{\lambda\mu \nu}_{\bf AAA} (k_1, k_2 )
+  D^{}_{B_iYY} \, g^{}_{B_i} g^{\,2}_{Y} \frac{i}{ \pi^{2} } \, \frac{2}{3} \,
\epsilon^{\lambda\mu\nu \sigma}(k^{}_{1} - k^{}_{2})^{}_{\sigma}
\nonumber\\
{\bf V}^{ \mu \nu\lambda}_{YB_iY} &=&4  D^{}_{B_iYY} \, g^{}_{B_i} g^{\,2}_{Y}
\, \Delta^{\mu\nu\lambda}_{\bf AAA} (k_2, k_3 )
+  D^{}_{B_iYY} \, g^{}_{B_i} g^{\,2}_{Y} \frac{i}{ \pi^{2} } \, \frac{2}{3} \,
\epsilon^{\mu\nu\lambda \sigma}(k^{}_{2} - k^{}_{3})^{}_{\sigma}
\nonumber\\
{\bf V}^{\nu\lambda\mu}_{YYB_i} &=&4  D^{}_{B_iYY} \, g^{}_{B_i} g^{\,2}_{Y}
\, \Delta^{\nu\lambda\mu}_{\bf AAA} (k_3, k1)
+  D^{}_{B_iYY} \, g^{}_{B_i} g^{\,2}_{Y} \frac{i}{ \pi^{2} } \, \frac{2}{3} \,
\epsilon^{\nu\lambda\mu \sigma}(k^{}_{3} - k^{}_{1})^{}_{\sigma}
\nonumber\\
{\bf V}^{\lambda \mu \nu}_{Y B_i B_j} &=& 4 D^{}_{YB_i B_j} \, g^{}_{Y} g_{B_i}g_{B_j}
 \, \Delta^{\lambda \mu \nu}_{\bf AAA} (k^{}_{1}, k^{}_{2} )
-  D^{}_{YB_iB_j} \, g^{}_{Y} g_{B_i}g_{B_j}  \frac{i}{  \pi^{2} } \, \frac{1}{3} \,
\epsilon^{\lambda \mu \nu \sigma}(k^{}_{1} - k^{}_{2})^{}_{\sigma},\nonumber\\
\label{vertices}
\ea
where we have rotated them onto the $Z_l\gamma\gamma$ vertex.
For the non abelian case ($WB_iW$ and $WWB_i$), the calculation is similar, so we omit the details.

Finally the anomalous contributions plus the CS interactions are given by
\ba
&&\langle B_iYY\rangle|_{m_f=0}+\langle B_iWW\rangle|_{m_f=0}=
\nonumber\\
&&\hspace{1.8cm}
+g_{B_i} g_Y^2\sum_f\left[Q_{B_i,f}^{L} (Q_{Y,f}^{L})^2-Q_{B_i,f}^{R} (Q_{Y,f}^{R})^2\right]
\frac{1}{2}\Delta_{AAA}^{\lambda\mu\nu}(0)R^{B_iYY}_{Z_l\g\g} Z^{\lambda}_l A_{\g}^{\mu}A_{\g}^{\nu}
\nonumber\\
&&\hspace{1.8cm}
+ g_{B_i} g_2^2\sum_f Q_{B_i,f}^{L}(T^{3}_{L,f})^2
\frac{1}{2}\Delta_{AAA}(0)^{\lambda\mu\nu}R^{B_iWW}_{Z_l\g\g} Z^{\lambda}_{l}A_{\g}^{\mu}A_{\g}^{\nu}
\nonumber\\
&&\hspace{1.8cm}
-i \left[g_{B_i} g_{Y}^2\frac{4}{3}a_n D_{B_iYY}R^{B_iYY}_{Z_l\g\g}
+g_{B_i} g_{2}^2\frac{4}{3}a_n D_{B_i}^{(L)}R^{B_iWW}_{Z_l\g\g}\right]
\varepsilon^{\lambda\mu\nu\alpha}
\left(k_{1,\alpha}-k_{2,\alpha}\right) Z^{\lambda}_{l}A_{\g}^{\mu}A_{\g}^{\nu},
\nonumber\\
\ea
which allows to move the anomaly on the axial current
and we simply get
\ba
&&\langle Z_l\g\g\rangle|_{m_f=0}=\sum_i
g_{B_i} g_Y^2\sum_f\left[Q_{B_i,f}^{L} (Q_{Y,f}^{L})^2-Q_{B_i,f}^{R} (Q_{Y,f}^{R})^2\right]
\frac{1}{2}\Delta_{AVV}^{\lambda\mu\nu}(0)R^{B_iYY}_{Z_l\g\g} Z^{\lambda}_l A_{\g}^{\mu}A_{\g}^{\nu}
\nonumber\\
&&\hspace{1.8cm}
+\sum_i g_{B_i} g_2^2\sum_fQ_{B_i,f}^{L}(T^{3}_{L,f})^2
\frac{1}{2}\Delta_{AVV}^{\lambda\mu\nu}(0)R^{B_iWW}_{Z_l\g\g} Z^{\lambda}_lA_{\g}^{\mu}A_{\g}^{\nu}\,,
\ea
where we transfer all the anomaly on the vertex labelled by the $\lambda$ index, obtaining that
the Ward identities on the photons are satisfied.

At this point, it is convenient to introduce the chiral asymmetry
\ba
\theta^{Y B_i B_j}_{f}=\left[(Q_{Y,f}^{L}) (Q_{B_i,f}^{L}) (Q_{B_j,f}^{L})-(Q_{Y,f}^{R}) (Q_{B_i,f}^{R}) (Q_{B_j,f}^{R})\right]
\ea
and express the coefficients in front of the CS counterterms as follows
\ba
&&D_{B_iYY}=-\frac{1}{8}\sum_f\theta^{B_iYY}_{f}
\nonumber\\
&&D_{B_iWW}=-\frac{1}{8}\sum_f\theta^{B_iWW}_{f}
\nonumber\\
&&D_{YB_iB_j}=-\frac{1}{8}\sum_f\theta^{YB_iB_j}_{f}\,.
\ea

After some manipulations we obtain the expression of the
$\langle Z_l\g\g\rangle$ vertex in the $m_f =0 $ phase which is given by
\ba
\langle Z_l\g\g\rangle|_{m_f=0}=-\frac{1}{2}\Delta_{AVV}^{\lambda\mu\nu}(0)Z^{\lambda}_{l}A_{\g}^{\mu}A_{\g}^{\nu}
\sum_i\sum_f\left[g_{B_i} g_Y^2\theta_f^{B_iYY}R^{B_iYY}_{Z_l\g\g}+ g_{B_i} g_2^2\theta_f^{B_iWW}R^{B_iWW}_{Z_l\g\g}\right],
\nonumber\\
\label{masslessamp}
\ea
where for $\Delta_{AVV}(0)$ we write
\ba
&&\Delta_{AVV}(0)^{\lambda\mu\nu}(k_1,k_2,0)=\frac{1}{\pi^2}\int_0^1 dx \int_{0}^{1-x}dy
\frac{1}{\Delta(0)}\nonumber\\
&&\hspace{2cm}
\left\{\varepsilon[k_1,\lambda,\mu,\nu]
\left[y(y-1)k_2^2 -x y k_1\cdot k_2\right]
\right.\nonumber\\
&&\hspace{2cm}\left.
+\varepsilon[k_2,\lambda,\mu,\nu]
\left[x(1-x)k_1^2 +x y k_1\cdot k_2\right]
\right.\nonumber\\
&&\hspace{2cm}\left.
+\varepsilon[k_1,k_2,\lambda,\nu]
\left[x(x-1)k_1^{\mu} -x y k_{2}^{\mu} \right]
\right.\nonumber\\
&&\hspace{2cm}\left.
+\varepsilon[k_1,k_2,\lambda,\mu]
\left[x y k_1^{\nu} +(1-y)y k_{2}^{\nu} \right]
\right\}\,,
\nonumber\\
\nonumber\\
&&\Delta(0)=x(x-1)k_1^2+y(y-1)k_2^2-2 x y k_1\cdot k_2.
\label{avv}
\ea

At this stage we should keep in mind that if all the external particles
are on-shell, the total amplitude vanishes because of the Landau-Yang theorem.
In other words the $Z_l$'s can't decay on shell into two on-shell photons.
However it is possible to have two on-shell photons if the initial
state is characterized by an anomalous process as well, such as gluon fusion.
This does not contradict the Landau-Yang theorem since the Z-pole disappears 
\cite{CGM} in the presence of an anomalous $Z^{\prime}$ exchange \cite{CGM}.

\subsection{The $m_f\neq 0$ phase}

Now we move to the analysis of the vertices away from the chiral limit.
Also in this case we separate the mass-dependent from the mass-independent contributions.

\subsubsection{Chirality preserving vertices }
We start analyzing the vertices away from the chiral limit by separating
the chiral preserving contributions from the remaining ones.
The general expression of LLL is given by
\ba
&&\langle LLL\rangle|_{m_f\neq0}=A_1\varepsilon[k_1,\lambda,\mu,\nu]+A_2\varepsilon[k_2,\lambda,\mu,\nu]
+A_3 k_1^{\nu}\varepsilon[k_1,k_2,\lambda,\mu]+A_4 k_2^{\nu}\varepsilon[k_1,k_2,\lambda,\mu]
\nonumber\\
&&\hspace{1.5cm}+A_5 k_1^{\mu}\varepsilon[k_1,k_2,\lambda,\nu]
+A_6 k_2^{\mu}\varepsilon[k_1,k_2,\lambda,\nu]
\ea
where we have removed, for simplicity, the dependence on the charges and the coupling constants.

The divergent structures $A_1$ and $A_2$ are given by
\ba
&&A_1= 8 i \left[{\cal I}_{3 0}(k_1,k_2)-{\cal I}_{2 0}(k_1,k_2) \right]k_1^2
+ 16 i \left[{\cal I}_{1 1}(k_1,k_2) -{\cal I}_{2 1}(k_1,k_2)\right]k_1\cdot k_2
\nonumber\\
&&\hspace{1.5cm}
+8 i \left[{\cal I}_{0 1}(k_1,k_2)-{\cal I}_{0 2}(k_1,k_2)
+{\cal I}_{1 2}(k_1,k_2)\right] k_2^2
\nonumber\\
&&\hspace{1.5cm}
+4 i \left[3{\cal D}_{1 0}(k_1,k_2)-2{\cal D}_{0 0}(k_1,k_2) \right]
\ea
where
\ba
&&{\cal I}_{s t}(k_1,k_2)=\int_0^1 dx \int_0^{1-x} dy \int \frac{d^4q}{(2\pi)^4}\frac{x^s y^t}
{\left[q^2-x(1-x)k_1^2-y(1-y)k_2^2 - 2 x y k_1\cdot k_2+m_f^2\right]^3}\nonumber\\
&&{\cal D}_{s t}(k_1,k_2)=\int_0^1 dx \int_0^{1-x} dy \int \frac{d^4q}{(2\pi)^4}\frac{q^2 x^s y^t}
{\left[q^2-x(1-x)k_1^2-y(1-y)k_2^2 - 2 x y k_1\cdot k_2+m_f^2\right]^3}.\nonumber\\
\ea
and one can verify that $A_1(k_1,k_2)=-A_2(k_2,k_1)$.
All the mass dependence is contained only in the denominators
of the propagators appearing in the Feynman parametrization.

The finite structures $A_3\dots A_6$ are the following
\ba
&&A_3(k_1,k_2)=-16 i {\cal I}_{1 1}(k_1,k_2)=-A_6(k_2,k_1) \nonumber \\
&& A_4(k_1,k_2)=16 i\left[{\cal I}_{0 2}(k_1,k_2)-{\cal I}_{0 1}(k_1,k_2)\right]=-A_5(k_2,k_1)
\ea
where still we need to perform the trivial finite integrals over the momentum $q$.

The decomposition of $\langle LLL\rangle_f$ into massless and massive
components gives
\ba
&&\langle LLL\rangle_f=\langle LLL(m_f\neq 0)\rangle-\langle LLL\rangle(0)\nonumber\\
&&\langle LLL\rangle(0)=\langle LLL(m_f=0)\rangle\nonumber\\
&&\langle LLL(m_f\neq 0)\rangle=\langle LLL\rangle_f+\langle LLL\rangle(0),
\label{separation}
\ea
where we have isolated the massless contributions.
As we have seen before, the CS terms act only on the massless
part of the triangles (having used Eq. (\ref{smargia}))  and
reproduce the massless contribution calculated in Eq. (\ref{masslessamp}).
Since the mass terms are proportional to the tensors
$\varepsilon[k_1,\lambda,\mu,\nu]$ and $\varepsilon[k_2,\lambda,\mu,\nu]$
they can be included in the singular structures $A_1$ and $A_2$
of $\langle LLL\rangle|_{m_f\neq 0}$
\ba
&&\bar{A_1}=A_1+
i m_f^2 (Q_{Y,f}^{R})^2(Q_{Y,f}^{L}) \left[-8 {\cal I}_{0 0}(q^2,k_1,k_2)+24 {\cal I}_{1 0}(q^2,k_1,k_2)
\right]
\nonumber\\
&&\hspace{2cm}
+i m_f^2 (Q_{Y,f}^{L})^2(Q_{Y,f}^{R})\left[8 {\cal I}_{0 0}(q^2,k_1,k_2)-24 {\cal I}_{1 0}(q^2,k_1,k_2)
\right]
\nonumber\\
&&\hspace{2cm}
-8 i m_f^2 Q_{Y,f}^{R}(T^{L}_{3,f})^2 {\cal I}_{1 0}(q^2,k_1,k_2)
\nonumber\\
&&\hspace{2cm}
-i m_f^2 \sum_i Q_{B_i,f}^{R}Q_{Y,f}^{L}Q_{Y,f}^{R}\left[8 {\cal I}_{1 0}(q^2,k_1,k_2)+4 {\cal I}_{0 0}(q^2,k_1,k_2)
\right]
\nonumber\\
&&\hspace{2cm}
+i m_f^2 \sum_iQ_{B_i,f}^{L}Q_{Y,f}^{R}Q_{Y,f}^{L}\left[8 {\cal I}_{1 0}(q^2,k_1,k_2)+4 {\cal I}_{0 0}(q^2,k_1,k_2)
\right]
\nonumber\\
&&\hspace{2cm}
-8 i m_f^2 \sum_iQ_{B_i,f}^{R}(Q_{Y,f}^{L})^2 {\cal I}_{1 0}(q^2,k_1,k_2)
+8 i m_f^2 \sum_iQ_{B_i,f}^{L}(Q_{Y,f}^{R})^2 {\cal I}_{1 0}(q^2,k_1,k_2)
\nonumber\\
&&\hspace{2cm}
-8 i m_f^2 \sum_iQ_{B_i,f}^{R}(T^{L}_{3,f})^2 {\cal I}_{1 0}(q^2,k_1,k_2).
\ea
At this point we have to consider also the chirality flipping terms.
For simplicity we discuss only the case of the $YYY$ vertex, the others being similar.

\subsubsection{Chirality flipping vertices}

These contributions are extracted rather straighforwardly and contribute to the total vertex amplitude with mass corrections that modify $A_1$ and $A_2$.
We discuss this point first for the $\langle YYY\rangle$, and then quote the result for the entire
contribution to $Z \gamma \gamma$.

For YYY we obtain
\ba
&&(Q_{Y,f}^{R})^2(Q_{Y,f}^{L})\left[\langle RRL\rangle+\langle LRR\rangle+\langle RLR\rangle\right]=
\nonumber\\
&&\hspace{1.5cm}
(Q_{Y,f}^{R})^2(Q_{Y,f}^{L})\left[
8 i m_f^2{\cal I}_{0 0}(k_1,k_2)\left(\varepsilon[k_2,\lambda,\mu,\nu]
-\varepsilon[k_1,\lambda,\mu,\nu]\right)
\right.\nonumber\\
&&\hspace{1.5cm}\left.
+24 i m_f^2\left({\cal I}_{1 0}(k_1,k_2)\varepsilon[k_1,\lambda,\mu,\nu]
-{\cal I}_{0 1}(k_1,k_2)\varepsilon[k_2,\lambda,\mu,\nu]\right)\right],
\ea
and the analysis can be extended to the other trilinear contributions and can be simplified using the relations
\ba
\left[\langle RRL\rangle+\langle LRR\rangle+\langle RLR\rangle\right]=
-\left[\langle LLR\rangle+\langle RLL\rangle+\langle LRL\rangle\right].
\ea
The final result is given by
\ba
&&\textrm{mass terms}=i m_f^2 g_Y^3(Q_{Y,f}^{R})^2(Q_{Y,f}^{L})\left[
8{\cal I}_{0 0}(k_1,k_2)\left(\varepsilon[k_2,\lambda,\mu,\nu]
-\varepsilon[k_1,\lambda,\mu,\nu]\right)
\right.\nonumber\\
&&\hspace{1.5cm}\left.
+24\left({\cal I}_{1 0}(k_1,k_2)\varepsilon[k_1,\lambda,\mu,\nu]
-{\cal I}_{0 1}(k_1,k_2)\varepsilon[k_2,\lambda,\mu,\nu]\right)\right]
\nonumber\\
&&\hspace{1.5cm}
-i m_f^2 g_Y^3(Q_{Y,f}^{R})^2(Q_{Y,f}^{L})\left[
8{\cal I}_{0 0}(k_1,k_2)\left(\varepsilon[k_2,\lambda,\mu,\nu]
-\varepsilon[k_1,\lambda,\mu,\nu]\right)
\right.\nonumber\\
&&\hspace{1.5cm}\left.
+24\left({\cal I}_{1 0}(k_1,k_2)\varepsilon[k_1,\lambda,\mu,\nu]
-{\cal I}_{0 1}(k_1,k_2)\varepsilon[k_2,\lambda,\mu,\nu]\right)\right]
\nonumber\\
&&\hspace{1.5cm}
+8 i m_f^2 g_Y g_2^2 Q_{Y,f}^{R}(T_{3,f}^{L})^2
\left({\cal I}_{0 1}(k_1,k_2)\varepsilon[k_2,\lambda,\mu,\nu]
-{\cal I}_{1 0}(k_1,k_2)\varepsilon[k_1,\lambda,\mu,\nu]\right)
\nonumber\\
&&\hspace{1.5cm}
+i m_f^2 \sum_i g_{B_i} g_Y^2 Q_{B_i,f}^{L}Q_{Y,f}^{R}Q_{Y,f}^{L}\left[(8{\cal I}_{0 1}(q^2,k_1,k_2)
-4{\cal I}_{0 0}(k_1,k_2))\varepsilon[k_2,\lambda,\mu,\nu]
\right.\nonumber\\
&&\hspace{2cm}\left.
+(8{\cal I}_{1 0}(k_1,k_2)
+4{\cal I}_{0 0}(k_1,k_2))\varepsilon[k_1,\lambda,\mu,\nu]\right]
\nonumber\\
&&\hspace{1.5cm}
-i m_f^2 \sum_i g_{B_i} g_Y^2 Q_{B_i,f}^{R}Q_{Y,f}^{L}Q_{Y,f}^{R}\left[(8{\cal I}_{0 1}
(k_1,k_2)
-4{\cal I}_{0 0}(k_1,k_2))\varepsilon[k_2,\lambda,\mu,\nu]
\right.\nonumber\\
&&\hspace{2cm}\left.
+(8{\cal I}_{1 0}(k_1,k_2)
+4{\cal I}_{0 0}(k_1,k_2))\varepsilon[k_1,\lambda,\mu,\nu]\right]
\nonumber\\
&&\hspace{1.5cm}
+i m_f^2\sum_i g_{B_i} g_Y^2 Q_{B_i,f}^{R}(Q_{Y,f}^{L})^2
8 \left({\cal I}_{0 1}(k_1,k_2)\varepsilon[k_2,\lambda,\mu,\nu]
-{\cal I}_{1 0}(k_1,k_2)\varepsilon[k_1,\lambda,\mu,\nu]\right)
\nonumber\\
&&\hspace{1.5cm}
-i m_f^2\sum_i g_{B_i} g_Y^2 Q_{B_i,f}^{L}(Q_{Y,f}^{R})^2
8 \left({\cal I}_{0 1}(k_1,k_2)\varepsilon[k_2,\lambda,\mu,\nu]
-{\cal I}_{1 0}(k_1,k_2)\varepsilon[k_1,\lambda,\mu,\nu]\right)
\nonumber\\
&&\hspace{1.5cm}
+8 i m_f^2\sum_i g_{B_i} g_2^2  Q_{B_i,f}^{R}(T_{3,f}^{L})^2
\left({\cal I}_{0 1}(k_1,k_2)\varepsilon[k_2,\lambda,\mu,\nu]
-{\cal I}_{1 0}(k_1,k_2)\varepsilon[k_1,\lambda,\mu,\nu]\right)
\nonumber\\
\ea
and is finite. To conclude our derivation in this special case, we can summarize our findings as follows.

\begin{figure}[t]
{\centering \resizebox*{12cm}{!}{\rotatebox{0}
{\includegraphics{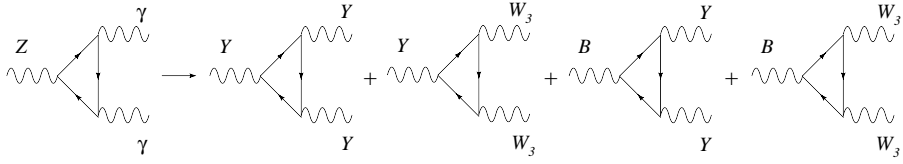}}}\par}
\caption{Interaction basis contributions to the $Z\gamma \gamma$ vertex.
In the SM only the first two diagrams survive. The CS terms, in this case, are absorbed
so that only the $B$ vertex is anomalous. In the chiral limit in the SM the first two diagrams vanish.}
\label{ZGG1}
\end{figure}

In a triangle diagram of the form, say, AVV, if we impose a vector Ward identity
on the two V lines we redefine the divergent invariant amplitudes
$A_1$ and $A_2$ ($A_2=- A_1$) in terms of the remaining amplitudes
$A_3,...,A_6$,  which are convergent. The chirality flip contributions such
as $LLR $ turn out to be finite, but are proportional to $A_1$ and $A_2$, and
disappear once we impose the WI's on the V lines.
This observation clarifies why in the $Z\gamma \gamma$ vertex of the SM
the mass dependence of  the numerators disappears and the traces can be computed as in the chiral limit.
Including the mass dependent contributions we obtain (see Fig.~\ref{ZGG1} for the $m_f\neq 0$ phase)
\ba
&&\langle Z_l\g\g\rangle|_{m_f\neq0}\,=\langle Z_l\g\g\rangle|_{m_f=0}
-\sum_f \frac{1}{8}\langle LLL\rangle_f^{\lambda\mu\nu}
\left\{
g_Y^3\theta_{f}^{YYY}\bar{R}^{YYY}_{Z_l\g\g}
+g_2^3\theta_{f}^{WWW}\bar{R}^{WWW}_{Z_l\g\g}
\right.\nonumber\\
&&\hspace{1.8cm}\left.
+g_2^2 g_Y\theta_{f}^{YWW} R^{YWW}_{Z_l\g\g}
+g_2 g_Y^2\theta_{f}^{YYW} R^{YYW}_{Z_l\g\g}
+ \sum_i g_{B_i} g_2 g_Y\theta_{f}^{B_iYW}R^{B_iYW}_{Z_l\g\g}
\right.\nonumber\\
&&\hspace{1.8cm}\left.
+ \sum_i g_{B_i} g_Y^2\theta_{f}^{B_iYY}R^{B_iYY}_{Z_l\g\g}
+ \sum_i g_{B_i} g_2^2\theta_{f}^{B_iWW}R^{B_iWW}_{Z_l\g\g}
\right\}Z^{\lambda}_l A_{\g}^{\mu}A_{\g}^{\nu}
\nonumber\\
&&\hspace{1.8cm}
+m_f^2~ (\textrm{chirally flipped terms})
\label{complete}
\ea
where $\langle LLL\rangle_f^{\lambda\mu\nu}$ is now defined by Eq.(\ref{separation}).
In Eq.(\ref{complete}) we have also defined the following chiral asymmetries
\ba
&&\theta_{f}^{WWW}=(T^{3}_{L,f})^{3}
\nonumber\\
&&\theta_{f}^{YYW}=\left[(Q^{L}_{Y,f})^2 T^{3}_{L,f}\right]
\nonumber\\
&&\theta_{f}^{B_iYW}=\left[Q^{B_i,f}Q^{L}_{Y,f} T^{3}_{L,f}\right]
\ea

It is important to note that Eq.(\ref{complete}) is still expressed as
in Rosenberg (see \cite{Rosenberg}, \cite{CIM1}), with the usual finite cubic terms in the
momenta $k_1$ and $k_2$, the two singular invariant amplitudes ($A_1$ and $A_2$) and the mass contributions.

At this stage, to get the physical amplitude,
we must impose e.m. current conservation on the external photons
\ba
&k_1^{\mu}\langle Z_l\g\g\rangle|_{m_f\neq0}^{\lambda\mu\nu}=0\nonumber\\
&k_2^{\nu}\langle Z_l\g\g\rangle|_{m_f\neq0}^{\lambda\mu\nu}=0\,.
\ea
Using these conditions, again we can re-express the coefficient $\bar{A_1},\bar{A_2}$ in terms of
$A_3,\dots,A_6$ and we drop the explicit mass dependence in the numerators of the expression of
the physical amplitude.

Thus, applying the Ward identities on the triangle $\langle LLL\rangle_f$, it reduces
to the combination $\Delta_{AVV}(m_f)-\Delta_{AVV}(0)$ which must be added to the first term in the curly brackets of
Eq.(\ref{complete}), thereby giving our final result for the physical amplitude
\ba
&&\langle Z_l\g\g\rangle|_{m_f\neq 0}=
-\frac{1}{2}Z^{\lambda}_l A_{\g}^{\mu}A_{\g}^{\nu}
\sum_f\left[
g_Y^3\theta_f^{YYY}\bar{R}^{YYY}_{Z_l\g\g}
+g_2^3\theta_f^{WWW}\bar{R}^{WWW}_{Z_l\g\g}
+g_Y g_2^2\theta_f^{YWW}R^{YWW}_{Z_l\g\g}
\right.\nonumber\\
&&\hspace{2cm}\left.
+ g_Y^2 g_2\theta_f^{YYW}R^{YYW}_{Z_l\g\g}
+\sum_i g_{B_i} g_Y g_2\theta_f^{B_iYW}R^{B_i Y W}_{Z_l\g\g}
\right.\nonumber\\
&&\hspace{2cm}\left.
+\sum_i g_{B_i} g_Y^2\theta_f^{B_iYY}R^{B_iYY}_{Z_l\g\g}
+ g_{B_i} g_2^2\theta_f^{B_iWW}R^{B_iWW}_{Z_l\g\g}\right]
\Delta_{AVV}^{\lambda\mu\nu}(m_f\neq 0).\nonumber\\
\ea
We have defined
\beq
\bar{R}^{YYY}_{Z_l\g\g}=(O^A)_{YZ_l}(O^{A})_{Y\g}^{2},
\hspace{1cm}
\bar{R}^{WWW}_{Z_l\g\g}=(O^A)_{W_3Z_l}(O^{A})_{W_3\g}^{2},
\eeq
and the triangle $\Delta_{AVV}(m_f\neq 0)$ is given by
\ba
&&\Delta_{AVV}(m_f\neq 0,k_1,k_2)^{\lambda\mu\nu}=\frac{1}{\pi^2}\int_0^1 dx \int_{0}^{1-x}dy
\frac{1}{\Delta(m_f)}\nonumber\\
&&\hspace{2cm}
\left\{\varepsilon[k_1,\lambda,\mu,\nu]
\left[y(y-1)k_2^2 -x y k_1\cdot k_2\right]
\right.\nonumber\\
&&\hspace{2cm}\left.
+\varepsilon[k_2,\lambda,\mu,\nu]
\left[x(1-x)k_1^2 +x y k_1\cdot k_2\right]
\right.\nonumber\\
&&\hspace{2cm}\left.
+\varepsilon[k_1,k_2,\lambda,\nu]
\left[x(x-1)k_1^{\mu} -x y k_{2}^{\mu} \right]
\right.\nonumber\\
&&\hspace{2cm}\left.
+\varepsilon[k_1,k_2,\lambda,\mu]
\left[x y k_1^{\nu} +(1-y)y k_{2}^{\nu} \right]
\right\}\,,
\nonumber\\
\nonumber\\
&&\Delta(m_f)=m_f^2+x(x-1)k_1^2+y(y-1)k_2^2-2 x y k_1\cdot k_2\,.
\ea

\subsubsection{The SM limit}

It is straightforward to obtain the corresponding expression in the SM
from the previous result. As usual we obtain, beside the 
tensor structures of the Rosenberg expansion, all the chirally flipped 
terms which are proportional to a mass term times a tensor
$k_{1,2}^{\alpha}\varepsilon[\alpha,\lambda,\mu,\nu]$.
As we have seen before in the previous sections all these terms
can be re-absorbed once we impose the conservation of the electromagnetic current.

Then, setting the anomalous pieces to zero by taking $g_{B_i}\rightarrow 0$, we are left with
the usual Z boson ($Z_l\rightarrow Z$), and we have
\clearpage
\ba
&&\langle Z\g\g\rangle|_{m_f\neq 0}=-g_Z e^2\sum_f
\left[Q_Z^{L,f}(Q_f^{L})^2-Q_Z^{R,f}(Q_f^{R})^2\right]
\frac{1}{2}\Delta_{AVV}^{\lambda\mu\nu}(m_f\neq 0) Z^{\lambda}A_{\g}^{\mu}A_{\g}^{\nu}
\nonumber\\
&&\hspace{1.8cm}
=-\sum_f \frac{1}{2} \Delta_{AVV}^{\lambda\mu\nu}(m_f\neq 0)
\left\{
g_Y^3\theta_f^{YYY}\bar{R}^{YYY}
+g_2^2 g_Y \theta_f^{YWW} R^{YWW}_{Z\g\g}
\right.\nonumber\\
&&\hspace{1.8cm}\left.
+g_2^3 \theta_f^{WWW} \bar{R}^{WWW}_{Z\g\g}
+g_Y^2g_2\theta_f^{YYW}R^{YYW}_{Z\g\g}
\right\}
Z^{\lambda}A_{\g}^{\mu}A_{\g}^{\nu},
\nonumber\\
\ea
where the coefficients $\bar{R}^{YYY}_{Z\g\g},\bar{R}^{WWW}_{Z\g\g}$ 
are defined in the previous section.
It is not difficult to recognize that in the first line we have
\ba
\langle Z\g\g\rangle|_{m_f\neq 0}=-g_Z e^2 \frac{1}{2}\sum_f(Q_f)^2
\left[Q_Z^{L,f}-Q_Z^{R,f}\right]\Delta_{AVV}^{\lambda\mu\nu}(m_f\neq 0)
Z^{\lambda}A_{\g}^{\mu}A_{\g}^{\nu}
\ea
and since
\ba
&&\left[Q_Z^{L,f}-Q_Z^{R,f}\right]=2 g_{A,f}^{Z}\nonumber\\
&&g_Z\approx\frac{g_2}{\cos{\theta}_W}
\ea
finally we obtain
\ba
\langle Z\g\g\rangle|_{m_f\neq 0}=-\frac{g_2}{\cos{\theta}_W} e^2 \sum_f(Q_f)^2
g_{A,f}^{Z}\Delta_{AVV}^{\lambda\mu\nu}(m_f\neq 0)
Z^{\lambda}A_{\g}^{\mu}A_{\g}^{\nu}\,,
\ea
which is exactly the SM vertex \cite{Boud}.

\begin{figure}[t]
{\centering \resizebox*{12cm}{!}{\rotatebox{0}
{\includegraphics{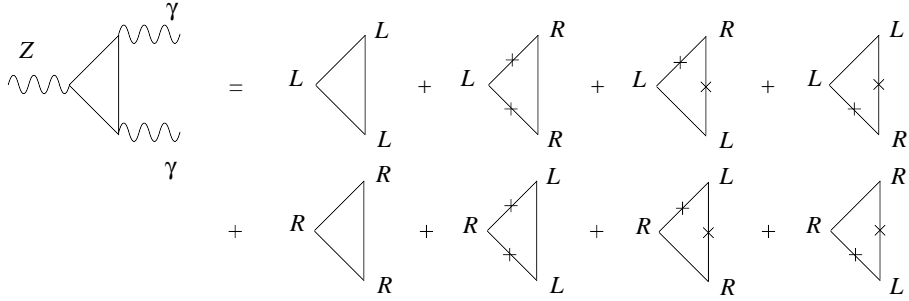}}}\par}
\caption{Chiral triangle contributions to the $Z\g\g$ vertex.}
\label{Zgaga}
\end{figure}

\section{The $\g ZZ$ vertex}

Before coming to analyze the most general cases involving two or three anomalous
$Z^{\prime}$s, it is more convenient to start with the $\gamma Z Z$ interaction
with two identical $Z^{\prime}$s in the final state and use the result in this
simpler case for the general analysis.

\subsection{The vertex in the chiral limit}

We proceed in the same manner as before. In the $m_f= 0$ phase, the terms in the interaction
eigenstates basis we need to consider are
\ba
&&\frac{1}{3!}Tr\left[Q_{Y}^3\right]\langle YYY \rangle=\frac{1}{3!} Tr\left[Q_{Y}^3\right]
\left[3 (O^{A}_{YZ})^2 O^{A}_{Y\gamma}\right]\langle \g ZZ\rangle+\dots
\nonumber\\
&&\frac{1}{2!}Tr\left[Q_{Y} T_{3}^2\right]\langle YWW \rangle=\frac{1}{2!}Tr\left[Q_{Y} T_{3}^2\right]
\left[2 O^{A}_{WZ} O^{A}_{W\gamma} O^{A}_{YZ}+(O^{A}_{WZ})^2 O^{A}_{Y\gamma}\right]\langle \g ZZ\rangle
+\dots
\nonumber\\
&&\frac{1}{2!}Tr\left[Q_{Y} Q_{B}^2\right]\langle YBB \rangle=\frac{1}{2!}Tr\left[Q_{Y} Q_B^2\right]
\left[O^{A}_{Y\g} (O^{A}_{BZ})^2 \right]\langle \g ZZ\rangle
+\dots
\nonumber\\
&&\frac{1}{2!}Tr\left[Q_{B} Q_{Y}^2\right]\langle BYY \rangle=\frac{1}{2!}Tr\left[Q_{B} Q_{Y}^2\right]
\left[2 O^{A}_{BZ} O^{A}_{YZ} O^{A}_{Y\gamma}\right]\langle \g ZZ\rangle+\dots
\nonumber\\
&&\frac{1}{2!}Tr\left[Q_{B} T_{3}^2\right]\langle BWW \rangle=\frac{1}{2!}Tr\left[Q_{B} T_{3}^2\right]
\left[2 O^{A}_{BZ} O^{A}_{WZ} O^{A}_{W\gamma}\right]\langle \g ZZ\rangle+\dots
\ea
We define for future reference the following expressions for the rotation matrices
\ba
&&R^{YYY}_{\g ZZ}=\left[3 (O^{A}_{YZ})^2 O^{A}_{Y\gamma}\right]
\nonumber\\
&&R^{WWW}_{\g ZZ}=\left[3 (O^{A}_{W_3 Z})^2 O^{A}_{W_3\gamma}\right]
\nonumber\\
&&R^{WYY}_{\g ZZ}=\left[2 O^{A}_{W_3 Z} O^{A}_{Y \gamma} O^{A}_{YZ}+(O^{A}_{W_3 \g}) (O^{A}_{YZ})^2\right]
\nonumber\\
&&R^{YWW}_{\g ZZ}=\left[2 O^{A}_{W_3 Z} O^{A}_{W_3 \gamma} O^{A}_{YZ}+(O^{A}_{W_3 Z})^2 O^{A}_{Y\gamma}\right]
\nonumber\\
&&R^{BYY}_{\g ZZ}=\left[2 O^{A}_{BZ} O^{A}_{YZ} O^{A}_{Y\gamma}\right]
\nonumber\\
&&R^{BBY}_{\g ZZ}=\left[O^{A}_{Y\g} (O^{A}_{BZ})^2 \right]
\nonumber\\
&&R^{BBW}_{\g ZZ}=\left[O^{A}_{W_3\g} (O^{A}_{BZ})^2 \right]
\nonumber\\
&&R^{BWW}_{\g ZZ}=\left[2 O^{A}_{BZ} O^{A}_{W_3 Z} O^{A}_{W_3\gamma}\right]
\nonumber\\
&&R^{BYW}_{\g ZZ}=\left[O^{A}_{BZ} O^{A}_{W_3 Z} O^{A}_{Y\gamma}
+O^{A}_{BZ} O^{A}_{W_3 \g} O^{A}_{YZ}\right].
\ea
The chiral decomposition proceeds similarly to the case of $Z\g\g$ (see Fig.~\ref{Zgaga}).
Also in this situation the tensor $\langle LLL\rangle_f^{\lambda\mu\nu}$ is characterized by the two independent
momenta $k_{1,\mu}$ and $k_{2,\nu}$ of the two outgoing $Z^{\prime}$s.
Since the $LLL$ triangle is still ill-defined, we must distribute the anomaly in a certain way.
This is driven by the symmetry of the theory, and in this case the STI's play a crucial role even in the ($m_f=0$) unbroken chiral phase of the theory.
In order to define the $\langle LLL\rangle^{\lambda\mu\nu}|_{m_f=0}$ diagram we choose
a symmetric assignment of the anomaly

\ba
&&k_{1,\mu}\langle LLL\rangle^{\lambda\mu\nu}|_{m_f=0}=\frac{a_n}{3}\varepsilon[k_1,k_2,\lambda,\nu]
\nonumber\\
&&k_{2,\nu}\langle LLL\rangle^{\lambda\mu\nu}|_{m_f=0}=-\frac{a_n}{3}\varepsilon[k_1,k_2,\lambda,\mu]
\nonumber\\
&&k_{\lambda}\langle LLL\rangle^{\lambda\mu\nu}|_{m_f=0}=\frac{a_n}{3}\varepsilon[k_1,k_2,\mu,\nu]\,.
\ea
These conditions together with the Bose symmetry on the two $Z^\prime$s
\ba
\langle LLL\rangle^{\lambda\mu\nu}|_{m_f=0}(k,k_1,k_2)=
\langle LLL\rangle^{\lambda\nu\mu}|_{m_f=0}(k,k_2,k_1)\,
\ea
allow us to remove the singular coefficients proportional to
the two linear tensor structures of the amplitude. The complete
tensor structure of the $\g ZZ$ vertex in this case can be written in
terms of the usual invariant amplitudes $A_1,...A_6$
\ba
&&A_3=-16 \left({\cal I}_{10}(k_1,k_2)-{\cal I}_{20}(k_1,k_2)\right)\nonumber\\
&&A_4=+16 {\cal I}_{11}(k_1,k_2)\nonumber\\
&&A_5=-16 {\cal I}_{11}(k_1,k_2)\nonumber\\
&&A_6=-16 \left({\cal I}_{01}(k_1,k_2)-{\cal I}_{02}(k_1,k_2)\right)\nonumber\\
&&A_1=-k_1\cdot k_2 A_5 -k_2^2 A_6 + \frac{a_n}{3}\nonumber\\
&&A_2=-k_1\cdot k_2 A_4 -k_1^2 A_3 - \frac{a_n}{3}.
\ea
We have the constraints
\ba
&&k_{\lambda}\langle LLL\rangle^{\lambda\mu\nu}|_{m_f=0}=\frac{a_n}{3}\varepsilon\left[k_1,k_2,\mu,\nu\right]
\Rightarrow A_1-A_2=\frac{a_n}{3}
\ea
and the relation written in Eq. (\ref{smargia}).
In this case the CS terms coming from the lagrangean in the interaction eigenstates basis
are defined as follows
\ba
&&V_{CS}=\sum_f\left\{-g_B g_Y^2\frac{1}{8}\theta_f^{YBY}R_{\g ZZ}^{YBY}
\frac{a_n}{3}\varepsilon^{\mu\nu\lambda\alpha}(k_{2,\alpha} - k_{3,\alpha})
-g_B g_Y^2\frac{1}{8} \theta_f^{YYB} R_{\g ZZ}^{YYB}
\frac{a_n}{3}\varepsilon^{\nu\lambda\mu\alpha}(k_{3,\alpha}-k_{1,\alpha})
\right.\nonumber\\
&&\hspace{2cm}\left.
+ g_Y g_B^2 \frac{1}{8} \theta_{f}^{YBB}R_{ZZ\g}^{YBB}
\frac{a_n}{6}\varepsilon^{\lambda\mu\nu\alpha}(k_{1,\alpha}-k_{2,\alpha})
-g_B g_2^2\frac{1}{8}\theta^{WBW}_f R_{ZZ\g}^{WBW}
\frac{a_n}{3}\varepsilon^{\mu\nu\lambda\alpha}(k_{2,\alpha}-k_{3,\alpha})
\right.\nonumber\\
&&\hspace{2cm}\left.
-g_B g_2^2\frac{1}{8}\theta^{WWB}_f R_{ZZ\g}^{WWB}
\frac{a_n}{3}\varepsilon^{\nu\lambda\mu\alpha}(k_{3,\alpha}-k_{1,\alpha})
\right\}\,.
\ea

Then, collecting all the terms, 
the expression in the $m_f=0$ phase for the $\g ZZ$ process can be written as
\ba
&&\langle \g ZZ\rangle|_{m_f=0}=
-\frac{1}{2}A_{\g}^{\lambda}Z^{\mu}Z^{\nu} \sum_f \left\{
g_B g_Y^2\theta_f^{YBY}R_{\g ZZ}^{YBY}
\left[\Delta_{AAA}^{\mu\nu\lambda}(0)
-\frac{a_n}{3}\varepsilon^{\mu\nu\lambda\alpha}(k_{2,\alpha} - k_{3,\alpha})\right]
\right.\nonumber\\
&&\hspace{2cm}\left.
+g_B g_Y^2 \theta_f^{YYB} R_{\g ZZ}^{YYB}
\left[\Delta_{AAA}^{\nu\lambda\mu}(0)
-\frac{a_n}{3}\varepsilon^{\nu\lambda\mu\alpha}(k_{3,\alpha}-k_{1,\alpha})\right]
\right.\nonumber\\
&&\hspace{2cm}\left.
+ g_Y g_B^2  \theta_{f}^{YBB}R_{ZZ\g}^{YBB}
\left[\Delta_{AAA}^{\lambda\mu\nu}(0)
+\frac{a_n}{6}\varepsilon^{\lambda\mu\nu\alpha}(k_{1,\alpha}-k_{2,\alpha})\right]
\right.\nonumber\\
&&\hspace{2cm}\left.
+g_B g_2^2\theta^{WBW}_f R_{ZZ\g}^{WBW}
\left[\Delta_{AAA}^{\mu\nu\lambda}(0)
-\frac{a_n}{3}\varepsilon^{\mu\nu\lambda\alpha}(k_{2,\alpha}-k_{3,\alpha})\right]
\right.\nonumber\\
&&\hspace{2cm}\left.
+g_B g_2^2\theta^{WWB}_f R_{ZZ\g}^{WWB}
\left[\Delta_{AAA}^{\nu\lambda\mu}(0)
-\frac{a_n}{3}\varepsilon^{\nu\lambda\mu\alpha}(k_{3,\alpha}-k_{1,\alpha})\right]
\right\}\,,
\nonumber\\
\ea
and after some manipulations, we obtain
\ba
&&\langle \g ZZ\rangle|_{m_f=0}=
-\frac{1}{2}\left[\Delta_{VAV}^{\lambda\mu\nu}(0)+\Delta_{VVA}^{\lambda\mu\nu}(0)\right]
A_{\g}^{\lambda}Z^{\mu}Z^{\nu}
\sum_f\left\{g_B g_Y^2 \theta_f^{BYY}R^{BYY}
\right.\nonumber\\
&&\hspace{3cm}\left.
+g_Y g_B^2\theta_f^{YBB} \bar{R}^{YBB}
+g_B g_2^2\theta_f^{BWW}R^{BWW}
\right\},
\nonumber\\
\ea
where we have used
\ba
&&\theta_f^{YBB}=Q_{Y,f}^{L}(Q_{B,f}^{L})^2-Q_{Y,f}^{R}(Q_{B,f}^{R})^2
\nonumber\\
&&\bar{R}^{BBY}_{\g Z Z}=\frac{1}{2} R^{BBY}_{\g Z Z}.
\ea
If we define
\beq
T^{\lambda\mu\nu}(0)=\left[\Delta_{VAV}^{\lambda\mu\nu}(0)+\Delta_{VVA}^{\lambda\mu\nu}(0)\right]
\eeq
we can write an explicit expression for $T^{\lambda\mu\nu}$, which is given by
\ba
&&T^{\lambda\mu\nu}(0)=\frac{1}{\pi^2}\int_0^{1}d x\int_0^{1-x}d y\frac{1}{\Delta(0)}
\left\{
\varepsilon^{\alpha\lambda\mu\nu}k_{1,\alpha}\left[(1-x)x k_1^2 + y(y-1)k_2^2\right]
\right.\nonumber\\
&&\left.\hspace{4cm}
+\varepsilon^{\alpha\lambda\mu\nu}k_{2,\alpha}\left[(1-x)x k_1^2 +y(y-1)k_2^2\right]
\right.\nonumber\\
&&\left.\hspace{4cm}
+\varepsilon[k_1,k_2,\lambda,\nu]\left[2(x-1)x k_{1,\mu} - 2x y k_{2,\mu}\right]
\right.\nonumber\\
&&\left.\hspace{4cm}
+\varepsilon[k_1,k_2,\lambda,\mu]\left[2(1-y)y k_{2,\nu} + 2x y k_{1,\nu}\right]
\right\}\,,
\ea
and it is straightforward to observe that the electromagnetic current conservation is satisfied
on the photon line
\ba
&&k_{1,\mu}T^{\lambda\mu\nu}=\frac{1}{2\pi^2}\varepsilon\left[k_1,k_2,\lambda,\nu\right]
\nonumber\\
&&k_{2,\nu}T^{\lambda\mu\nu}=-\frac{1}{2\pi^2}\varepsilon\left[k_1,k_2,\lambda,\mu\right]
\nonumber\\
&&(k_{1,\lambda}+k_{2,\lambda})T^{\lambda\mu\nu}=0.\,
\ea

\subsection{$\g ZZ$: The $m_f\neq 0$ phase}

In the $m_f\neq 0$ phase we must add to the previous chirally conserved contributions
all the chirally flipped interactions of the type $\langle LLR\rangle$ and similar,  which are
proportional to $m_f^2$. As we have already seen in the $Z\g\g$ case, all the
mass terms have a tensor structure of the type
$m_f^2\varepsilon^{\alpha\lambda\mu\nu}k_{1,2,\alpha}$ and we can always define the
coefficients $\bar{A}_1$ and $\bar{A}_2$ so that they include all the mass terms.
Again, they are expressed in terms of the finite quantities $A_3,\dots, A_6$
by imposing the physical restriction, i.e. the e.m. current conservation on the photon line, and the anomalous
Ward identities on the two $Z^\prime$s lines.
Since the CS interactions act only on the massless part of the triangles,
they are absorbed by splitting the tensor $\langle LLL\rangle^{\lambda\mu\nu}$ as
\ba
&&\langle LLL\rangle^{\lambda\mu\nu}|_f=\langle LLL\rangle^{\lambda\mu\nu}|_{m_f=0}
~+~\langle LLL\rangle^{\lambda\mu\nu}(m_f);
\nonumber\\
&&\langle LLL\rangle^{\lambda\mu\nu}(m_f)=\langle LLL\rangle^{\lambda\mu\nu}|_{m_f\neq 0}~-~
\langle LLL\rangle^{\lambda\mu\nu}|_{m_f=0}.
\nonumber\\
\ea
Then, the structure of the amplitude will be
\ba
&&\frac{1}{2!}\langle \g ZZ\rangle|_{m_f\neq 0}=
\bar{A}_1\varepsilon[k_1,\lambda,\mu,\nu]+\bar{A}_2\varepsilon[k_2,\lambda,\mu,\nu]
+ A_3 k_1^{\mu}\varepsilon[k_1,k_2,\lambda,\nu]   \nonumber\\
&&\hspace{2cm}+A_4 k_2^{\mu}\varepsilon[k_1,k_2,\lambda,\nu]
+A_5 k_1^{\nu}\varepsilon[k_1,k_2,\lambda,\mu]+A_6 k_2^{\nu}\varepsilon[k_1,k_2,\lambda,\nu]
\ea
and using the explicit expressions of the coefficients we obtain
\ba
&&\langle \g ZZ\rangle|_{m_f\neq 0}=-\sum_f\left[
g_Y^3\theta^{YYY}_f \bar{R}_{\g ZZ}^{YYY}
+g_2^3\theta^{WWW}_f \bar{R}_{\g ZZ}^{WWW}
\right.\nonumber\\
&&\hspace{4cm}\left.
+g_Y g_2^2\theta^{YWW}_f R_{\g ZZ}^{YWW}
+g_Y^2 g_2\theta^{YYW}_f R_{\g ZZ}^{YYW}
\right.\nonumber\\
&&\hspace{4cm}\left.
+g_B g_Y^2\theta^{BYY}_f R_{\g ZZ}^{BYY}
+g_Y g_B^2\theta^{YBB}_f \bar{R}_{\g ZZ}^{YBB}
\right.\nonumber\\
&&\hspace{4cm}\left.
+g_B^2 g_2\theta^{WBB}_f \bar{R}_{\g ZZ}^{WBB}
+g_B g_2^2\theta^{BWW}_f R_{\g ZZ}^{BWW}
\right.\nonumber\\
&&\hspace{4cm}\left.
+g_B^2 g_2 g_Y\theta^{BYW}_f R_{\g ZZ}^{BYW}
\right]\frac{1}{2}T^{\lambda\mu\nu}(m_f\neq 0)A_{\g}Z^{\mu}Z^{\nu}\, ,
\ea
where we have defined
\ba
&&T^{\lambda\mu\nu}(m_f\neq 0)=\left[\Delta_{VAV}^{\lambda\mu\nu}(m_f\neq 0)
+\Delta_{VVA}^{\lambda\mu\nu}(m_f\neq 0)\right]\,,
\nonumber\\
&&\theta^{WBB}_f=(Q_{B,f}^{L})^2 T^{3}_{L,f}\,,
\nonumber\\
&&\bar{R}_{\g ZZ}^{WBB}=\frac{1}{2}R_{\g ZZ}^{WBB}\,,
\ea
with
\ba
&&T^{\lambda\mu\nu}(m_f\neq 0)=\frac{1}{\pi^2}\int_0^{1}d x\int_0^{1-x}d y\frac{1}{\Delta(m_f)}
\left\{
\varepsilon^{\alpha\lambda\mu\nu}k_{1,\alpha}\left[(1-x)x k_1^2 -y(1-y)k_2^2\right]
\right.\nonumber\\
&&\left.\hspace{4cm}
+\varepsilon^{\alpha\lambda\mu\nu}k_{2,\alpha}\left[(1-x)x k_1^2 -y(1-y)k_2^2\right]
\right.\nonumber\\
&&\left.\hspace{4cm}
+\varepsilon[k_1,k_2,\lambda,\nu]\left[2(x-1)x k_{1,\mu} -2 x y k_{2,\mu}\right]
\right.\nonumber\\
&&\left.\hspace{4cm}
+\varepsilon[k_1,k_2,\lambda,\mu]\left[2(1-y)y k_{2,\nu} +2 x y k_{2,\mu}\right]
\right\}\,.
\ea
We can immediately see that the expected broken Ward identities
\ba
&&k_{1,\mu}T^{\lambda\mu\nu}=\frac{1}{\pi^2}\varepsilon\left[k_1,k_2,\lambda,\nu\right]
\left\{\frac{1}{2} -m_f^2\int_0^{1}d x\int_0^{1-x}d y\frac{1}{\Delta(m_f)}\right\}
\nonumber\\
&&k_{2,\nu}T^{\lambda\mu\nu}=-\frac{1}{\pi^2}\varepsilon\left[k_1,k_2,\lambda,\nu\right]
\left\{\frac{1}{2} -m_f^2\int_0^{1}d x\int_0^{1-x}d y\frac{1}{\Delta(m_f)}\right\}
\nonumber\\
&&(k_{1,\lambda}+k_{2,\lambda})T^{\lambda\mu\nu}=0
\ea
are indeed satisfied.

\section{ Trilinear interactions in multiple $U(1)$ models}

Building on the computation of the $Z\gamma \gamma$ and
$\gamma ZZ$ presented in the sections above, we formulate here some 
general prescriptions that can be used in the analysis of anomalous 
abelian models when several $U(1)$'s are present and which help to simplify the process
of building the structure of the anomalous vertices in the mass eigenstates basis. 
The general case is already encountered when the anomalous gauge structure contains
three anomalous $U(1)$'s besides the usual gauge group of the SM.
We prefer to work with this specific choice in order to simplify the formalism,
though the discussion and the results are valid in general.

We denote respectively with $W_3, A_Y, B_1, B_2,B_3$ the weak, the hypercharge 
gauge boson and their 3 anomalous partners. At this point
we consider the anomalous triangle diagrams of the model and
observe that we can either
\begin{itemize}
\item[1)]
distribute the anomaly equally among all the corresponding generators
$(T_3,Y,Y_{B_1},Y_{B_2},Y_{B_3})$ and compensate for
the violation of the Ward identity on the non anomalous vertices with suitable CS interactions
\end{itemize}
or
\begin{itemize}
\item[2)]
re-define the trilinear vertices {\em ab initio } so that some partial 
anomalies are removed from the $Y-W_3$ generators in the diagrams containing 
mixed anomalies. Also in this case some CS counterterms may remain.

\end{itemize}
We recall that the anomaly-free generators are not accompanied by axions.
The difference between the first and the second method is in 
the treatment of the CS terms: in the first case they all appear explicitly
as separate contributions, while in the second one they can be absorbed,
at least in part, into the definition of the vertices. In one case or the other the final result is the same.
In particular one has to be careful on how to handle the distribution of the partial anomalies
(in the physical basis) especially when a certain vertex does not have any Bose symmetry,
such as for three different gauge bosons, and this is
not constrained by specific relations.
In this section we will go back again to the examples that we have
discussed in detail above and illustrate how to proceed in the most general case.

Consider the $Z \gamma \gamma$ case in the chiral limit. For instance, a vertex of the form $B_2YY$ will
be projected into the $Z\gamma\gamma$ vertex with a combination of rotation matrices of the form
$R^{B_2YY}_{Z\gamma\gamma}$, generating a partial contribution which is typically of the form
$\langle LLL\rangle R^{B_2YY}_{Z\gamma\gamma}$. At this point, in the $B_2YY$ diagram, 
which is interpreted as a
$\langle LLL \rangle \sim \Delta_{AAA}$ contribution, we move the anomaly on the
$B_2$-vertex by absorbing one CS term, thereby changing the $\langle LLL\rangle$  
vertex into an ${\bf AVV}$ vertex.

We do the same for all the trilinear contributions such as $B_3YY$, $B_1 W W$ 
and so on, similarly to what we have discussed in the previous sections. 
For instance $B_3YY$, which is also proportional to an ${\bf AAA}$ diagram, is turned
into an ${\bf AVV}$ diagram by a suitable CS term. The $Z\gamma\gamma$ is 
identified by adding up all the projections. This is the second approach.

The alternative procedure, which is the basic content of the first prescription 
mentioned above, consists in keeping the $B_2YY$ vertex as an ${\bf AAA}$ vertex, 
while the CS counterterm, which is needed to remove the anomaly from the Y vertex, 
has to be kept separate. Also in this case
the contribution of $B_2YY$ to $Z\gamma\gamma $ is of the form
$\langle LLL\rangle R^{B_2YY}_{Z\gamma\gamma}$, with
$\langle LLL\rangle\sim \Delta_{AAA}$, and the CS term that accompanies 
this contribution is also rotated into the same $Z\gamma\gamma$ vertex.

Using the second approach in the final construction of the $Z\gamma\gamma$ 
vertex we add up all the projections and obtain as a result a single $AVV$ 
diagram, as one would have naively expect using QED Ward identities on the 
photon lines. Instead, following the first we are forced to describe the 
same vertex as a sum of two contributions:
a fermionic triangle (which has partial anomalies on the two photon lines)
plus the CS counterterm, the sum of which is again of the form ${\bf AVV}$.

However, when possible, it is convenient to use a single diagram to describe 
a certain interaction, especially if the vertex has specific Bose symmetries, 
as in the case of the $Z\gamma \gamma$ vertex.

For instance, we could have easily inferred the result in the 
$Z\gamma\gamma$ case with no difficulty at all, since the partial 
anomaly on the photon lines is zero and the total anomaly, which is 
a constant, has to be necessarily attached to the $Z$ line and not to the photons.

A similar result holds for the
$ZZZ$ vertex where the anomaly has to be assigned symmetrically. 
Notice that, in prescription 2)  when several
extra $U(1)$'s are present, the vertices in the interaction eigenstate 
basis such as $B_1B_2B_3$ or
$B_1B_1B_2$ should be kept in their {\bf AAA} form, since the presence 
of axions $(b_1,b_2,b_3)$ is sufficient to guarantee the gauge invariance 
of each anomalous gauge boson line.

A final example concerns the case when 3 different anomalous gauge bosons 
are present, for instance $Z Z^{\prime} Z^{\prime\prime}$. In this case 
the distribution of the partial anomalies  can be easily inferred by 
combining all the projections of the trilinear vertices  
$B_1 Y Y,B_1 W W, B_1 B_2 B_3, B_1 B_2 B_3, B_2 B_3 B_3...$ etc.
into $Z Z^{\prime} Z^{\prime\prime}$. The absorption of the CS terms 
here is also straightforward, since vertices such as $B_1YY$, $YB_1Y$ 
and $YYB_1$ are rewritten as ${\bf AVV}$, ${\bf VAV}$ and $\bf{VVA}$ contributions respectively.
On the other hand, terms such as $B_2B_1B_1$ or $B_1 B_2 B_3$ are kept 
in their ${\bf AAA}$ form with an equal share of partial anomalies.
Notice that in this case the final vertex, also in the second approach 
where the CS terms are partially absorbed, does not result in a single 
diagram  as in the $Z\gamma\gamma$ case, but in a combination of several contributions.

\subsection{Moving away from the chiral limit with several anomalous $U(1)$'s}

Chiral symmetry breaking, as we have seen in the examples discussed before, 
introduces a higher level of complications in the analysis of these vertices. 
Also in this case we try to find a prescription to fix the trilinear anomalous 
gauge interactions away from the
chiral limit. As we have seen from the treatment of the previous sections, 
the presence
of mass terms in any triangle graph is confined to the denominator of their 
Feynman parameterization, once the Ward identities are imposed on each vertex. 
This implies that all the mixed terms of the form
$LLR$ or $RRL$ containing quadratic mass insertions can be omitted in any 
diagram and the final result for any anomalous contributions such as $B_1B_2 B_3$ or $B_1YY$ involves only an
$\langle LLL \rangle$ fermionic triangle where the mass from the Dirac traces is removed.

For instance, let's consider again the derivation of the $\gamma ZZ$ vertex in this case.
We project the trilinear gauge interactions of the effective action written in the eigenstate
basis into the $\gamma ZZ $ vertex (see Fig.~\ref{gammaZ1Z2}) as before and, typically, we encounter vertices
such as $Y B_1 Y$ or $B_1 Y Y$ (and so on) that need to be rotated.
We remove the masses from the numerator of these vertices and reduce each 
of them to a standard $\langle LLL\rangle$  form, having omitted the mixing terms  $LLR$, $RRL$, etc.
Also in this case a vertex such as $B_1YY$ is turned into an ${\bf AVV}$ 
by absorbing a corresponding CS interaction, while its broken Ward identities will be of the form
\beqn
k_{1\mu}\Delta^{\lambda\mu\nu}(\beta,k_1,k_2)&=& 0\nonumber\\
k_{2\nu}\Delta^{\lambda\mu\nu}(\beta,k_1,k_2)&=& 0\nonumber\\
k_\lambda\Delta^{\lambda\mu\nu}(\beta,k_1,k_2)&=& a_n(\beta) \varepsilon^{\mu\nu\alpha\beta}
k_1^\alpha k_2^\beta +  2 m_{f} \Delta^{\mu \nu},
\label{bbshift}
\eeqn
with a broken WI on the ${\bf A}$ line and exact ones on the remaining ${\bf V}$ 
lines corresponding to the two $Y$ generators.
Similarly, when we consider the projection of a term such as $B_1B_2B_3$ into
the $Z^{\prime} Z^{\prime\prime} Z $ vertex, we impose a symmetric distribution 
of the anomaly and broken WI's on the three external lines
\beqn
k_{1\mu}\Delta^{\lambda\mu\nu}(k_1,k_2)&=& \frac{a_n}{3} \varepsilon^{\lambda\nu\alpha\beta}
k_1^\alpha k_2^\beta +   2 m_{f} \Delta^{\lambda \nu}       ,\nonumber\\
k_{2\nu}\Delta^{\lambda\mu\nu}(k_1,k_2)&=& \frac{a_n}{3} \varepsilon^{\lambda\mu\alpha\beta}
k_2^\alpha k_1^\beta +   2 m_{f} \Delta^{\lambda \mu}          ,\nonumber\\
k_\lambda \Delta^{\lambda\mu\nu}(k_1,k_2)&=& \frac{a_n}{3} \varepsilon^{\mu\nu\alpha\beta}
k_1^\alpha k_2^\beta +   2 m_{f} \Delta^{\mu \nu}.
\label{bbshift1}
\eeqn
The total vertex is therefore obtained by adding up all these projections 
together with 3 CS contributions to redistribute the anomalies. Next we are 
going to discuss the explicit way of doing this.

\section{The $\langle \g Z_l Z_m\rangle$ vertex}

\begin{figure}[t]
{\centering \resizebox*{10cm}{!}{\rotatebox{0}
{\includegraphics{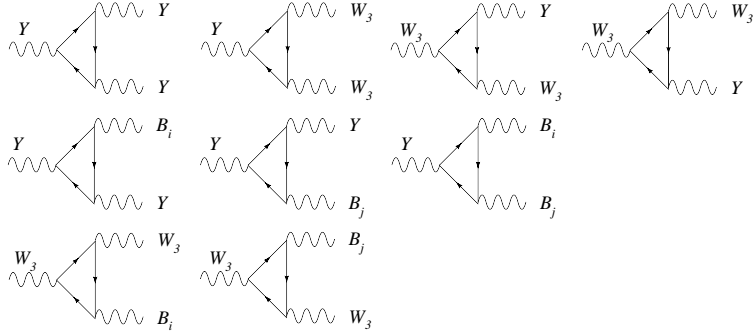}}}\par}
\caption{Triangle contributions to the $\langle \g Z_l Z_m\rangle$ vertex in the chiral phase.
Notice that the first four contributions vanish because of the SM charge assignment.}
\label{gammaZ1Z2}
\end{figure}

\begin{figure}[t]
{\centering \resizebox*{6cm}{!}{\rotatebox{0}
{\includegraphics{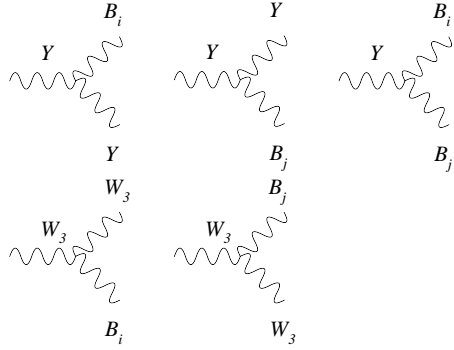}}}\par}
\caption{Chern-Simons counterterms of the $\langle \g Z_l Z_m\rangle$ vertex.}
\label{CSgammaZ1Z2}
\end{figure}

At this stage we can generalize the construction of $\langle \g Z Z\rangle$
to a general $\langle \g Z_l Z_m\rangle$ vertex.
The contributions coming from the interaction eigenstates basis to the
$\langle \g Z_l Z_m \rangle$ in the chiral limit are given by
\ba
&&\frac{1}{3!}Tr\left[Q_{Y}^3\right]\langle YYY \rangle=\frac{1}{3!} Tr\left[Q_{Y}^3\right]
R^{YYY}_{\g Z_l Z_m} \langle \g Z_l Z_m\rangle+\dots
\nonumber\\
&&\frac{1}{2!}Tr\left[Q_{Y} T_{3}^2\right]\langle YWW \rangle=\frac{1}{2!}Tr\left[Q_{Y} T_{3}^2\right]
R^{YWW}_{\g Z_l Z_m} \langle \g Z_l Z_m\rangle+\dots
\nonumber\\
&&\frac{1}{2!}Tr\left[Q_{Y} T_{3}^2\right]\langle WYW \rangle=\frac{1}{2!}Tr\left[Q_{Y} T_{3}^2\right]
R^{WYW}_{\g Z_l Z_m} \langle \g Z_l Z_m\rangle+\dots
\nonumber\\
&&\frac{1}{2!}Tr\left[Q_{Y} T_{3}^2\right]\langle WWY \rangle=\frac{1}{2!}Tr\left[Q_{Y} T_{3}^2\right]
R^{WWY}_{\g Z_l Z_m} \langle \g Z_l Z_m\rangle+\dots
\nonumber\\
&&\frac{1}{2!}Tr\left[Q_{B_j} T_{3}^2\right]\langle W B_j W \rangle=\frac{1}{2!}Tr\left[Q_{B_j} T_{3}^2\right]
R^{W B_{j} W}_{\g Z_l Z_m} \langle \g Z_l Z_m\rangle+\dots
\nonumber\\
&&\frac{1}{2!}Tr\left[Q_{B_j} T_{3}^2\right]\langle W W B_j \rangle=\frac{1}{2!}Tr\left[Q_{B_j} T_{3}^2\right]
R^{WWB_{j}}_{\g Z_l Z_m} \langle \g Z_l Z_m\rangle+\dots
\nonumber\\
&&\frac{1}{2!}Tr\left[Q_{B_j} Q_{Y}^2\right]\langle Y B_j Y \rangle=\frac{1}{2!}Tr\left[Q_{B_j} Q_{Y}^2\right]
R^{YB_{j}Y}_{\g Z_l Z_m} \langle \g Z_l Z_m\rangle+\dots
\nonumber\\
&&\frac{1}{2!}Tr\left[Q_{B_j} Q_{Y}^2\right]\langle YY B_j \rangle=\frac{1}{2!}Tr\left[Q_{B_j} Q_{Y}^2\right]
R^{YYB_{j}}_{\g Z_l Z_m} \langle \g Z_l Z_m\rangle+\dots
\nonumber\\
&&Tr\left[Q_{Y} Q_{B_j}Q_{B_k}\right]\langle Y B_j B_k \rangle=
Tr\left[Q_{Y} Q_{B_j}Q_{B_k}\right]
R^{Y B_{j}B_{k}}_{\g Z_l Z_m} \langle \g Z_l Z_m\rangle+\dots
\nonumber\\
\ea
and they are pictured in Fig.~\ref{gammaZ1Z2}.
The rotation matrices are defined as
\ba
&&R^{YYY}_{\g Z_l Z_m}=\left[3 O^{A}_{YZ_l}O^{A}_{YZ_m} O^{A}_{Y\gamma}\right]
\nonumber\\
&&R^{WWW}_{\g Z_l Z_m}=\left[3 O^{A}_{W_3 Z_l}O^{A}_{W_3Z_m} O^{A}_{W_3\gamma}\right]
\nonumber\\
&&R^{YWW}_{\g Z_l Z_m}=\left[O^{A}_{WZ_l} O^{A}_{W\gamma} O^{A}_{YZ_m}
+O^{A}_{WZ_m} O^{A}_{W\gamma} O^{A}_{YZ_l}+O^{A}_{WZ_l}O^{A}_{WZ_m} O^{A}_{Y\gamma}\right]
\nonumber\\
&&R^{WYY}_{\g Z_l Z_m}=\left[(O^{A}_{W_3 Z_l} O^{A}_{YZ_m}
+O^{A}_{W_3 Z_m} O^{A}_{Y Z_l})O^{A}_{Y\gamma}
+O^{A}_{W_3\gamma}O^{A}_{Y Z_{m}} O^{A}_{Y Z_{l}}\right]
\nonumber\\
&&R^{B_{j}YY}_{\g Z_l Z_m}=\left[O^{A}_{B_j Z_l} O^{A}_{YZ_m} O^{A}_{Y\gamma}
+O^{A}_{B_j Z_m} O^{A}_{YZ_l} O^{A}_{Y\gamma}\right]
\nonumber\\
&&R^{B_{j}YW}_{\g Z_l Z_m}=\left[(O^{A}_{B_j Z_l} O^{A}_{Y Z_m}+O^{A}_{B_j Z_m} O^{A}_{Y Z_l}) O^{A}_{W_3\gamma}
+(O^{A}_{B_j Z_m} O^{A}_{W_3 Z_l}+O^{A}_{B_j Z_l} O^{A}_{W_3 Z_m}) O^{A}_{Y\gamma}\right]
\nonumber\\
&&R^{Y B_i B_{j}}_{\g Z_l Z_m}=\left[(O^{A}_{B_i Z_l} O^{A}_{B_j Z_m}+O^{A}_{B_i Z_m} O^{A}_{B_j Z_l}) O^{A}_{Y\g}\right]
\nonumber\\
&&R^{W B_i B_{j}}_{\g Z_l Z_m}=\left[(O^{A}_{B_i Z_l} O^{A}_{B_j Z_m}+O^{A}_{B_i Z_m} O^{A}_{B_j Z_l}) O^{A}_{W_3\g}\right]
\nonumber\\
&&R^{B_j WW}_{\g Z_l Z_m}=\left[O^{A}_{B_j Z_l} O^{A}_{W Z_m} O^{A}_{W\gamma}+O^{A}_{B_j Z_m} O^{A}_{W Z_l} O^{A}_{W\gamma}\right]
\nonumber\\
\ea
while all the possible CS counterterms are listed in Fig.~\ref{CSgammaZ1Z2}
and their explicit expression in the rotated basis is given by
\ba
&&V_{CS,l m}=\sum_f\left\{
-\sum_{i}\frac{1}{8}\theta^{Y B_i Y}_{f}
\frac{a_n}{3}\varepsilon^{\lambda\mu\nu\alpha}(k_{2,\alpha}- k_{3,\alpha})
R^{YB_iY}_{\g Z_lZ_m}A_{\g}^{\lambda} Z_l^{\mu} Z_m^{\nu}
\right.\nonumber\\
&&\hspace{2cm}\left.
-\sum_{j}\frac{1}{8}\theta^{Y Y B_j}_{f}
\frac{a_n}{3}\varepsilon^{\lambda\mu\nu\alpha}(k_{3,\alpha}- k_{1,\alpha})
R^{Y Y B_j}_{\g Z_lZ_m}A_{\g}^{\lambda} Z_l^{\mu} Z_m^{\nu}
\right.\nonumber\\
&&\hspace{2cm}\left.
+\sum_{i,j}\frac{1}{8}\theta^{Y B_i B_j}_{f}
\frac{a_n}{6}\varepsilon^{\lambda\mu\nu\alpha}(k_{1,\alpha}- k_{2,\alpha})
R^{Y B_i B_j}_{\g Z_lZ_m}A_{\g}^{\lambda} Z_l^{\mu} Z_m^{\nu}
\right.\nonumber\\
&&\hspace{2cm}\left.
-\sum_{i}\frac{1}{8}\theta^{W B_i W}_{f}
\frac{a_n}{3}\varepsilon^{\lambda\mu\nu\alpha}(k_{2,\alpha}- k_{3,\alpha})
R^{WB_iW}_{\g Z_lZ_m}A_{\g}^{\lambda} Z_l^{\mu} Z_m^{\nu}
\right.\nonumber\\
&&\hspace{2cm}\left.
-\sum_{j}\frac{1}{8}\theta^{WW B_j}_{f}
\frac{a_n}{3}\varepsilon^{\lambda\mu\nu\alpha}(k_{3,\alpha}- k_{1,\alpha})
R^{WW B_j}_{\g Z_lZ_m}A_{\g}^{\lambda} Z_l^{\mu} Z_m^{\nu}
\right\}\,,
\ea
where we have defined $k_{3,\alpha}=-k_{\alpha}$, with $k_{\alpha}=(k_1+k_2)_{\alpha}$
the incoming momenta of the triangle. Using Eq.~(\ref{vertices}) it is easy to 
write the expression of the amplitude for the $\langle \g Z_l Z_m\rangle$ interaction
in the $m_f=0$ phase, and to separate the chiral components exactly as we have 
done for the $\langle\g Z Z\rangle$ vertex.
Again, the tensorial structure that we can factorize out is $\langle LLL \rangle^{\lambda\mu\nu}(0)$
\ba
&&\langle \g Z_l Z_m\rangle|_{m_f=0}=
\sum_f \frac{1}{8}\langle LLL\rangle^{\lambda\mu\nu}(0)A_{\g}^{\lambda}Z_{l}^{\mu}Z_{m}^{\nu}
\left\{
\sum_i g_Y^2 g_{B_i} \theta_f^{Y B_i Y}R_{\g Z_l Z_m}^{Y B_i Y}
+\sum_j g_Y^2 g_{B_j}\theta_f^{YY B_j}R_{\g Z_l Z_m}^{YYB_j}
\right.\nonumber\\
&&\hspace{2cm}\left.
+\sum_{i,j}g_Y g_{B_i}g_{B_j}\theta_f^{Y B_i B_j }R_{\g Z_l Z_m}^{Y B_i B_j}
+\sum_i g_2^2g_{B_i}\theta_f^{WB_iW}
R_{\g Z_l Z_m}^{WB_iW}
+\sum_{j}g_2^2 g_{B_j}\theta_f^{WWB_j}R_{\g Z_l Z_m}^{WWB_j}
\right\}\,.
\nonumber\\
\ea
Also in this case we use Eq. (\ref{smargia}) and proceed from a symmetric 
distribution of the anomalies and absorb the equations the CS interactions so to obtain
\ba
&&-\langle \g Z_l Z_m\rangle|_{m_f=0}=
\sum_i g_Y^2 g_{B_i} \sum_f \frac{1}{2}\theta_f^{Y B_i Y}
\Delta_{VAV}^{\lambda\mu\nu}(0)
R_{\g Z_l Z_m}^{Y B_i Y} A_{\g}^{\lambda}Z_{l}^{\mu}Z_{m}^{\nu}
\nonumber\\
&&\hspace{2cm}
+\sum_j g_Y^2 g_{B_j} \sum_f\frac{1}{2}\theta_f^{YY B_j }
\Delta_{VVA}^{\lambda\mu\nu}(0)
R_{\g Z_l Z_m}^{YYB_j} A_{\g}^{\lambda}Z_{l}^{\mu}Z_{m}^{\nu}
\nonumber\\
&&\hspace{2cm}
+\sum_{i,j}g_Y g_{B_i}g_{B_j}\sum_f \theta_f^{Y B_i B_j }
\frac{1}{2}\left[\Delta_{VA V}^{\lambda\mu\nu}(0)+\Delta_{VVA}^{\lambda\mu\nu}(0)\right]
R_{\g Z_l Z_m}^{Y B_i B_j} A_{\g}^{\lambda}Z_{l}^{\mu}Z_{m}^{\nu}
\nonumber\\
&&\hspace{2cm}
+\sum_i g_2^2g_{B_i} \sum_f \theta_f^{WB_iW}
\frac{1}{2}\Delta_{VAV}^{\lambda\mu\nu}(0)
R_{\g Z_l Z_m}^{WB_iW} A_{\g}^{\lambda}Z_{l}^{\mu}Z_{m}^{\nu}
\nonumber\\
&&\hspace{2cm}
+\sum_{j}g_2^2 g_{B_j} \sum_f\theta_f^{WWB_j}
\frac{1}{2}\Delta_{VVA}^{\lambda\mu\nu}(0)
R_{\g Z_l Z_m}^{WWB_j} A_{\g}^{\lambda}Z_{l}^{\mu}Z_{m}^{\nu}\,.
\nonumber\\
\label{masslessgaZZ}
\ea
At this point one can readily observe that a simple rearrangement
of the summations over the $i,j$ index leads us to factor out the structure {\bf VAV} plus {\bf VVA}
since we have the same rotation matrices.
Finally, in the $m_f=0$ phase we have
\ba
&&\langle \g Z_l Z_m\rangle|_{m_f=0}=-
\sum_f\frac{1}{2}\left[
\Delta_{VAV}^{\lambda\mu\nu}(0)+\Delta_{VVA}^{\lambda\mu\nu}(0)\right]
A_{\g}^{\lambda}Z_{l}^{\mu}Z_{m}^{\nu}
\times
\nonumber\\
&&\hspace{2cm}
\sum_i \left\{
g_Y^2 g_{B_i}\theta_f^{B_i Y Y} R_{\g Z_l Z_m}^{Y Y B_i}
+\sum_{j}g_Y g_{B_i}g_{B_j}\theta_f^{Y B_i B_j }R_{\g Z_l Z_m}^{Y B_i B_j}
+g_2^2g_{B_i}\theta_f^{WWB_i}R_{\g Z_l Z_m}^{WW B_i}
\right\}\,.
\nonumber\\
\label{massless_gaZZ}
\ea
If the CS terms are instead not absorbed we have

\ba
&&\langle \g Z_l Z_m\rangle|_{m_f=0}=V_{CS,lm}
-\sum_f\frac{1}{2}
\Delta_{AAA}^{\lambda\mu\nu}(0)A_{\g}^{\lambda}Z_{l}^{\mu}Z_{m}^{\nu}
\times
\nonumber\\
&&\hspace{2cm}
\sum_i \left\{
g_Y^2 g_{B_i}\theta_f^{B_i Y Y} R_{\g Z_l Z_m}^{Y Y B_i}
+\sum_{j}g_Y g_{B_i}g_{B_j}\theta_f^{Y B_i B_j }R_{\g Z_l Z_m}^{Y B_i B_j}
+g_2^2g_{B_i}\theta_f^{WWB_i}R_{\g Z_l Z_m}^{WW B_i}
\right\}\,,\nonumber\\
\ea
which is equivalent to that obtained in (\ref{massless_gaZZ}).

\subsection{Amplitude in the $m_f\neq 0$ phase}

Once we have fixed the structure of the triangle in the $m_f=0$ phase,
its extension to the massive case can be obtained
using the relation
\ba
\langle LLL\rangle(m_f\neq 0)=-\left[\Delta_{AVV}(m_f\neq 0)+\Delta_{VAV}(m_f\neq 0)
+\Delta_{VVA}(m_f\neq 0)+\Delta_{AAA}(m_f\neq 0)\right]
\nonumber\\
\ea
and the expression of the vertex will be
\ba
&&\langle \g Z_l Z_m\rangle|_{m_f\neq 0}=\frac{1}{8}\sum_f\langle LLL\rangle^{\lambda\mu\nu}(m_f\neq 0)
A_{\g}^{\lambda}Z_{l}^{\mu}Z_{m}^{\nu}
\left\{
g_Y^3  \theta_f^{Y Y Y}R_{\g Z_l Z_m}^{Y Y Y}
\right.\nonumber\\
&&\hspace{2cm}\left.
+g_2^3\theta_f^{WWW}R_{\g Z_l Z_m}^{WWW}
+g_Y g_2^2 \theta_f^{Y W W}R_{\g Z_l Z_m}^{Y W W}
\right.\nonumber\\
&&\hspace{2cm}\left.
+g_Y^2 g_2 \theta_f^{W YY}R_{\g Z_l Z_m}^{WYY}
+\sum_i g_Y^2 g_{B_i} \theta_f^{YYB_i}R_{\g Z_l Z_m}^{YYB_i}
\right.\nonumber\\
&&\hspace{2cm}\left.
+\sum_i g_Y g_2 g_{B_i} \theta_f^{B_iYW}R_{\g Z_l Z_m}^{B_iYW}
+\sum_{i,j}g_Y g_{B_i}g_{B_j} \theta_f^{Y B_i B_j}R_{\g Z_l Z_m}^{Y B_i B_j}
\right.\nonumber\\
&&\hspace{2cm}\left.
+\sum_{i,j}g_2 g_{B_i}g_{B_j} \theta_f^{W B_i B_j}R_{\g Z_l Z_m}^{W B_i B_j}
+\sum_{i}g_2^2 g_{B_i} \theta_f^{WWB_i}R_{\g Z_l Z_m}^{WWB_i}
\right\}
\nonumber\\
&&\hspace{2cm}+m_f^2\left[\langle LRL\rangle+\langle RRL \rangle+\dots \right].
\ea
By imposing the following broken Ward identities on the tensor structure
\ba
&&k_{1}^{\mu}\left(\langle \g Z_l Z_m\rangle^{\lambda\mu\nu}+V_{CS}^{\lambda\mu\nu}\right)
=\frac{a_n}{2}\varepsilon^{\lambda\nu\alpha\beta}
k_{1,\alpha}k_{2,\beta}+2 m_f\Delta^{\lambda\nu}
\nonumber\\
&&k_{2}^{\nu}\left(\langle \g Z_l Z_m\rangle^{\lambda\mu\nu}+V_{CS}^{\lambda\mu\nu}\right)
=-\frac{a_n}{2}\varepsilon^{\lambda\mu\alpha\beta}
k_{1,\alpha}k_{2,\beta}-2 m_f\Delta^{\lambda\mu}
\nonumber\\
&&k^{\lambda}\left(\langle \g Z_l Z_m \rangle^{\lambda\mu\nu}+V_{CS}^{\lambda\mu\nu}\right)=0
\ea
we arrange all the mass terms into the coefficients $\bar{A}_1$ and $\bar{A}_2$ of
the Rosenberg parametrization of $\langle LLL\rangle^{\lambda\mu\nu}$ and we absorbe all the
singular pieces. Since all the CS interactions act only on the massless part of
the {\bf LLL} structure, we are left with an expression which is similar
to Eq.~(\ref{masslessgaZZ}) but with the addition
of the triangle contributions coming from the Standard Model
where the mass is contained only in the denominators.
Organizing all the partial contributions we arrive at the final expression in which the structure
{\bf VAV} plus {\bf VVA} is factorized out
\ba
&&\langle \g Z_l Z_m\rangle|_{m_f\neq 0}=-
\sum_f\frac{1}{2}\left[
\Delta_{VAV}^{\lambda\mu\nu}(m_f\neq 0)+\Delta_{VVA}^{\lambda\mu\nu}(m_f\neq 0)\right]
A_{\g}^{\lambda}Z_{l}^{\mu}Z_{m}^{\nu}
\times
\nonumber\\
&&\hspace{2cm}
\left\{
g_Y^3 \theta_f^{Y Y Y}\bar{R}_{\g Z_l Z_m}^{Y Y Y}
+g_2^3 \theta_f^{WWW}\bar{R}_{\g Z_l Z_m}^{WWW}
\right.\nonumber\\
&&\hspace{2cm}\left.
+g_Y g_2^2 \theta_f^{Y W W}R_{\g Z_l Z_m}^{Y W W}
+g_Y^2 g_2 \theta_f^{W Y Y }R_{\g Z_l Z_m}^{W YY}
\right.\nonumber\\
&&\hspace{2cm}\left.
+\sum_i g_Y^2 g_{B_i}\theta_f^{B_i Y Y} R_{\g Z_l Z_m}^{B_i Y Y }
+\sum_i g_Y g_2 g_{B_i}\theta_f^{B_i Y W} R_{\g Z_l Z_m}^{B_i Y W }
\right.\nonumber\\
&&\hspace{2cm}\left.
+\sum_{i,j}g_Y g_{B_i}g_{B_j}\theta_f^{Y B_i B_j }R_{\g Z_l Z_m}^{Y B_i B_j}
+\sum_{i,j}g_2 g_{B_i}g_{B_j}\theta_f^{W B_i B_j }R_{\g Z_l Z_m}^{W B_i B_j}
\right.\nonumber\\
&&\hspace{2cm}\left.
+\sum_i g_2^2g_{B_i}\theta_f^{WWB_i}R_{\g Z_l Z_m}^{B_iWW }
\right\}.
\nonumber\\
\label{final_massivegaZZ}
\ea

\section{The $\langle Z_l Z_m Z_r\rangle$ vertex}

\begin{figure}[t]
{\centering \resizebox*{10cm}{!}{\rotatebox{0}
{\includegraphics{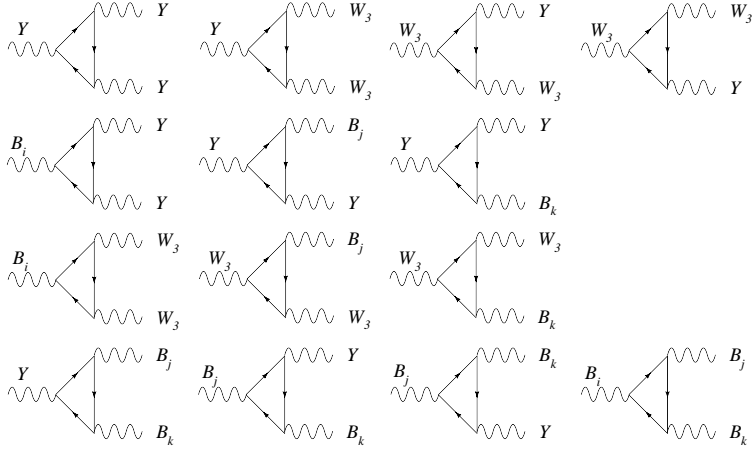}}}\par}
\caption{Triangle contributions to the $\langle Z_l Z_m Z_r\rangle$ vertex.
As before, in the $m_f=0$ phase all the SM contributions vanish 
because of the charge assignment.}
\label{Z1Z2Z3}
\end{figure}

\begin{figure}[t]
{\centering \resizebox*{8cm}{!}{\rotatebox{0}
{\includegraphics{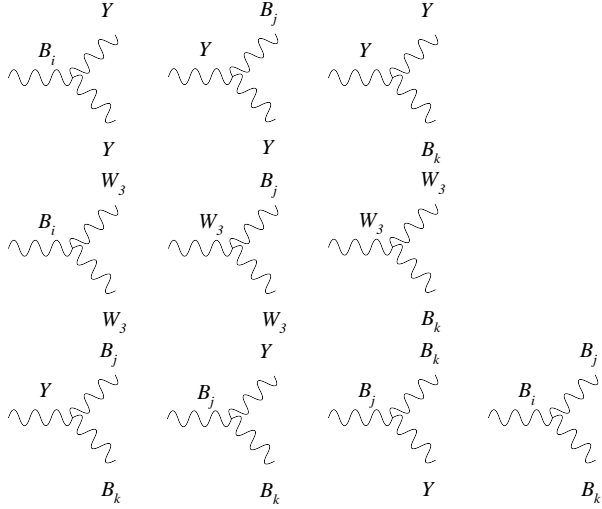}}}\par}
\caption{Chern-Simons contributions to the $\langle Z_l Z_m Z_r\rangle$ vertex}
\label{CS_Z1Z2Z3}
\end{figure}

Moving to the more general trilinear vertex is rather straightforward.
We can easily identify all the contributions coming from the
interaction eigenstates basis to the $\langle Z_l Z_m Z_r\rangle$.
In the chiral limit these are
\ba
&&\frac{1}{3!}Tr\left[Q_{Y}^3\right]\langle YYY \rangle=\frac{1}{3!} Tr\left[Q_{Y}^3\right]
R^{YYY}_{Z_l Z_m Z_r} \langle Z_l Z_m Z_r\rangle+\dots
\nonumber\\
&&\frac{1}{2!}Tr\left[Q_{Y} T_{3}^2\right]\langle YWW \rangle=\frac{1}{2!}Tr\left[Q_{Y} T_{3}^2\right]
R^{YWW}_{Z_l Z_m Z_r} \langle Z_l Z_m Z_r\rangle+\dots
\nonumber\\
&&\frac{1}{2!}Tr\left[Q_{Y} T_{3}^2\right]\langle WYW \rangle=\frac{1}{2!}Tr\left[Q_{Y} T_{3}^2\right]
R^{WYW}_{Z_l Z_m Z_r} \langle Z_l Z_m Z_r\rangle+\dots
\nonumber\\
&&\frac{1}{2!}Tr\left[Q_{Y} T_{3}^2\right]\langle WWY \rangle=\frac{1}{2!}Tr\left[Q_{Y} T_{3}^2\right]
R^{WWY}_{Z_l Z_m Z_r} \langle Z_l Z_m Z_r\rangle+\dots
\nonumber\\
&&\frac{1}{2!}Tr\left[Q_{B_j} T_{3}^2\right]\langle B_j WW \rangle=\frac{1}{2!}Tr\left[Q_{B_j} T_{3}^2\right]
R^{B_{j} WW}_{Z_l Z_m Z_r} \langle Z_l Z_m Z_r\rangle+\dots
\nonumber\\
&&\frac{1}{2!}Tr\left[Q_{B_j} T_{3}^2\right]\langle W B_j W \rangle=\frac{1}{2!}Tr\left[Q_{B_j} T_{3}^2\right]
R^{W B_{j} W}_{Z_l Z_m Z_r} \langle Z_l Z_m Z_r\rangle+\dots
\nonumber\\
&&\frac{1}{2!}Tr\left[Q_{B_j} T_{3}^2\right]\langle W W B_j \rangle=\frac{1}{2!}Tr\left[Q_{B_j} T_{3}^2\right]
R^{WWB_{j}}_{Z_l Z_m Z_r} \langle Z_l Z_m Z_r\rangle+\dots
\nonumber\\
&&\frac{1}{2!}Tr\left[Q_{B_j} Q_{Y}^2\right]\langle B_j YY \rangle=\frac{1}{2!}Tr\left[Q_{B_j} Q_{Y}^2\right]
R^{B_{j}YY}_{Z_l Z_m Z_r} \langle Z_l Z_m Z_r\rangle+\dots
\nonumber\\
&&\frac{1}{2!}Tr\left[Q_{B_j} Q_{Y}^2\right]\langle Y B_j Y \rangle=\frac{1}{2!}Tr\left[Q_{B_j} Q_{Y}^2\right]
R^{YB_{j}Y}_{Z_l Z_m Z_r} \langle Z_l Z_m Z_r\rangle+\dots
\nonumber\\
&&\frac{1}{2!}Tr\left[Q_{B_j} Q_{Y}^2\right]\langle YY B_j \rangle=\frac{1}{2!}Tr\left[Q_{B_j} Q_{Y}^2\right]
R^{YYB_{j}}_{Z_l Z_m Z_r} \langle Z_l Z_m Z_r\rangle+\dots
\nonumber\\
&&Tr\left[Q_{Y} Q_{B_j}Q_{B_k}\right]\langle Y B_j B_k \rangle=
Tr\left[Q_{Y} Q_{B_j}Q_{B_k}\right]
R^{Y B_{j}B_{k}}_{Z_l Z_m Z_r} \langle Z_l Z_m Z_r\rangle+\dots
\nonumber\\
&&Tr\left[Q_{Y} Q_{B_j}Q_{B_k}\right]\langle B_j Y  B_k \rangle=
Tr\left[Q_{Y} Q_{B_j}Q_{B_k}\right]
R^{B_{j} Y B_{k}}_{Z_l Z_m Z_r} \langle Z_l Z_m Z_r\rangle+\dots
\nonumber\\
&&Tr\left[Q_{Y} Q_{B_j}Q_{B_k}\right]\langle B_j B_k Y  \rangle=
Tr\left[Q_{Y} Q_{B_j}Q_{B_k}\right]
R^{B_{j}B_{k} Y }_{Z_l Z_m Z_r} \langle Z_l Z_m Z_r\rangle+\dots
\nonumber\\
&&Tr\left[Q_{B_i} Q_{B_j}Q_{B_k}\right]\langle B_i B_j B_k  \rangle=
Tr\left[Q_{B_i} Q_{B_j}Q_{B_k}\right]
R^{B_i B_j B_k}_{Z_l Z_m Z_r} \langle Z_l Z_m Z_r\rangle+\dots
\nonumber\\
\ea
and are listed in Fig.~\ref{Z1Z2Z3}.
The rotation matrices, in this case, are defined as
\ba
&&R^{YYY}_{Z_l Z_m Z_r}=\left[3O^{A}_{YZ_l}O^{A}_{YZ_m}O^{A}_{YZ_r}\right]
\nonumber\\
&&R^{WWW}_{Z_l Z_m Z_r}=\left[3O^{A}_{W_3Z_l}O^{A}_{W_3Z_m}O^{A}_{W_3Z_r}\right]
\nonumber\\
&&R^{YWW}_{Z_l Z_m Z_r}=\left[O^{A}_{YZ_l}O^{A}_{WZ_m}O^{A}_{WZ_r}
+O^{A}_{YZ_m}O^{A}_{WZ_l}O^{A}_{WZ_r}+O^{A}_{YZ_r}O^{A}_{WZ_l}O^{A}_{WZ_m}\right]
\nonumber\\
&&R^{WYY}_{Z_l Z_m Z_r}=\left[O^{A}_{W_3 Z_l}O^{A}_{YZ_m}O^{A}_{YZ_r}
+O^{A}_{W_3 Z_m}O^{A}_{YZ_l}O^{A}_{YZ_r}+O^{A}_{W_3Z_r}O^{A}_{YZ_l}O^{A}_{YZ_m}\right]
\nonumber\\
&&R^{B_{j}YY}_{Z_l Z_m Z_r}=\left[O^{A}_{B_jZ_l}O^{A}_{YZ_m}O^{A}_{YZ_r}
+O^{A}_{B_jZ_m}O^{A}_{YZ_l}O^{A}_{YZ_r}+O^{A}_{B_jZ_r}O^{A}_{YZ_m}O^{A}_{YZ_l}\right]
\nonumber\\
&&R^{B_{j}YW}_{Z_l Z_m Z_r}=\left[O^{A}_{B_jZ_l}(O^{A}_{YZ_m}O^{A}_{W_3 Z_r} +O^{A}_{YZ_r}O^{A}_{W_3 Z_m})
+O^{A}_{B_jZ_m}(O^{A}_{YZ_l}O^{A}_{W_3Z_r}+O^{A}_{W_3 Z_l}O^{A}_{YZ_r})
\right.\nonumber\\
&&\hspace{1.5cm}\left.
+O^{A}_{B_jZ_r}(O^{A}_{YZ_m}O^{A}_{W_3Z_l}+O^{A}_{YZ_l}O^{A}_{W_3Z_m})\right]
\nonumber\\
&&R^{B_j B_k Y}_{Z_l Z_m Z_r}=\left[(O^{A}_{B_jZ_m}O^{A}_{B_kZ_r}+O^{A}_{B_jZ_r}O^{A}_{B_kZ_m}) O^{A}_{YZ_l}
+(O^{A}_{B_jZ_r}O^{A}_{B_kZ_l}+O^{A}_{B_jZ_l}O^{A}_{B_kZ_r})O^{A}_{YZ_m}
\right.\nonumber\\
&&\hspace{1.5cm}\left.
+(O^{A}_{B_jZ_l}O^{A}_{B_kZ_m}+O^{A}_{B_jZ_m}O^{A}_{B_kZ_l})O^{A}_{YZ_r}\right]
\nonumber\\
&&R^{B_j B_k W}_{Z_l Z_m Z_r}=\left[(O^{A}_{B_jZ_m}O^{A}_{B_kZ_r}+O^{A}_{B_jZ_r}O^{A}_{B_kZ_m}) O^{A}_{W_3 Z_l}
+(O^{A}_{B_jZ_r}O^{A}_{B_kZ_l}+O^{A}_{B_jZ_l}O^{A}_{B_kZ_r})O^{A}_{W_3Z_m}
\right.\nonumber\\
&&\hspace{1.5cm}\left.
+(O^{A}_{B_jZ_l}O^{A}_{B_kZ_m}+O^{A}_{B_jZ_m}O^{A}_{B_kZ_l})O^{A}_{W_3Z_r}\right]
\nonumber\\
&&R^{B_j WW}_{Z_l Z_m Z_r}=\left[O^{A}_{B_jZ_l}O^{A}_{W_3Z_m}O^{A}_{W_3Z_r}+
O^{A}_{B_jZ_m}O^{A}_{W_3 Z_l}O^{A}_{W_3 Z_r}+O^{A}_{B_jZ_r}O^{A}_{W_3Z_m}O^{A}_{W_3 Z_l}\right]
\nonumber\\
&&R^{B_i B_j B_k}_{Z_l Z_m Z_r}=\left[
(O^{A}_{B_jZ_m}O^{A}_{B_k Z_r}+O^{A}_{B_jZ_r}O^{A}_{B_kZ_m}) O^{A}_{B_i Z_l}
+(O^{A}_{B_jZ_r}O^{A}_{B_k Z_l}+O^{A}_{B_jZ_l}O^{A}_{B_kZ_r})O^{A}_{B_i Z_m}
\right.\nonumber\\
&&\hspace{1.5cm}\left.
+(O^{A}_{B_jZ_l}O^{A}_{B_kZ_m}+O^{A}_{B_jZ_m}O^{A}_{B_kZ_l})O^{A}_{B_i Z_r}
\right].
\ea
Regarding the CS interactions (see Fig.~(\ref{CS_Z1Z2Z3})), we observe that
we have a CS term corresponding to the anomalous vertex of the type
$B_iB_jB_k$ which is non-zero, and we can formally
write this trilinear interaction as
\ba
&&V_{CS,\,lmr}^{ijk}=g_{B_i}g_{B_j}g_{B_k}a_n
\theta^{ijk}_{lmr} R^{i j k}_{l m r}Z_l^{\lambda} Z_m^{\mu}Z_r^{\nu}
\left[\kappa_i\left(\varepsilon[k_1,\lambda,\mu,\nu]-\varepsilon[k_2,\lambda,\mu,\nu]\right)
\right.\nonumber\\
&&\left.\hspace{2cm}
+\kappa_j \left(\varepsilon[k_2,\lambda,\mu,\nu]-\varepsilon[k_3,\lambda,\mu,\nu]\right)
+\kappa_k \left(\varepsilon[k_3,\lambda,\mu,\nu]-\varepsilon[k_1,\lambda,\mu,\nu]\right)
\right],
\nonumber\\
\ea
where for brevity we have defined $R^{i jk}_{lmr}=R^{B_i B_j B_k}_{Z_m Z_l Z_r}$, and so on.

The coefficients $\theta^{ijk}_{lmr}$ are the charge asymmetries, and the coefficients
$\kappa_{i,j,k}$, are real numbers that tell us how the anomaly will be distributed on the ${\bf AAA}$
triangles. Both are driven by the generalized Ward identities of the theory.
In this generalized case the CS interactions are not all re-absorbed in the definition of the fermionic triangles. In fact in this case there is no symmetry in the diagram that forces a symmetric assignment of the anomaly, and the CS terms in the $B_i B_j B_k$ interaction can re-distribute the partial anomalies.
In this case the expression of the $B_iB_jB_k$ vertex in the momentum space
is given by
\ba
&&{\bf V}^{\lambda \mu \nu}_{B_i B_j B_k} = 4 D^{}_{B_i B_j B_k} \, g_{B_i}g_{B_j} g_{B_k}
 \, \Delta^{\lambda \mu \nu}_{{\bf A A A}} (m_f=0,k^{}_{1}, k^{}_{2} )
\nonumber\\
&&\hspace{1.5cm}
+D^{}_{B_i B_j B_k} \, g_{B_i}g_{B_j} g_{B_k}\frac{i}{\pi^{2} }\left[
\frac{2\kappa_i}{9} \varepsilon^{\lambda\mu\nu\alpha}(k_{1,\alpha}-k_{2,\alpha})
\right.\nonumber\\
&&\left.\hspace{2cm}
+ \frac{2\kappa_j}{9} \varepsilon^{\lambda\mu\nu\alpha}(k_{2,\alpha}-k_{3,\alpha})
+\frac{2\kappa_k}{9} \varepsilon^{\lambda\mu\nu\alpha}(k_{3,\alpha}-k_{1,\alpha})\right].
\nonumber\\
\ea
We recall that in the treatment of $YB_jB_k$ and other similar triangles we
still have two contributions for each triangle, due to the two orientations of the fermion
number in the loop, so that our previous expression, obtained for the case of the $YBB$ vertex, still holds.
Also in this case we are allowed to absorb the CS interaction in the anomalous vertex.
On the other hand, for the $B_iB_jB_k$ vertex we have
\ba
&&3\Delta_{AAA}^{\lambda\mu\nu}(0,k_1,k_2)
-\frac{a_n^i}{3} \varepsilon^{\lambda\mu\nu\alpha}(k_{1,\alpha}-k_{2,\alpha})
-\frac{a_n^j}{3} \varepsilon^{\lambda\mu\nu\alpha}(k_{2,\alpha}-k_{3,\alpha})
-\frac{a_n^k}{3} \varepsilon^{\lambda\mu\nu\alpha}(k_{3,\alpha}-k_{1,\alpha})
\nonumber\\
&&=3\Delta_{A_i A_j A_k}^{\lambda\mu\nu}(0,k_1,k_2)\,,
\ea
where we have used the notation $\Delta(m_f=0,k_1,k_2)=\Delta(0,k_1,k_2)$ and $a_n^i=\kappa^i a_n$.
Using these equations we can write the $\langle Z_l Z_m Z_r\rangle$ triangle in the following way
\ba
&&\langle Z_l Z_m Z_r\rangle|_{m_f=0}=-\frac{1}{3}
\left[\Delta_{VA V}^{\lambda\mu\nu}(0)+\Delta_{VVA}^{\lambda\mu\nu}(0)+\Delta_{A VV}^{\lambda\mu\nu}(0)\right]
Z_{l}^{\lambda}Z_{m}^{\mu}Z_{r}^{\nu}\times
\nonumber\\
&&\hspace{2cm}
\sum_f\sum_i\left\{
g_Y^2 g_{B_i} \theta_f^{YYB_i}R_{Z_l Z_m Z_r}^{YYB_i}
+\sum_j g_Y g_{B_i}g_{B_j}\theta_f^{B_i B_j Y}
R_{Z_l Z_m Z_r}^{Y B_j B_k}
+g_{B_i} g_2^2 \theta_f^{B_i WW}
R_{Z_l Z_m Z_r}^{B_i WW}
\right\}
\nonumber\\
&&\hspace{2cm}
+\sum_f \sum_{i,j,k}g_{B_i}g_{B_j}g_{B_k}\theta_f^{B_i B_j B_k}
\frac{1}{2}
\Delta_{A_i A_j A_k}^{\lambda\mu\nu}(0)
R_{Z_l Z_m Z_r}^{B_i B_j B_k } Z_{l}^{\lambda}Z_{m}^{\mu}Z_{r}^{\nu}.
\nonumber\\
\ea
From this last result we can observe that
the anomaly distribution on the last piece is, in general, not of the type
$\Delta_{A A A}^{\lambda\mu\nu}(0)$, i.e. symmetric.
If we want to factorize out a $\Delta_{AAA}^{\lambda\mu\nu}(0)$ triangle,
we should think of this amplitude as a factorized $\Delta_{AAA}^{\lambda\mu\nu}(0)$
contribution plus an external suitable CS interaction which is not re-absorbed
and such that it changes the partial anomalies from the symmetric distribution $\Delta_{AAA}^{\lambda\mu\nu}(0)$
to the non-symmetric one $\Delta_{A_i A_j A_k}^{\lambda\mu\nu}(0)$.
These two points of view are completely equivalent and give the same result.

Finally, the analytic expression for each tensor contribution in the $m_f=0$ phase is given below.
The {\bf AVV} vertex has been shown in Eq. (\ref{avv}) while for {\bf VAV} we have
\ba
&&\Delta_{VAV}^{\lambda\mu\nu}(0)=\frac{1}{\pi^2}\int_0^1 dx\int_{0}^{1-x}dy\frac{1}{\Delta(0)}
\left\{
\varepsilon[k_1,\lambda,\mu,\nu](k_2\cdot k_2 y(y-1)- x y k_1\cdot k_2)
\right.\nonumber\\
&&\left.\hspace{2cm}
+\varepsilon[k_2,\lambda,\mu,\nu](k_2\cdot k_2 y(y-1)- x y k_1\cdot k_2)
\right.\nonumber\\
&&\left.\hspace{2cm}
+\varepsilon[k_1,k_2,\lambda,\nu](k_1^{\mu} x(x-1)- x y k_2^{\mu})
\right.\nonumber\\
&&\left.\hspace{2cm}
+\varepsilon[k_1,k_2,\lambda,\mu](k_2^{\nu} y(1-y)+ x y k_1^{\nu})
\right\}\,,
\ea
where the denominator is defined as $\Delta(0)=k_1^2(x-1)x + y (y-1) k_2^2 + 2 x y k_1\cdot k_2$.

Then, for the {\bf VVA} contribution we obtain
\ba
&&\Delta_{VVA}^{\lambda\mu\nu}(0)=\frac{1}{\pi^2}\int_0^1 dx\int_{0}^{1-x}dy\frac{1}{\Delta(0)}
\left\{
\varepsilon[k_1,\lambda,\mu,\nu](k_1\cdot k_1 x(1-x)+ x y k_1\cdot k_2)
\right.\nonumber\\
&&\left.\hspace{2cm}
+\varepsilon[k_2,\lambda,\mu,\nu](k_1\cdot k_1 x(1-x)+ x y k_1\cdot k_2)
\right.\nonumber\\
&&\left.\hspace{2cm}
+\varepsilon[k_1,k_2,\lambda,\nu](k_1^{\mu} x(x-1)- x y k_2^{\mu})
\right.\nonumber\\
&&\left.\hspace{2cm}
+\varepsilon[k_1,k_2,\lambda,\mu](k_2^{\nu} y(1-y)+ x y k_1^{\nu})
\right\}\,,
\ea
and finally the contribution for {\bf AAA} is $\Delta_{AAA}(0)=1/3(\Delta_{AVV}(0)+\Delta_{VAV}(0)+\Delta_{VVA}(0))$
\ba
&&\Delta_{AAA}^{\lambda\mu\nu}(0)=\frac{1}{3\pi^2}\int_0^1 dx\int_{0}^{1-x}dy\frac{1}{\Delta(0)}
\left\{
\varepsilon[k_1,\lambda,\mu,\nu]\left( 2 y(y-1) k_2^2 - x y k_1\cdot k_2 + x(1-x) k_1^2\right)
\right.\nonumber\\
&&\left.\hspace{2cm}
+\varepsilon[k_2,\lambda,\mu,\nu]\left(2(1-x)x k_1^2 + x y k_1\cdot k_2 +y(y-1)k_2^2\right)
\right.\nonumber\\
&&\left.\hspace{2cm}
+\varepsilon[k_1,k_2,\lambda,\nu](k_1^{\mu} x(x-1)- x y k_2^{\mu})
\right.\nonumber\\
&&\left.\hspace{2cm}
+\varepsilon[k_1,k_2,\lambda,\mu](k_2^{\nu} y(1-y)+ x y k_1^{\nu})
\right\}.
\ea

\section{The $m_f\neq 0$ phase of the $\langle Z_lZ_mZ_r \rangle$ triangle}

To obtain the contribution in the $m_f\neq 0$ phase we must include
again all the contributions $\langle YYY\rangle$ and $\langle YWW\rangle$
coming from the SM.
Since the final tensor structure of the triangle
is driven by the STI's, we start by assuming the following symmetric distribution
of the anomalies on the $\Delta_{AAA}$ triangle
\ba
&&k_1^{\mu}\Delta_{AAA}^{\lambda\mu\nu}(m_f\neq 0,k_1,k_2)=
\frac{a_n}{3}\varepsilon^{\lambda\nu\alpha\beta}k_{1\alpha}k_{2\beta}
+2 m_f \frac{1}{3}\Delta^{\lambda\nu}
\nonumber\\
&&k_2^{\nu}\Delta_{AAA}^{\lambda\mu\nu}(m_f\neq 0,k_1,k_2)=
-\frac{a_n}{3}\varepsilon^{\lambda\mu\alpha\beta}k_{1\alpha}k_{2\beta}
-2 m_f \frac{1}{3}\Delta^{\lambda\mu}
\nonumber\\
&&k^{\lambda}\Delta_{AAA}^{\lambda\mu\nu}(m_f\neq 0,k_1,k_2)=
\frac{a_n}{3}\varepsilon^{\mu\nu\alpha\beta}k_{1\alpha}k_{2\beta}
+2 m_f \frac{1}{3}\Delta^{\mu\nu}\,,
\ea
where
\ba
\Delta^{\lambda\nu}=-\frac{m_f}{\pi^2}\varepsilon^{\lambda\nu\alpha\beta}k_{1\alpha}k_{2\beta}
\int_0^1\int_0^{1-x}dx dy\frac{1}{\Delta(m_f)}.
\ea

These relations define the {\bf AAA} structure in the massive case.
The explicit form of this triangle is given by
\ba
&&\Delta_{AAA}^{\lambda\mu\nu}(m_f \neq 0)=\frac{1}{\pi^2}\int_0^1 dx\int_{0}^{1-x}dy\frac{1}{\Delta(m_f)}
\left\{
\right.\nonumber\\
&&\left.\hspace{2cm}
\varepsilon[k_1,\lambda,\mu,\nu]\left[-\frac{\Delta(m_f)-m_f^2}{3} + k_2\cdot k_2 y(y-1)- x y k_1\cdot k_2\right]
\right.\nonumber\\
&&\left.\hspace{2cm}
+\varepsilon[k_2,\lambda,\mu,\nu]\left[\frac{\Delta(m_f)-m_f^2}{3}- k_1\cdot k_1 x(x-1)+ x y k_1\cdot k_2\right]
\right.\nonumber\\
&&\left.\hspace{2cm}
+\varepsilon[k_1,k_2,\lambda,\nu](k_1^{\mu} x(x-1)- x y k_2^{\mu})
\right.\nonumber\\
&&\left.\hspace{2cm}
+\varepsilon[k_1,k_2,\lambda,\mu](k_2^{\nu} y(1-y)+ x y k_1^{\nu})
\right\}\,,
\ea
where $\Delta(m_f)=m_f^2 + (y-1)y k_2^2 +(x-1)x k_1^2 -2 xyk_1\cdot k_2$.

Then, the final expression in the $m_f\neq 0$ phase is
\ba
&&\langle Z_l Z_m Z_r\rangle|_{m_f\neq 0}=-
Z_{l}^{\lambda}Z_{m}^{\mu}Z_{r}^{\nu}\times
\sum_f\Delta_{AAA}^{\lambda\mu\nu}(m_f\neq 0)
\sum_i\left\{
g_Y^3 \theta_f^{YYY}R_{Z_l Z_m Z_r}^{YYY}
+g_2^3 \theta_f^{WWW}R_{Z_l Z_m Z_r}^{WWW}
\right.\nonumber\\
&&\hspace{3cm}\left.
+g_Y g_2^2 \theta_f^{YWW}R_{Z_l Z_m Z_r}^{YWW}
+g_Y^2 g_2 \theta_f^{YYW}R_{Z_l Z_m Z_r}^{YYW}
\right.\nonumber\\
&&\hspace{3cm}\left.
+g_Y^2 g_{B_i} \theta_f^{YYB_i}R_{Z_l Z_m Z_r}^{YYB_i}
+g_Y g_2 g_{B_i} \theta_f^{B_iYW}R_{Z_l Z_m Z_r}^{B_iYW}
\right.\nonumber\\
&&\hspace{3cm}\left.
+\sum_j g_Y g_{B_i}g_{B_j}\theta_f^{B_i B_j Y}R_{Z_l Z_m Z_r}^{Y B_j B_k}
+\sum_j g_2 g_{B_i}g_{B_j}\theta_f^{B_i B_j W}R_{Z_l Z_m Z_r}^{B_j B_k W}
\right.\nonumber\\
&&\hspace{3cm}\left.
+g_{B_i} g_2^2 \theta_f^{B_i WW}R_{Z_l Z_m Z_r}^{B_i WW}
+\sum_{j,k}g_{B_i}g_{B_j}g_{B_k}\theta_f^{B_i B_j B_k} R_{Z_l Z_m Z_r}^{B_i B_j B_k }
\right\} + V_{CS}\,.
\nonumber\\
\ea

\begin{figure}[t]
{\centering \resizebox*{14cm}{!}{\rotatebox{0}
{\includegraphics{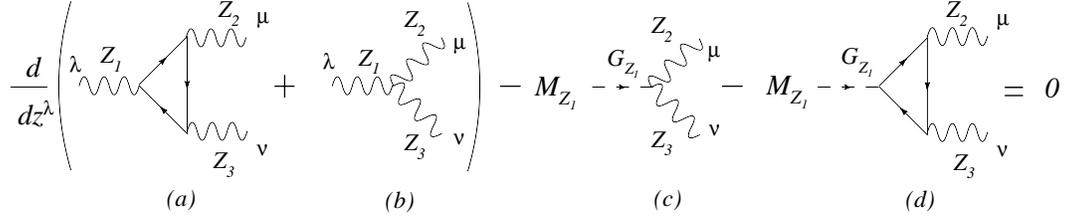}}}\par}
\caption {STI for the $Z_1$ vertex in a trilinear anomalous vertex with several $U(1)$'s. The CS counterterm is
not absorbed and redistributes the anomaly according to the specific model.}
\label{anomX}
\end{figure}

The diagrammatic structure of the STI for this general vertex is shown in Fig. \ref{anomX}, where an irreducible CS vertex (the second contribution in the bracket) is now present.

\section{Discussions}
The possibility of detecting anomalous gauge interactions at the LHC remains an interesting avenue that requires further analysis. The topic is clearly very interesting and may be a way to shed
light on physics beyond the SM in a rather simple framework, though, at a hadron collider these studies
are naturally classified as difficult ones. There are some points, however, that need clarification when anomalous contributions are taken into account.
The first concerns the real mechanism of cancellation of the anomalies, if it is not realized by a charge assignment, and in particular whether it is
of GS or of WZ type. In the two cases the high energy behaviour of  a certain class of processes is rather different,  and the WZ theory, which induces
an axion-like particle in the spectrum, is in practice an effective theory with a unitarity bound, which has now been quantified \cite{CGM}.  The second
point concerns the size of these anomalous interactions compared against the QCD background, which needs to be
determined  to next-to-next-to-leading-order  (NNLO) in the strong coupling, at least for those processes involving anomalous gluon interactions
with  the extra $Z^\prime$.  These points are under investigations and we hope to return with some quantitative predictions in the near future.

\section{Conclusions}
In this work we have analyzed those trilinear gauge interactions that appear in the context of anomalous abelian extensions of the SM with several extra $U(1)$'s. We have discussed the defining conditions on the effective action, starting from the
St\"uckelberg phase of this model, down to the electroweak phase, where Higgs-axion mixing takes place. In particular, we have shown that it is possible to simplify the study of the model in a suitable gauge, where the Higgs-axion mixing is removed from the effective action. The theory is conveniently defined, after electroweak symmetry breaking, by a set of generalized Ward identities and the counterterms
can be fixed  in any of the two phases. We have also derived the expressions of these vertices using the equivalence of the effective action in the interaction and in the mass eigenstate basis, and used this result to formulate general rules for the computation of the vertices which allow to simplify this construction.
Using the various anomalous models that have been constructed in the previous literature in the last decade or so, it is now possible to explicitly proceed with a more direct phenomenological analysis
of these theories, which remain an interesting avenue for future experimental searches of anomalous gauge interactions
at the LHC.

\vspace{1cm}

\centerline{\bf Acknowledgments}

We thank Simone Morelli for discussions and for checking our results and to Nikos Irges for discussions and a long term collaboration to this analysis.
We are particularly grateful to Alan R. White for the many discussions and clarifications
along the way on the role of the chiral anomaly and to Marco Roncadelli, Z. Berezhiani and all
the participants to the annual  INFN meetings at Gran Sasso for discussions about
axion physics and for the friendly and fruitful atmosphere; C.C. also thanks
Z. Berezhiani for hospitality, and  F. Fucito, P. Anastasopoulous, Y. Stanev,
A. Lionetto,  A. Racioppi and A. Di Ciaccio for discussions and hospitality
during a recent visit at Tor Vergata. C.C. thanks E. Kiritsis,
T. Tomaras, T. Petkou and N. Tsamis for hospitality at the Univ. of Crete.
The work of C.C. was supported (in part)
by the European Union through the Marie Curie Research and Training Network
``Universenet'' (MRTN-CT-2006-035863) and by The Interreg II Crete-Cyprus Program.

\section{Appendix. Gauge variations}
In this and in the following appendices we fill in the steps that take to the construction of the Faddeev-Popov lagrangean of the model.

To define the ghost lagrangean we need to compute the gauge variations.
Therefore let's consider the variation
\bea
\delta W^{3}_{\mu} = \partial_{\mu} \alpha_3 - g_2 \varepsilon^{3bc} W^{b}_{\mu} \alpha_{c},  \qquad
\delta Y_{\mu} = \partial_{\mu}  \theta_{Y},  \qquad  \delta B_{\mu} = \partial_{\mu}  \theta_{B},
\eea
where the parameters have been rotated as the corresponding fields
using the same matrix $O_A$
\bea
\theta_{\g} &=& O^{A}_{11} \alpha_{3} + O^{A}_{12} \theta_{Y},    \\
\theta_{Z} &=& O^{A}_{21} \alpha_{3} + O^{A}_{22} \theta_{Y} + O^{A}_{23} \theta_{B},  \\
\theta_{Z^{\prime}} &=& O^{A}_{31} \alpha_{3} + O^{A}_{32} \theta_{Y} + O^{A}_{33} \theta_{B}.
\eea
In the neutral sector we obtain the variations
\bea
\delta A_{\g \, \mu} &=&  O^{A}_{11}   \, \delta W^{3}_{\mu} + O^{A}_{12}  \,  \delta Y_{\mu}     \nonumber\\
&=&   \partial_{\mu} \theta_{\g}+ i\, O^{A}_{11}  \, g_2  \left( \alpha^{-} W^{+}_{\mu}
 - \alpha^{+} W^{-}_{\mu}  \right),
 \label{deltaAgauge} \\
\delta Z_{\mu} &=& O^{A}_{21} \, \delta W^{3}_{\mu} + O^{A}_{22}  \, \delta Y_{\mu}  + O^{A}_{23} \,  \delta B_{\mu}  \nonumber\\
&=&  \partial_{\mu} \theta_{Z} + i \, O^{A}_{21} \,  g_2 \left(\alpha^{-} W^{+}_{\mu}
 - \alpha^{+} W^{-}_{\mu} \right),
 \label{deltaZgauge}\nonumber\\
 \eea
 \bea
\delta Z^{\prime}_{\mu}  &=&  O^{A}_{31} \,  \delta W^{3}_{\mu} + O^{A}_{32} \,  \delta Y_{\mu}
+ O^{A}_{33}  \, \delta B_{\mu}  \nonumber\\
&=&  \partial_{\mu} \theta_{Z^\prime} + i  \,  O^{A}_{31}  \,  g_2 \left( \alpha^{-} W^{+}_{\mu}
- \alpha^{+} W^{-}_{\mu}  \right),
\label{deltaZpgauge}
\eea
and for the charged fields we obtain
\bea
\delta W^{\pm}_{\mu} &=&
\partial_{\mu} \alpha^{\pm} \mp i g_{2} W^{\pm}_{\mu} \left( O^{A}_{11} \theta_{\g} + O^{A}_{21} \theta_{Z}
+ O^{A}_{31} \theta_{Z^{\prime}} \right) \nn \\
&\pm& i g_2\left( O^{A}_{11} A_{\g \mu} + O^{A}_{21} Z_{\mu}
+ O^{A}_{31} Z^{\prime}_{\mu} \right)  \alpha^{\pm}.
\label{deltaWgauge}
\eea
After a lengthy computation we obtain
\bea
\delta H^{+}_{u}&=& - i \frac{g_2}{\sqrt 2} v_u \alpha^{+} - i \Big[ \frac{\alpha_{A}}{2} \left( g_2 O^{A}_{11}
+ g_Y O^{A}_{12} + g_{B} q^{B}_{u} O^{A}_{13} \right) \nn \\
&+&  \frac{\alpha_{Z}}{2} \left( g_2 O^{A}_{21}
+ g_Y O^{A}_{22} + g_{B} q^{B}_{u} O^{A}_{23} \right)     \nonumber\\
& +&  \frac{\alpha_{Z^\prime}}{2} \left( g_2 O^{A}_{31}
+ g_Y O^{A}_{32} + g_{B} q^{B}_{u} O^{A}_{33} \right) \Big] H^{+}_{u}
- i \frac{g_2}{2} \left( H^{0}_{uR} + i H^{0}_{uI} \right)  \alpha^{+}\nn \\
\eea
and using  the expressions for $H^{+}_{u}$, $H^{0}_{uR}$, $H^{0}_{uI}$ derived in \cite{CIM2}
we obtain
\bea
\delta H^{+}_{u}
&=&   - i \frac{g_2}{\sqrt 2} v_u \alpha^{+} - i \left[ \frac{\alpha_{A}}{2} \left( g_2 O^{A}_{11}
+ g_Y O^{A}_{12} + g_{B} q^{B}_{u} O^{A}_{13} \right) \right. \nn \\
&+&  \frac{\alpha_{Z}}{2} \left( g_2 O^{A}_{21}
+ g_Y O^{A}_{22} + g_{B} q^{B}_{u} O^{A}_{23} \right)     \nonumber\\
& +&  \left. \frac{\alpha_{Z^\prime}}{2}
\left( g_2 O^{A}_{31} + g_Y O^{A}_{32} + g_{B} q^{B}_{u} O^{A}_{33} \right) \right]
(\sin{\beta} G^{+} - \cos{\beta} H^{+})    \nonumber\\
&-& i \frac{g_2}{2} \Big[ (\sin{\alpha} \, h^{0} - \cos{\alpha} \, H^{0}) \Big. \nn \\
&+& \Big. i  \biggl(O^{\chi}_{11} \chi
+ \frac{O^{\chi}_{12} c'_2 - O^{\chi}_{13} c'_1}{c_1 c'_2 - c'_1 c_2} G^{Z}
 + \frac{- O^{\chi}_{12} c_2 + O^{\chi}_{13} c_1}{c_1 c'_2 - c'_1 c_2} G^{Z'}   \biggr) \Big]  \alpha^{+}.  \nonumber\\
\eea
Similarly, for the field $H^{+}_{d}$ we get
\bea
\delta H^{+}_{d}
&=&  - i \frac{g_2}{\sqrt 2} v_d \alpha^{+} - i \left[ \frac{\alpha_{A}}{2} \left( g_2 O^{A}_{11}
+ g_Y O^{A}_{12} + g_{B} q^{B}_{d} O^{A}_{13} \right) \right. \nn \\
&+&   \frac{\alpha_{Z}}{2} \left( g_2 O^{A}_{21}
+ g_Y O^{A}_{22} + g_{B} q^{B}_{d} O^{A}_{23} \right)     \nonumber\\
&+& \left.   \frac{\alpha_{Z^\prime}}{2} \left( g_2 O^{A}_{31}
+ g_Y O^{A}_{32} + g_{B} q^{B}_{d} O^{A}_{33} \right) \right] (\cos{\beta} G^{+} + \sin{\beta} H^{+})    \nonumber\\
&-& i \frac{g_2}{2} \left[ (\cos{\alpha} \, h^{0} + \sin{\alpha} \, H^{0}) \right. \nn \\
&+& \left. i \left(O^{\chi}_{21} \chi
+ \frac{O^{\chi}_{22} c'_2 - O^{\chi}_{23} c'_1}{c_1 c'_2 - c'_1 c_2} G^{Z}
 + \frac{- O^{\chi}_{22} c_2 + O^{\chi}_{23} c_1}{c_1 c'_2 - c'_1 c_2} G^{Z'}   \right)\right]  \alpha^{+}.    \nonumber\\
\eea
Using the relations obtained for the charged Higgs in \cite{CIM2}
we get for the charged goldstones
\bea
\delta G^{+} &=& \sin \beta \delta H^{+}_{u} + \cos \beta \delta H^{+}_{d}   \nonumber\\
\delta G^{-} &=& \sin \beta \delta H^{-}_{u} + \cos \beta \delta H^{-}_{d}.
\eea
In the Higgs sector we have
\bea
\delta H^{0}_{uI}
&=&  - \frac{g_2}{2} \left( \alpha^{-} (\sin \beta G^{+} - \cos \beta H^{+} )
+ \alpha^{+} (\sin \beta G^{-} - \cos \beta H^{-} ) \right)   \nonumber\\
& +& \frac{v_u}{\sqrt 2} \left[
\left( g_2 O^{A}_{21} - g_Y O^{A}_{22} - g_{B} q^{B}_{u} O^{A}_{23} \right) \alpha_Z \right. \nn \\
&+& \left. \left( g_2 O^{A}_{31} - g_Y O^{A}_{32} - g_{B} q^{B}_{u} O^{A}_{33} \right) \alpha_{Z^\prime}\right]    \nonumber\\
&+&  \left[
\left( g_2 O^{A}_{21} - g_Y O^{A}_{22} - g_{B} q^{B}_{u} O^{A}_{23} \right) \alpha_Z    \right.  \nonumber\\
&+& \left.    \left( g_2 O^{A}_{31} - g_Y O^{A}_{32} - g_{B} q^{B}_{u} O^{A}_{33} \right) \alpha_{Z^\prime} \right]
\frac{(\sin \alpha h^{0} - \cos \alpha H^{0})}{2},    \nonumber\\
\eea
and
\bea
\delta H^{0}_{dI}
&=&  - \frac{g_2}{2} \left( \alpha^{-} (\cos \beta G^{-} + \sin \beta H^{+}) + \alpha^{+} (\cos \beta G^{-} + \sin \beta H^{-})
\right)   \nonumber\\
&+& \frac{v_d}{\sqrt 2} \left[
\left( g_2 O^{A}_{21} - g_Y O^{A}_{22} - g_{B} q^{B}_{d} O^{A}_{23} \right) \alpha_Z \right. \nn \\
&+& \left. \left( g_2 O^{A}_{31} - g_Y O^{A}_{32} - g_{B} q^{B}_{d} O^{A}_{33} \right) \alpha_{Z^\prime}\right]    \nonumber\\
&+&  \left[
\left( g_2 O^{A}_{21} - g_Y O^{A}_{22} - g_{B} q^{B}_{d} O^{A}_{23} \right) \alpha_Z  \right.  \nonumber\\
&+&  \left. \left( g_2 O^{A}_{31} - g_Y O^{A}_{32} - g_{B} q^{B}_{d} O^{A}_{33} \right)
\alpha_{Z^\prime}\right] \frac{(\cos \alpha h^{0} + \sin \alpha H^{0}) }{2},  \nonumber\\
\eea
while for the neutral goldstones we have
%
\bea
\delta G^{0}_{1} &=& O^{\chi}_{12} \delta  H^{0}_{uI} + O^{\chi}_{22}  \delta H^{0}_{dI} + O^{\chi}_{32} \delta  b,  \\
\delta G^{0}_{2} &=& O^{\chi}_{13} \delta  H^{0}_{uI} + O^{\chi}_{23} \delta  H^{0}_{dI} + O^{\chi}_{33} \delta b.
\eea
Finally, we determine the variations of the two goldstones
%
\bea
\delta G^{Z} &=& c_1 \delta G^{0}_{1} + c_2 \delta G^{0}_{2},  \\
\delta G^{Z^\prime} &=& c^{\prime}_1 \delta G^{0}_{1} + c^{\prime}_2 \delta G^{0}_{2},
\eea
and the gauge variation of the St\"uckelberg $b$ in the base of the mass eigenstates
\bea
\delta b &=& - M_{1}  \theta_{B}  \nonumber\\
&=&  - M_{1} \left( O^{A}_{23}   \theta_{Z} + O^{A}_{33}  \theta_{Z^\prime} \right).
\eea

%

\section{Appendix: The FP lagrangean}
 This is explicitly given by
\bea
\mathcal{L}_{FP}&=& -\bar{c}^{Z} \frac{\d \mathcal{F}^Z}{\d \theta_Z} c^Z
-\bar{c}^{Z} \frac{\d \mathcal{F}^Z}{\d \theta_{Z'}} c^{Z'}
-\bar{c}^{Z} \frac{\d \mathcal{F}^Z}{\d \theta_\g} c^{\g}
-\bar{c}^{Z} \frac{\d \mathcal{F}^Z}{\d \theta_{+}} c^+
-\bar{c}^{Z} \frac{\d \mathcal{F}^Z}{\d \theta_{-}} c^{-} \nonumber\\
& -& \bar{c}^{Z'} \frac{\d \mathcal{F}^{Z'}}{\d \theta_Z} c^Z
-\bar{c}^{Z'} \frac{\d \mathcal{F}^{Z'}}{\d \theta_{Z'}} c^{Z'}
-\bar{c}^{Z'} \frac{\d \mathcal{F}^{Z'}}{\d \theta_\g} c^{\g}
-\bar{c}^{Z'} \frac{\d \mathcal{F}^{Z'}}{\d \theta_{+}} c^+
-\bar{c}^{Z'} \frac{\d \mathcal{F}^{Z'}}{\d \theta_{-}} c^{-} \nonumber\\
&-& \bar{c}^{\g} \frac{\d \mathcal{F}^{A_\g}}{\d \theta_Z} c^Z
-\bar{c}^{\g} \frac{\d \mathcal{F}^{A_\g}}{\d \theta_{Z'}} c^{Z'}
-\bar{c}^{\g} \frac{\d \mathcal{F}^{A_\g}}{\d \theta_\g} c^{\g}
-\bar{c}^{\g} \frac{\d \mathcal{F}^{A_\g}}{\d \theta_{+}} c^+
-\bar{c}^{\g} \frac{\d \mathcal{F}^{A_\g}}{\d \theta_{-}} c^{-} \nonumber\\
& -& \bar{c}^{\,+} \frac{\d \mathcal{F}^{W^+}}{\d \theta_Z} c^Z
-\bar{c}^{\,+} \frac{\d \mathcal{F}^{W^+}}{\d \theta_{Z'}} c^{Z'}
-\bar{c}^{\,+} \frac{\d \mathcal{F}^{W^+}}{\d \theta_\g} c^{\g}
-\bar{c}^{\,+} \frac{\d \mathcal{F}^{W^+}}{\d \theta_{+}} c^+
-\bar{c}^{\,+} \frac{\d \mathcal{F}^{W^+}}{\d \theta_{-}} c^{-} \nonumber\\
&-& \bar{c}^{\,-} \frac{\d \mathcal{F}^{W^-}}{\d \theta_Z} c^Z
-\bar{c}^{\,-} \frac{\d \mathcal{F}^{W^-}}{\d \theta_{Z'}} c^{Z'}
-\bar{c}^{\,-} \frac{\d \mathcal{F}^{W^-}}{\d \theta_\g} c^{\g}
-\bar{c}^{\,-} \frac{\d \mathcal{F}^{W^-}}{\d \theta_{+}} c^+
-\bar{c}^{\,-} \frac{\d \mathcal{F}^{W^-}}{\d \theta_{-}} c^{-}, \nonumber\\
\eea
where we have computed
\bea
\frac{\d \mathcal{F}^Z}{\d \theta_Z} =
\partial_{\mu} \frac{\d Z^{\mu}} {\d \theta_Z} - \xi_Z M_Z\frac{\d G^Z}{\d \theta_Z};  \qquad \qquad
\frac{\d Z^{\mu}}{\d \theta_Z} =
\partial^{\mu};
\eea
\vspace{-0.5cm}
\bea
\frac{\d G^Z}{\d \theta_Z} =
c_1 \frac{\delta G^{0}_{1}} {\d \theta_Z} + c_2 \frac{\delta G^{0}_{2}}{\d \theta_Z} =
c_1 \Biggl( O^{\chi}_{12} \frac{\delta  H^{0}_{uI}} {\d \theta_Z} + O^{\chi}_{22}
\frac{\delta H^{0}_{dI}} {\d \theta_Z} + O^{\chi}_{32} \frac{\delta  b} {\d \theta_Z} \Biggr)  \nonumber \\
+ c_2 \Biggl( O^{\chi}_{13} \frac{\delta  H^{0}_{uI}} {\d \theta_Z}
+ O^{\chi}_{23} \frac{\delta  H^{0}_{dI}} {\d\theta_Z} + O^{\chi}_{33} \frac{\delta b} {\d \theta_Z} \Biggl),
\eea
\bea
\frac{\delta  H^{0}_{uI}}{\d \theta_Z} &=&
\Biggl[ \frac{v_u}{\sqrt {2}} + \frac{(\sin \alpha h^{0} - \cos \alpha H^{0})}{2} \Biggl] f_u, \\
\frac{\delta  H^{0}_{dI}}{\d \theta_Z} &=&
\Biggl[ \frac{v_d}{\sqrt {2}} + \frac{(\cos \alpha h^{0} + \sin \alpha H^{0})}{2} \Biggl] f_d,
\eea
\bea
f_{u,d} = g_2 O^A_{21} - g_Y O^A_{22} - g_B q_{u,d}^B O^A_{23},
& \quad &
\frac{\delta  b} {\d \theta_Z} =  - M_1 O^A_{23}.
\eea
\bea
\frac{\d \mathcal{F}^Z}{\d \theta_{Z^{\prime}}} = \partial_{\mu}\frac{\d Z^{\mu}}{\d \theta_{Z^{\prime}}}- \xi_Z M_Z\frac{\d G^Z}{\d \theta_{Z^{\prime}}};  \qquad \qquad
\frac{\d Z^{\mu}} {\d \theta_{Z'}} = 0;
\eea
\vspace{-0.5cm}
\bea
\frac{\d G^Z} {\d \theta_{Z^{\prime}}} = c_1 \frac{\delta G^{0}_{1}} {\d \theta_{Z^{\prime}}} + c_2 \frac{\delta G^{0}_{2}} {\d \theta_{Z'}}=
c_1 \Biggl( O^{\chi}_{12} \frac{\delta  H^{0}_{uI}} {\d \theta_{Z^{\prime}}} + O^{\chi}_{22}  \frac{\delta H^{0}_{dI}} {\d \theta_{Z^{\prime}}}
 + O^{\chi}_{32} \frac{\delta  b} {\d \theta_{Z^{\prime}}} \Biggr) \nonumber \\
+ c_2 \Biggl( O^{\chi}_{13} \frac{\delta  H^{0}_{uI}} {\d \theta_{Z^{\prime}}}
+ O^{\chi}_{23} \frac{\delta  H^{0}_{dI}} {\d \theta_{Z^{\prime}}} + O^{\chi}_{33} \frac{\delta b} {\d \theta_{Z^{\prime}}} \Biggl);
\eea
\bea
\frac{\delta  H^{0}_{uI}}{\d \theta_{Z^{\prime}}} &=&
\Biggl[ \frac{v_u}{\sqrt {2}} + \frac{(\sin \alpha h^{0} - \cos \alpha H^{0})}{2} \Biggl] f_{u}^B; \\
\frac{\delta  H^{0}_{dI}}{\d \theta_{Z^{\prime}}} &=&
\Biggl[ \frac{v_d}{\sqrt {2}} + \frac{(\cos \alpha h^{0} + \sin \alpha H^{0})}{2} \Biggl] f_{d}^B;
\eea
\bea
f_{u,d}^B = g_2 O^A_{31} - g_Y O^A_{32} - g_B q_{u,d}^B O^A_{33};
& \quad &
\frac{\delta  b} {\d \theta_{Z^{\prime}}} =  - M_1 O^A_{33}.
\eea
%
%
\bea
\frac{\d \mathcal{F}^Z}{\d \theta_{\g}} = \partial_{\mu}\frac{\d Z^{\mu}}{\d \theta_{\g}} - \xi_Z M_Z\frac{\d G^Z}{\d \theta_{\g}};
\qquad \qquad
\frac{\d Z^{\mu}} {\d \theta_{\g}} = 0;
\qquad \qquad
\frac{\d G^Z} {\d \theta_{\g}}= 0;
\eea

\bea
\frac{\d \mathcal{F}^Z} {\d \theta_{+}} = \partial_{\mu}\frac{\d Z^{\mu}}{\d \theta_{+}} -  \xi_Z M_Z\frac{\d G^Z}{\d \theta_{+}};
\qquad \qquad
\frac{\d Z^{\mu}} {\d \theta_{+}} =  -i g_2 O^A_{21} W^{-\mu};
\eea
\vspace{-0.5cm}
\bea
\frac{\d G^Z} {\d \theta_{+}} = c_1 \frac{\delta G^{0}_{1}} {\d \theta_{+}} + c_2 \frac{\delta G^{0}_{2}} {\d \theta_{+}}=
c_1 \Biggl( O^{\chi}_{12} \frac{\delta  H^{0}_{uI}} {\d \theta_{+}} + O^{\chi}_{22}  \frac{\delta H^{0}_{dI}} {\d \theta_{+}}
 + O^{\chi}_{32} \frac{\delta  b} {\d \theta_{+}} \Biggr) \nonumber \\
+ c_2 \Biggl( O^{\chi}_{13} \frac{\delta  H^{0}_{uI}} {\d \theta_{+}}
+ O^{\chi}_{23} \frac{\delta  H^{0}_{dI}} {\d \theta_{+}} + O^{\chi}_{33} \frac{\delta b} {\d \theta_{+}} \Biggl);
\eea
\bea
\frac{\delta  H^{0}_{uI}}{\d \theta_{+}} &=&
- \frac{g_2}{2} (\sin \b G^{-} - \cos \b H^{-}); \\
\frac{\delta  H^{0}_{dI}}{\d \theta_{+}} &=&
- \frac{g_2}{2} (\cos \b G^{-} + \sin \b H^{-}); \\
\frac{\delta  b} {\d \theta_{+}} &=& 0.
\eea
\bea
\frac{\d \mathcal{F}^Z} {\d \theta_{-}} = \partial_{\mu}\frac{\d Z^{\mu}}{\d \theta_{-}}- \xi_Z M_Z\frac{\d G^Z}{\d \theta_{-}};
\qquad \qquad
\frac{\d Z^{\mu}} {\d \theta_{-}} = i g_2 O^A_{21} W^{+\mu};
\eea
\vspace{0.5cm}
\bea
\frac{\d G^Z} {\d \theta_{-}} = c_1 \frac{\delta G^{0}_{1}} {\d \theta_{-}} + c_2 \frac{\delta G^{0}_{2}} {\d \theta_{-}}=
c_1 \Biggl( O^{\chi}_{12} \frac{\delta  H^{0}_{uI}} {\d \theta_{-}} + O^{\chi}_{22}  \frac{\delta H^{0}_{dI}} {\d \theta_{-}}
 + O^{\chi}_{32} \frac{\delta  b} {\d \theta_{-}} \Biggr) \nonumber \\
+ c_2 \Biggl( O^{\chi}_{13} \frac{\delta  H^{0}_{uI}} {\d \theta_{-}}
+ O^{\chi}_{23} \frac{\delta  H^{0}_{dI}} {\d \theta_{-}} + O^{\chi}_{33} \frac{\delta b} {\d \theta_{-}} \Biggl);
\eea
\bea
\frac{\delta  H^{0}_{uI}}{\d \theta_{-}} &=&
- \frac{g_2}{2} (\sin \b G^{+} - \cos \b H^{+}); \\
\frac{\delta  H^{0}_{dI}}{\d \theta_{-}} &=&
- \frac{g_2}{2} (\cos \b G^{+} + \sin \b H^{+});\\
\frac{\delta  b} {\d \theta_{-}} &=& 0.
\eea
For the gauge boson $Z^{\prime}$ we obtain

\bea
\frac{\d \mathcal{F}^{Z^{\prime}}}{\d \theta_Z} = \partial_{\mu}\frac{\d Z^{\prime \mu}}{\d \theta_Z}-
\xi_{Z^{\prime}} M_{Z^{\prime}}\frac{\d G^{Z'}}{\d \theta_Z};
\qquad \qquad
\frac{\d Z^{\prime \mu}}{\d \theta_Z}= 0;
\eea
\bea
\frac{\d G^{Z'}}{\d \theta_Z}= c^{\prime}_1 \frac{\delta G^{0}_{1}} {\d \theta_Z} +
c^{\prime}_2 \frac{\delta G^{0}_{2}}{\d \theta_Z}=
c^{\prime}_1 \Biggl( O^{\chi}_{12} \frac{\delta  H^{0}_{uI}} {\d \theta_Z} + O^{\chi}_{22}  \frac{\delta H^{0}_{dI}} {\d \theta_Z}
 + O^{\chi}_{32} \frac{\delta  b} {\d \theta_Z} \Biggr)  \nonumber \\
+ c^{\prime}_2 \Biggl( O^{\chi}_{13} \frac{\delta  H^{0}_{uI}} {\d \theta_Z}
+ O^{\chi}_{23} \frac{\delta  H^{0}_{dI}} {\d\theta_Z} + O^{\chi}_{33} \frac{\delta b} {\d \theta_Z} \Biggl),
\eea

\bea
\frac{\d \mathcal{F}^{Z^{\prime}}}{\d \theta_{Z^{\prime}}} = \partial_{\mu}\frac{\d {Z^{\prime}}^{\mu}}{\d \theta_{Z^{\prime}}}-
\xi_{Z^{\prime}} M_{Z^{\prime}}\frac{\d G^{Z'}}{\d \theta_{Z^{\prime}}};
\qquad
\frac{\d Z^{\prime \mu}}{\d \theta_{Z^{\prime}}} = \partial^{\mu} ;
\qquad
\frac{\d G^{Z'}}{\d \theta_{Z^{\prime}}} =  c^{\prime}_1 \frac{\delta G^{0}_{1}}
{\d \theta_{Z^{\prime}}} + c^{\prime}_2 \frac{\delta G^{0}_{2}}{\d \theta_{Z^{\prime}}}.\nn \\
\eea

\bea
\frac{\d \mathcal{F}^{Z^{\prime}}}{\d \theta_{\g}} = \partial_{\mu}\frac{\d {Z^{\prime}}^{\mu}}{\d \theta_{\g}}-
\xi_{Z^{\prime}} M_{Z^{\prime}}\frac{\d G^{Z'}}{\d \theta_{\g}};
\qquad \qquad
\frac{\d Z^{\prime \mu}}{\d \theta_{\g}}= 0;
\qquad \qquad
\frac{\d G^{Z'}}{\d \theta_{\g}}= 0. \nn \\
\eea

\bea
&&\frac{\d \mathcal{F}^{Z^{\prime}}}{\d \theta_{+}} = \partial_{\mu}\frac{\d {Z^{\prime}}^{\mu}}{\d \theta_{+}}-
\xi_{Z^{\prime}} M_{Z^{\prime}}\frac{\d G^{Z'}}{\d \theta_{+}};
\qquad
\frac{\d Z^{\prime \mu}}{\d \theta_{+}}= -i g_2 O^A_{31} W^{-\mu}; \hspace{2cm}
\eea
\bea
&& \frac{\d G^{Z'}}{\d \theta_{+}} =  c^{\prime}_1 \frac{\delta G^{0}_{1}} {\d \theta_{+}} +
c^{\prime}_2 \frac{\delta G^{0}_{2}}{\d \theta_{+}};  \qquad
%
%
\frac{\d \mathcal{F}^{Z^{\prime}}}{\d \theta_{-}} = \partial_{\mu}\frac{\d {Z^{\prime \mu}}}{\d \theta_{-}}-
\xi_{Z^{\prime}} M_{Z^{\prime}}\frac{\d G^{Z'}}{\d \theta_{-}};  \hspace{1.5cm}
\eea
\bea
&& \frac{\d Z^{\prime\mu}}{\d \theta_{-}} = i g_2 O^A_{31} W^{+\mu};
\qquad
\frac{\d G^{Z'}}{\d \theta_{-}} =  c^{\prime}_1 \frac{\delta G^{0}_{1}} {\d \theta_{-}} +
c^{\prime}_2 \frac{\delta G^{0}_{2}}{\d \theta_{-}}.  \hspace{2cm}
\eea


\bea
\frac{\d \mathcal{F}^{A_{\g}}} {\d \theta_{Z}} = \partial_{\mu}\frac{\d A^{\mu}_{\g}} {\d \theta_{Z}};
&\qquad \qquad&
\frac{\d A^{\mu}_{\g}} {\d \theta_{Z}} = 0.
\eea
\bea
\frac{\d \mathcal{F}^{A_{\g}}} {\d \theta_{Z'}} = \partial_{\mu}\frac{\d A^{\mu}_{\g}} {\d \theta_{Z'}};
&\qquad \qquad&
\frac{\d A^{\mu}_{\g}} {\d \theta_{Z'}} = 0.
\eea
\bea
\frac{\d \mathcal{F}^{A_{\g}}} {\d \theta_{\g}} =  \partial_{\mu}\frac{\d A^{\mu}_{\g}} {\d \theta_{\g}};
&\qquad \qquad&
\frac{\d A^{\mu}_{\g}} {\d \theta_{\g}} = \partial^{\mu}.
\eea
\bea
\frac{\d \mathcal{F}^{A_{\g}}} {\d \theta_{+}} = \partial_{\mu}\frac{\d A^{\mu}_{\g}} {\d \theta_{+}};
&\qquad \qquad &
\frac{\d A^{\mu}_{\g}} {\d \theta_{+}} = -i g_2 O^A_{11} W^{-\mu}.
\eea
\bea
\frac{\d \mathcal{F}^{A_{\g}}} {\d \theta_{-}} = \partial_{\mu}\frac{\d A^{\mu}_{\g}} {\d \theta_{-}};
&\qquad \qquad&
\frac{\d A^{\mu}_{\g}} {\d \theta_{-}} = i g_2 O^A_{11} W^{+\mu}.
\eea
For $W^+$ in the FP lagrangean we have the contributions

\bea
&&\frac{\d \mathcal{F}^{W^{+\mu}}} {\d \theta_{Z}} = \partial_{\mu}\frac{\d {W^{+\mu}}} {\d \theta_{Z}}+
i \xi_{W} M_{W} \frac{\d G^+}{\d \theta_{Z}}; \\
&&\frac{\d {W^{+\mu}}} {\d \theta_{Z}} = -ig_2 O^A_{21} W^{+\mu};
\qquad
\frac{\d G^+}{\d \theta_{Z}} = \sin \b \frac{\d H^+_{u}} {\d \theta_{Z}} + \cos \b \frac{\d H^+_{d}}
{\d \theta_{Z}}; \nn \\
\eea
\bea
&& \frac{\d H^+_{u}} {\d \theta_{Z}} = - \frac{i}{2} f^{W}_{2u} (\sin \b G^+ - \cos \b H^+);\\
&&\frac{\d H^+_{d}} {\d \theta_{Z}} = - \frac{i}{2} f^{W}_{2d} (\cos \b G^+ + \sin \b H^+); \\
&&f^{W}_{2u,d} = g_2 O^A_{21} + g_Y O^A_{22} + g_B q^B_{u,d} O^A_{23}.
\eea
%
\bea
&&\frac{\d \mathcal{F}^{W^{+\mu}}} {\d \theta_{Z^{\prime}}} = \partial_{\mu}\frac{\d {W^{+\mu}}} {\d \theta_{Z^{\prime}}}+
i \xi_{W} M_{W} \frac{\d G^+}{\d \theta_{Z^{\prime}}}; \\
&&\frac{\d {W^{+\mu}}} {\d \theta_{Z^{\prime}}} =  -ig_2 O^A_{31} W^{+\mu};
\qquad \qquad \qquad
\frac{\d G^+}{\d \theta_{Z^{\prime}}} = \sin \b \frac{\d H^+_u} {\d \theta_{Z'}} +
\cos \b \frac{\d H^+_d} {\d \theta_{Z^{\prime}}};  \\
&&\frac{\d H^+_u} {\d \theta_{Z^{\prime}}} =  - \frac{i}{2} f^{W}_{3u} (\sin \b G^+ - \cos \b H^+);
 \qquad
\frac{\d H^+_d} {\d \theta_{Z^{\prime}}} = - \frac{i}{2} f^{W}_{3d} (\cos \b G^+ + \sin \b H^+); \nn \\
\\
&&f^{W}_{3u,d} = g_2 O^A_{31} + g_Y O^A_{32} + g_B q^B_{u,d} O^A_{33}.
\eea
%
\bea
&&\frac{\d \mathcal{F}^{W^{+\mu}}} {\d \theta_{\g}} = \partial_{\mu}\frac{\d {W^{+\mu}}} {\d \theta_{\g}}+
i \xi_{W} M_{W} \frac{\d G^+}{\d \theta_{\g}};\\
&& \frac{\d {W^{+\mu}}} {\d \theta_{\g}} = -ig_2 O^A_{11} W^{+\mu};
 \qquad
\frac{\d G^+}{\d \theta_{\g}} = \sin \b \frac{\d H^+_u} {\d \theta_{\g}} + \cos \b \frac{\d H^+_d} {\d \theta_{\g}}; \\
&& \frac{\d H^+_u} {\d \theta_{\g}} = - \frac{i}{2} f^{W}_{1u} (\sin \b G^+ - \cos \b H^+);
 \qquad
\frac{\d H^+_d} {\d \theta_{\g}} = - \frac{i}{2} f^{W^+}_{1d} (\cos \b G^+ + \sin \b H^+); \nn \\
\\
&& f^{W}_{1u,d} = g_2 O^A_{11} + g_Y O^A_{12} + g_B q^B_{u,d} O^A_{13}.
\eea
%
 \bea
&& \frac{\d \mathcal{F}^{W^{+\mu}}} {\d \theta_{+}} = \partial_{\mu}\frac{\d {W^{+\mu}}} {\d \theta_{+}} +
i \xi_{W} M_{W} \frac{\d G^+}{\d \theta_{+}};
\qquad
\frac{\d G^+}{\d \theta_{+}} = \sin \b \frac{\d H^+_u} {\d \theta_{+}} + \cos \b \frac{\d H^+_d} {\d \theta_{+}}; \nn \\
&& \frac{\d {W^{+\mu}}} {\d \theta_{+}} = \partial^{\mu} + ig_2(O^A_{11} A^{\mu}_{\g} + O^A_{21} Z^{\mu} +
O^A_{31} Z^{\prime \mu});
\eea
\bea
\frac{\d H^+_u} {\d \theta_{+}} &=& - \frac{i}{\sqrt {2}} g_2 v_u - \frac{i}{2} g_2  \biggl \{ (\sin \a h^0 - \cos \a H^0)+ \nonumber \\
&& i \biggl[ O^{\chi}_{11}+ \biggl( \frac{O^{\chi}_{12}c'_2 - O^{\chi}_{13}c'_1} {c_1 c'_2 - c'_1 c_2}\biggr)z+
\biggl( \frac{- O^{\chi}_{12}c_2 + O^{\chi}_{13}c_1} {c_1 c'_2 - c'_1 c_2} \biggr) z' \biggr] \biggr \}; \nn \\ \\
\frac{\d H^+_d} {\d \theta_{+}} &=& - \frac{i}{\sqrt {2}} g_2 v_d - \frac{i}{2} g_2  \biggl \{ (\cos \a h^0 + \sin \a H^0)+ \nonumber \\
&& i \biggl[ O^{\chi}_{21}+ \biggl( \frac{O^{\chi}_{22}c'_2 - O^{\chi}_{23}c'_1} {c_1 c'_2 - c'_1 c_2}\biggr)z+
\biggl( \frac{- O^{\chi}_{22}c_2 + O^{\chi}_{23}c_1} {c_1 c'_2 - c'_1 c_2} \biggr) z' \biggr] \biggr \}. \nn \\
\eea
\bea
&& \frac{\d \mathcal{F}^{W^{+\mu}}} {\d \theta_{-}} = \partial_{\mu}\frac{\d {W^{+\mu}}} {\d \theta_{-}} +
i \xi_{W} M_{W} \frac{\d G^+}{\d \theta_{-}}; \\
&& \frac{\d {W^{+\mu}}} {\d \theta_{-}} = 0;
\qquad
\frac{\d G^+}{\d \theta_{-}} = \sin \b \frac{\d H^+_u} {\d \theta_{-}} + \cos \b \frac{\d H^+_d} {\d \theta_{-}}; \\
&& \frac{\d H^+_u} {\d \theta_{-}} = 0;
\qquad
\frac{\d H^+_d} {\d \theta_{-}} = 0.
\eea
For  $W^-$ we get
\bea
&& \frac{\d \mathcal{F}^{W^{-\mu}}} {\d \theta_{Z}} = \partial_{\mu}\frac{\d {W^{-\mu}}} {\d \theta_{Z}} -
i \xi_{W} M_{W} \frac{\d G^-}{\d \theta_{Z}}; \\
&& \frac{\d {W^{-\mu}}} {\d \theta_{Z}} = ig_2 O^A_{21} W^{-\mu};
 \qquad
\frac{\d G^-}{\d \theta_{Z}} = \sin \b \frac{\d H^-_u} {\d \theta_{Z}} + \cos \b \frac{\d H^-_d} {\d \theta_{Z}}; \\
&& \frac{\d H^-_u} {\d \theta_{Z}} = \frac{i}{2} f^{W^+}_{2u} (\sin \b G^+ - \cos \b H^+);
\qquad
\frac{\d H^-_d} {\d \theta_{Z}} = \frac{i}{2} f^{W^+}_{2d} (\cos \b G^+ + \sin \b H^+). \nn \\
\eea
\bea
&& \frac{\d \mathcal{F}^{W^{-\mu}}} {\d \theta_{Z^{\prime}}} = \partial_{\mu}\frac{\d {W^{-\mu}}} {\d \theta_{Z^{\prime}}}-
i \xi_{W} M_{W} \frac{\d G^-}{\d \theta_{Z^{\prime}}}; \\
&& \frac{\d {W^{-\mu}}} {\d \theta_{Z^{\prime}}} = ig_2 O^A_{31} W^{-\mu};
\qquad
\frac{\d G^-}{\d \theta_{Z^{\prime}}} =  \sin \b \frac{\d H^-_u} {\d \theta_{Z^{\prime}}} +
\cos \b \frac{\d H^-_d} {\d \theta_{Z^{\prime}}}; \\
&& \frac{\d H^-_u} {\d \theta_{Z^{\prime}}} = \frac{i}{2} f^{W}_{3u} (\sin \b G^+ - \cos \b H^+);
 \qquad
\frac{\d H^-_d} {\d \theta_{Z^{\prime}}} = \frac{i}{2} f^{W}_{3d} (\cos \b G^+ + \sin \b H^+).\nn \\
\eea
\bea
&& \frac{\d \mathcal{F}^{W^{-\mu}}} {\d \theta_{\g}} = \partial_{\mu}\frac{\d {W^{-\mu}}} {\d \theta_{\g}}-
i \xi_{W} M_{W} \frac{\d G^-}{\d \theta_{\g}}; \\
&& \frac{\d {W^{-\mu}}} {\d \theta_{\g}} = ig_2 O^A_{11} W^{-\mu};
 \qquad
\frac{\d G^-}{\d \theta_{\g}} =  \sin \b \frac{\d H^-_u} {\d \theta_{\g}} + \cos \b \frac{\d H^-_d} {\d \theta_{\g}};
\nn \\ \\
&& \frac{\d H^-_u} {\d \theta_{\g}} = \frac{i}{2} f^{W}_{1u} (\sin \b G^+ - \cos \b H^+);
\qquad
\frac{\d H^-_d} {\d \theta_{\g}} = \frac{i}{2} f^{W^+}_{1d} (\cos \b G^+ + \sin \b H^+). \nn \\
\eea
\bea
&& \frac{\d \mathcal{F}^{W^{-\mu}}} {\d \theta_{+}} = \partial_{\mu}\frac{\d {W^{-\mu}}} {\d \theta_{+}}-
i \xi_{W} M_{W} \frac{\d G^-}{\d \theta_{+}}; \\
&& \frac{\d {W^{-\mu}}} {\d \theta_{+}} = 0;
 \qquad
\frac{\d G^-}{\d \theta_{+}} = \sin \b \frac{\d H^-_u} {\d \theta_{+}} + \cos \b \frac{\d H^-_d} {\d \theta_{+}}; \\
&&\frac{\d H^-_u} {\d \theta_{+}} = 0;
\qquad
\frac{\d H^-_d} {\d \theta_{+}} = 0.
\eea
\bea
&& \frac{\d \mathcal{F}^{W^{-\mu}}} {\d \theta_{-}} = \partial_{\mu}\frac{\d {W^{-\mu}}} {\d \theta_{-}}-
i \xi_{W} M_{W} \frac{\d G^-}{\d \theta_{-}}; \\
&& \frac{\d {W^{-\mu}}} {\d \theta_{-}} = \partial^{\mu} - ig_2(O^A_{11} A^{\mu}_{\g} + O^A_{21} Z^{\mu} + O^A_{31} {Z^{\prime}}^{\mu});\\
&&\frac{\d G^-}{\d \theta_{-}} =  \sin \b \frac{\d H^-_u} {\d \theta_{-}} + \cos \b \frac{\d H^-_d} {\d \theta_{-}};
\eea
\bea
\frac{\d H^-_u} {\d \theta_{-}} &=&  \frac{i}{\sqrt {2}} g_2 v_u + \frac{i}{2} g_2  \biggl \{ (\sin \a h^0 - \cos \a H^0)
\nonumber \\
&&- i \biggl[ O^{\chi}_{11}+ \biggl( \frac{O^{\chi}_{12}c'_2 - O^{\chi}_{13}c'_1} {c_1 c'_2 - c'_1 c_2}\biggr)z+
\biggl( \frac{- O^{\chi}_{12}c_2 + O^{\chi}_{13}c_1} {c_1 c'_2 - c'_1 c_2} \biggr) z' \biggr] \biggr \}; \nn \\ \\
\frac{\d H^-_d} {\d \theta_{-}} &=&  \frac{i}{\sqrt {2}} g_2 v_d + \frac{i}{2} g_2  \biggl \{ (\cos \a h^0 + \sin \a H^0) \nonumber \\
&&- i \biggl[ O^{\chi}_{21}+ \biggl( \frac{O^{\chi}_{22}c'_2 - O^{\chi}_{23}c'_1} {c_1 c'_2 - c'_1 c_2}\biggr)z+
\biggl( \frac{- O^{\chi}_{22}c_2 + O^{\chi}_{23}c_1} {c_1 c'_2 - c'_1 c_2} \biggr) z' \biggr] \biggr \} . \nn \\
\eea

\end{document}